\documentclass[11pt]{article}
\usepackage{eurosym}
\usepackage{amsfonts}
\usepackage{amssymb}
\usepackage{graphicx}
\usepackage{amsmath}
\usepackage{makeidx}
\usepackage{indentfirst}
\usepackage[T1]{fontenc}
\usepackage[utf8]{inputenc}

\newcounter{resultnum}[section]
\newcounter{conclusionnum}[section]
\newcounter{conditionnum}[section]
\newcounter{conjecturenum}[section]
\newcounter{examplenum}[section]
\newcounter{exercisenum}[section]
\newcounter{lemmanum}[section]
\newcounter{notationnum}[section]
\newcounter{theoremnum}[section]
\newcounter{definitionnum}[section]
\newcounter{corollarynum}[section]
\newcounter{remarknum}[section]
\newcounter{propositionnum}[section]
\newcounter{acknowledgementnum}[section]
\newcounter{algorithmnum}[section]
\newcounter{axiomnum}[section]
\newcounter{casenum}[section]
\newcounter{claimnum}[section]
\newcounter{summarynum}[section]
\newcounter{problemnum}[section]
\begin{document}

\title{Nonassociative gauge gravity theories with R-flux star products and Batalin-Vilkovisky quantization in algebraic quantum field theory}
\date{November 8, 2024}
\author{ \textbf{Sergiu I. Vacaru} \thanks{%
emails: sergiu.vacaru@fulbrightmail.org ; sergiu.vacaru@gmail.com } \\
{\small \textit{Department of Physics, California State University at
Fresno, Fresno, CA 93740, USA;\ and}}  \\
{\small \textit{Taras Shevchenko National University of Kyiv, Astronomical Observatory, Kyiv, Ukraine}}
}
\maketitle

\begin{abstract}
Nonassociative modifications of general relativity, GR, and quantum gravity, QG, models naturally arise as star product and R-flux deformations considered in string/ M-theory. Such nonassociative and noncommutative geometric and quantum information theories were formulated on phase spaces defined as cotangent Lorentz bundles enabled with nonassociative symmetric and nonsymmetric metrics and nonlinear and linear connection structures. We outline the analytic methods and proofs that corresponding geometric flow evolution and dynamical field equations can be decoupled and integrated in certain general off-diagonal forms.  New classes of solutions describing nonassociative black holes, wormholes, and locally anisotropic cosmological configurations are constructed using such methods. We develop the Batalin-Vilkovisky, BV, formalism for quantizing modified gravity theories, MGTs, involving twisted star products and semi-classical models of nonassociative gauge gravity with de Sitter/affine/ Poincar\'{e} double structure groups. Such theories can be projected on Lorentz spacetime manifolds in certain forms equivalent to GR or MGTs with torsion generalizations etc. We study the properties of the classical and quantum BV operators for nonassociative phase spaces and nonassociative gauge gravity. Recent results and methods from algebraic QFT are generalized to involve nonassociative star product deformations of the anomalous master Ward identity. Such constructions are elaborated in a nonassociative BV perspective and for developing non-perturbative methods in QG.

\end{abstract}

\tableofcontents



\section{Introduction}

\label{sec1}Quantization of gravity is one of the big open problems in
modern physics. In the most general and rigorous mathematical form, and in
the framework of a unification theory, the theory of quantum gravity, QG, is
approached in string/ M-theory \cite{string1,string2,string3}. Certain models of 
nonassociative and noncommutative modified gravity theories, MGTs, are
expected to arise in low-energy limits of string theories \cite%
{luest10,blumenhagen10,condeescu13,blumenhagen13}. For so-called R-flux
deformations, such MGTs are constructed using the concept of twisted star
product, $\star $, \cite{drinf89,connes97}. We follow an  approach when $\star $-products are  defined on nonassociative phase spaces in certain forms extending Einstein's gravity, i.e. the general relativity, GR,
theory \cite{blumenhagen16,aschieri17}. Nonassociative and noncommutative
algebraic, geometric and topological structures are studied in modern
mathematics. Various applications are considered  in quantum field theories, QFTs,
and used for elaborating new methods of quantization. 

\vskip4pt In a series of partner works \cite{partner02,partner04,partner05,partner06,partner07}, we proved that physically important nonassociative MGTs formulated on phase spaces are described by systems of nonlinear partial differential equations, PDEs, which can be decoupled and integrated in certain general  off-diagonal forms. This is possible if  the so-called anholonomic frame and connection deformation method, AFCDM, is applied. We constructed and studied physical properties of new classes of black holes, BHs; wormholes, WHs; and locally anisotropic cosmological solutions encoding nonassociative data. Nonassociative phase space off-diagonal solutions and respective physical models typically are not characterized by thermodynamic variables formulated in the framework of the Bekenstein-Hawking BH paradigm \cite{bek2,haw2}. This is because, in general, such configurations do not involve certain conventional horizons or holographic conditions. Nevertheless, we can formulate a new type of nonassociative geometric flow thermodynamics  \cite{partner04,partner05,partner06,partner07}. It generalizes G. Perelman's constructions defining a new type the statistical and geometric thermodynamics for Ricci flows of Riemannian metrics \cite{perelman1}.  Here, we note that in our  works the aim is not to formulate and proof any nonassociative variant  of the Poincar\'{e}--Thurston conjecture. This is a very difficult and ambiguous problem in modern mathematics  because of existence of different types of nonassociative and noncommutative calculi.  In (pseudo) Riemannian geometric the differential and integral calculus is defined in a unique way. We develop certain approaches when generalized G. Perelman thermodynamic variables are defined in MGTs and  used for characterizing off-diagonal solutions encoding, for instance, nonassociative data. Such models can be quantized by elaborating new methods of quantization. 

\vskip4pt Considering nontrivial nonassociative and noncommutative algebraic
and geometric structures on phase spaces, we substantially modify not only
the GR theory and the standard models in QFT.
Such modifications request the elaboration of a new mathematical formalism
and application of nonholonomic geometric and quantum information methods resulting in nonassociative and noncommutative models of QG. Here we note that the theories with a general twisted star product are
non-variational but can be formulated in abstract geometric forms. For
applications in modern cosmology and astrophysics, we can consider any convenient approach with 
variational, abstract geometric and effective field theories. The corresponding fundamental equations can be formulated in certain parametric forms (for instance, using linear decompositions on the
Planck  and string constants) which allows us to encode nonassociative and noncommutative data. 

\vskip4pt In a partner work \cite{partner07}, a nonassociative gauge gravity
model with double affine or de Sitter gauge structure groups on cotangent
Lorentz bundles were formulated. For projections on nonassociative phase
space bases, those constructions reproduce the results on nonassociative
MGTs from \cite{partner02,partner04,partner05,partner06}. Here we note that
possible implications of nonassociative geometric and classical gravity 
theories distinguishing specific properties of off-diagonal parametric and
physically important solutions were not studied in modern literature on QFT
and QG. Our nonassociative and nonholonomic gauge gravity approach is motivated by the fact
that using a corresponding class of nonholonomic distributions, the
nonassociative Yang-Mills, YM, equations can be decoupled and solved in very general forms and then quantized using standard methods for gauge field theories. The modified YM equations can be formulated as some equivalents of nonassociative star product deformed Einstein equations. The
projections of  nonassociative gravitational YM equations on Lorentz
spacetime manifolds transform into standard Einstein equations. Such equations which may
encode, or not, certain nonasssociative and noncommutative data defining certain effective sources. In a more
general context, we can elaborate on nonassociative gauge gravity models
with nontrivial torsion and nonmetricity fields and try to quantize such
nonassociative MGTs using (non) perturbative methods formulated for
quantizing gauge theories. 

\vskip4pt To quantize associative and commutative theories with local
symmetries (including gauge theories and various MGTs) is convenient to use
the Bechi, Rouet, Stora and Tyutin, BRST, method \cite{brst75,brst76,tyu94},
see recent results and references in \cite{barv18}. It is not clear how this
method can be extended in general variational or nonvariational forms for
nonassociative and noncommutative gauge theories. Nevertheless, we can
elaborate on an abstract geometric formalism for nonassociative gauge and
gravity fields and perform quantization of physically important quasi-stationary or locally
anisotropic cosmological configurations. Such configurations are determined by off-diagonal solutions
with corresponding types of nonlinear symmetries which are additional to the 
corresponding types of gauge symmetries or diffeomorphisms. This also
supposes adding effective and auxiliary fields (ghosts, anti-ghosts, etc.)
in nonassociative geometric form, and even an infinite number of differential and
integral nonassociative and noncommutative calculi (and various types of
geometries) can be elaborated. Using nonholonomic geometric methods,
nonassociative geometric flow and gravitational and matter field equations
can be postulated following algebraic and geometric principles, when
parametric decompositions resulting in effective thermofield and quantum
field and quantum evolution theories can be stated to possess a well-defined
variational calculus. This refers, in particular, to the models with
nonassociative star product R-flux if they are elaborated in parametric form
as a nonassociative gauge gravity when the constructions are similar to
standard BRST ones but with corresponding double affine or de Sitter
structure groups. 

\vskip4pt A generalization of the BRST is known as the Batalin and Vilkovisky, BV, formalism \cite{bv81}. We do not provide a comprehensive list of references on further developments and applications of the BV
formalism but cite \cite{rejzner20,brunetti22}. Those articles and references therei contain  recent reviews and rigorous mathematical approaches to perturbative algebraic QFT (in brief, pAQFT). In this work, the focus is on conceptual problems for developing the BV formalism and certain methods of pAQFT to quantizing nonassociative gauge gravity theories and MGTs of type \cite{partner07,partner04,partner06}. Our main goal is to show that the BV can be applied when   the principles of locality, deformation and homology (formulated in \cite{rejzner20,brunetti22} for Lorentz spacetimes) are extended on nonassociative phase spaces modelled on cotangent Lorentz bundles.  Here, we emphasize that the star products determined by R-flux  involve non-local constructions which became effectively local for certain parametric decompositions. In such an approach, we naturally generalize the BV formalism in abstract nonassociative  geometric form. 

\vskip4pt In this work, we develop a nonlinear functional approach to nonassociative gauge QFT and QG when explicit classes of generic off-diagonal parametric solutions are used for extending the BV formalism in
nonholonomic form on nonassociative phase spaces. The constructions may encode certain prescribed classes of nonassociative and noncommutative algebraic and nonlinear geometric symmetries, parametric decompositions, effective gravitational polarizations and deformation of horizons (if such ones are prescribed for some classes of solutions). Such configurations define nontrivial gravitational vacuum structures when certain quasi-classical limits are used for developing perturbative and non-perturbative schemes of quantization. This goes beyond traditional schemes of quantization with certain completely determined by Lagrange or Hamilton, S-matrix methods etc. Those approaches are not related directly to the properties of physically important systems of nonlinear PDEs and the properties of off-diagonal solutions depending on
various generating and integrating functions, physical constants, generating sources and corresponding (non) linear symmetries. We argue that nonassociative gauge gravity theories can be elaborated both in classical and quantum forms by reformulating the  BV formalism and the mentioned three principles (of locality, deformation and homology). This is possible for pAQFTs defined on nonholonomic cotangent Lorentz bundles, i.e. on nonassociative phase spaces, endowed with parametric star product structure.  

\vskip4pt The paper is structured as follows:\ In section \ref{sec02}, we provide an introduction to (classical) nonassociative gauge gravity with star product and R-flux deformations and state the respective conditions for
extracting associative and commutative gauge and gravity theories on phase spaces. We consider a general ansatz for generating quasi-stationary off-diagonal solutions in such theories and emphasize that corresponding configurations can be dualized (on a time-like coordinate) for constructing locally anisotropic cosmological models. The goal of section \ref{sec03} is to study the nonassociative dynamics and nonlinear symmetries of the classical gauge de Sitter gravity and the definition of nonassociative classical BV operator and the M\o ller maps. Section \ref{sec04} is devoted to the quantization of nonassociative gauge de Sitter gravity and related
renormalization procedures using the quantum BV operator. Perturbative and non-perturbative methods for nonassociative gauge gravity are elaborated and an example of how to perform the BV quantization of nonassociative 8-d BH configurations is analyzed. Conclusions and perspectives are considered in section \ref{sec05}. In the Appendix, we outline the necessary formulas for generating off-diagonal quasi-stationary solutions on nonassociative phase spaces.

\section{Nonassociative gauge gravity models with star product}

\label{sec02}In this section, we outline necessary results on nonassociative MGTs and gauge gravity with twisted star products determined by R-flux deformations in string theory. Details in are provided in \cite%
{partner02,partner04,partner06,partner07} and references therein.

\subsection{Nonassociative phase spaces modelled on cotangent Lorentz bundles%
}

We begin with the geometry of commutative phase spaces, which can be
constructed on tangent Lorentz bundles $\mathcal{M}=T\mathbf{V} $, or
cotangent Lorentz bundles $\ ^{\shortmid }\mathcal{M}=T^{\ast }\mathbf{V}$,
of a Lorentzian spacetime manifold $\mathbf{V}$ of signature $(+++-).$ In
this work, we shall consider only geometric and physical models defined on $%
\ ^{\shortmid }\mathcal{M}$. Respective spacetime and momentum
coordinates are denoted as $\ ^{\shortmid }u=(x,p)=\{\ ^{\shortmid
}u^{\alpha }=(x^{i},p_{a})\},$ for indices $i,j,k,...=1,2,3,4;
a,b,c,...=5,6,7,8$ and $x^{4}=t,$ or $y^{4}=t,$ is a time like coordinate as for a base Lorentz spacetime. A
N-connection structure is defined as a nonholonomic (equivalently,
anholonomic, i.e. non-integrable) splitting, $\ ^{\shortmid }\mathbf{N}%
:TT^{\ast }\mathbf{V}=hT^{\ast }\mathbf{V}\oplus cT^{\ast }\mathbf{V}$,
where $\oplus $ denotes the Whitney direct sum.\footnote{%
We use the duality label "$\ ^{\shortmid }$" and "boldface" symbols to state
the geometric constructions can be adapted to an N-connection splitting. For
such N-adapted models, tensors transform into d-tensors, vectors transform
into d-vectors and connections transform into d-connections, where "d" means
distinguished by a N-connection h-c-splitting. Here we note that we
developed in nonholonomic form for nonassociative geometry and gravity \cite%
{partner02} an abstract (index and coordinate-free) geometric formalism when any
N-connection can be defined equivalently as a nonholonomic, equivalently,
anholonomic, or non-integrable distribution. Our approach generalizes for
spaces with nontrivial N-connection structure the abstract and index
formalism for GR \cite{misner}. Such an N-connection and adapted distortions
of linear connections can be correspondingly introduced with the goal of
decoupling certain systems of nonlinear PDEs.} A metric structure $\
^{\shortmid }\mathbf{g}=(hg,cg)=\{\ ^{\shortmid }\mathbf{g}_{\alpha
\beta}=(g_{ij},\ ^{\shortmid }g^{ab})\}$ on $\ ^{\shortmid }\mathcal{M}$ is
of local signature $(+++-;+++-).$ Here, "h" states a corresponding horizontal splitting and "c" is for a co-fiber, i.e. co-vertical splitting of the geometric objects and necessary indices on a cotangent bundle. N-connections can be introduced on phase spaces and spacetimes in arbitrary geometric form or for some additional
assumptions that they are associated to certain nonlinear gauge fields; or related to
certain off-diagonal terms of metrics; or to some linear N-elongations of
local partial derivatives/ differentials and respective systems of reference. 

\vskip4pt If a phase space $\ ^{\shortmid }\mathcal{M}$ is enabled with a
N-connection structure, we can introduce the concept of distinguished
connection, d-connection $\ ^{\shortmid }\mathbf{D}=(h\ ^{\shortmid }D,c\
^{\shortmid}D).$ Such a d-connection is a  linear connection which
preserves a N-connection structure (in our case, nonholonmic 4+4 splitting) under
affine linear transports.\footnote{%
We note that a Levi Civita connection $\ ^{\shortmid }\mathbf{\nabla }$ (in brief, LC-connection; by
definition, it is metric compatible and torsionless) is not a d-connection
because it is not adapted to a N-connection structure. Nevertheless, an
N-adapted distortion formula $\ ^{\shortmid }\mathbf{D}=\ ^{\shortmid
}\nabla +\ ^{\shortmid }\mathbf{Z}$ can be defined; when $\ ^{\shortmid }%
\mathbf{Z}$ is the distortion d-tensor encoding contributions from
respective torsion of $\ ^{\shortmid }\mathbf{D},$ and non-metricity
d-tensor, $\ ^{\shortmid}\mathbf{Q}:=\mathbf{Dg},$ for $\
^{\shortmid}\nabla \ ^{\shortmid }\mathbf{g}=0$. The abstract and
coefficient formulas of d-adapted geometric objects (in general nonassociative form) can be found in \cite{partner02,partner04,partner06}.}
To apply the AFCDM for constructing exact and parametric solutions in MGTs
is convenient to work with the so-called canonical d-connection $\
^{\shortmid }\widehat{\mathbf{D}}.$ It is completely determined by the
coefficients of $\ ^{\shortmid }\mathbf{g}$ and $\ ^{\shortmid }\mathbf{N}$
to be metric compatible, $\ ^{\shortmid }\widehat{\mathbf{D}}\ ^{\shortmid }%
\mathbf{g}=0,$ but contains a nontrivial d-torsion structure induced as a
nonholonomic effect, when the canonical d-torsion tensor $\ ^{\shortmid}%
\widehat{\mathcal{T}}=\{hh\ ^{\shortmid }\widehat{\mathcal{T}}=0; cc\
^{\shortmid}\widehat{\mathcal{T}}=0,$ when $hc \ ^{\shortmid}\widehat{%
\mathcal{T}}\neq 0\}\neq 0.$\footnote{%
Such a d-torsion is different from that in the Einstein-Cartan theory or
other torsions in string and gauge gravity. In our case, the data $(\
^{\shortmid }\mathbf{g,}\ ^{\shortmid }\mathbf{N}, \ ^{\shortmid }\widehat{%
\mathbf{D}})$ can be considered as certain nonholonomic geometric variables
which allow us to solve physically important systems of nonlinear PDEs.
Using canonical distortion relations, all results can be re-defined
equivalently for LC-configurations $(\ ^{\shortmid }\mathbf{g}, \
^{\shortmid }\mathbf{\nabla })$ if it will be necessary.}

\vskip4pt To prove certain general decoupling and integration properties of
modified Einstein equations and generating generic off-diagonal solutions
encoding nonassociative or locally anisotropic data we had to develop also a
nonholonomic dyadic, 2-d, shell by shell oriented decomposition formalism 
\cite{partner02}. In such cases, we write $\ _{s}^{\shortmid }\mathcal{M}$
for a phase space $\ ^{\shortmid }\mathcal{M}$ enabled with a conventional
(2+2)+(2+2) splitting with four oriented shells $s=1,2,3,4.$ For such
s-decompositions, the N-connection is defined 
\begin{equation}
\ _{s}^{\shortmid }\mathbf{N}:\ \ _{s}T\mathbf{T}^{\ast }\mathbf{V}=\
^{1}hT^{\ast }V\oplus \ ^{2}vT^{\ast }V\oplus \ ^{3}cT^{\ast }V\oplus \
^{4}cT^{\ast }V,\mbox{  for }s=1,2,3,4.  \label{scon}
\end{equation}%
In a local coordinate basis, the N-connection (\ref{scon}) is characterized
by a corresponding set of coefficients $\ _{s}^{\shortmid }\mathbf{N}=\{\
^{\shortmid }N_{\ i_{s}a_{s}}(\ ^{\shortmid }u)\}$ and such coefficients
allow us to introduce certain N-elongated bases (N-/ s-adapted bases as
linear N-operators): 
\begin{eqnarray}
\ ^{\shortmid }\mathbf{e}_{\alpha _{s}}[\ ^{\shortmid }N_{\ i_{s}a_{s}}]
&=&(\ ^{\shortmid }\mathbf{e}_{i_{s}}=\ \frac{\partial }{\partial x^{i_{s}}}%
-\ ^{\shortmid }N_{\ i_{s}a_{s}}\frac{\partial }{\partial p_{a_{s}}},\ \
^{\shortmid }e^{b_{s}}=\frac{\partial }{\partial p_{b_{s}}})\mbox{ on }\
_{s}T\mathbf{T}_{\shortmid }^{\ast }\mathbf{V;}  \notag \\
\ ^{\shortmid }\mathbf{e}^{\alpha _{s}}[\ ^{\shortmid }N_{\ i_{s}a_{s}}]
&=&(\ ^{\shortmid }\mathbf{e}^{i_{s}}=dx^{i_{s}},\ ^{\shortmid }\mathbf{e}%
_{a_{s}}=d\ p_{a_{s}}+\ ^{\shortmid }N_{\ i_{s}a_{s}}dx^{i_{s}})\mbox{ on }\
\ _{s}T^{\ast }\mathbf{T}_{\shortmid }^{\ast }\mathbf{V.}  \label{nadapb}
\end{eqnarray}
Having prescribed a nonholonomic s-structure, we can express any metric or
d-metric as a s-metric 
\begin{equation*}
\ _{s}^{\shortmid }\mathbf{g}=\{\ ^{\shortmid }\mathbf{g}_{\alpha _{s}\beta
_{s}}\}= (h_{1}\ ^{\shortmid }\mathbf{g},~v_{2}\ ^{\shortmid }\mathbf{g},\
c_{3}\ ^{\shortmid }\mathbf{g,}c_{4}\ ^{\shortmid }\mathbf{g})\in T\mathbf{T}%
^{\ast }\mathbf{V}\otimes T\mathbf{T}^{\ast }\mathbf{V,}
\end{equation*}
when $\ ^{\shortmid }\mathbf{g}_{\alpha _{s}\beta _{s}}(\ _{s}^{\shortmid
}u)\ \ ^{\shortmid }\mathbf{e}^{\alpha _{s}}\otimes _{s}\ ^{\shortmid }%
\mathbf{e}^{\beta _{s}}= \{\ \ ^{\shortmid }\mathbf{g}_{\alpha _{s}\beta
_{s}}=(\ \ ^{\shortmid }\mathbf{g}_{i_{1}j_{1}},\ \ ^{\shortmid }\mathbf{g}%
_{a_{2}b_{2}},\ \ ^{\shortmid }\mathbf{g}^{a_{3}b_{3}},\ \ ^{\shortmid }%
\mathbf{g}^{a_{4}b_{4}})\},$ for $\ ^{\shortmid }\mathbf{e}^{\alpha _{s}}$ (%
\ref{nadapb}) chosen in s-adapted form. In this paper, we shall omit the
bulk of shell index formulas and explicit geometric proofs of classical
formulas, equations and solutions, which can be found in our partner works 
\cite{partner02,partner04,partner05,partner06,partner07}.

\vskip4pt A nonassociative (twisted) star product from \cite%
{blumenhagen16,aschieri17} can be defined in a form involving N-elongated
differential operators $\ ^{\shortmid }\mathbf{e}_{i_{s}}$ (\ref{nadapb})
acting on some functions $\ f(x,p)$ and $\ q(x,p)$ defined on $\
_{s}^{\shortmid }\mathcal{M},$ see details in \cite{partner02,partner04}.
Such a s-adapted star product $\star _{s}$ can be computed as 
\begin{eqnarray}
f\star _{s}q &:= &\cdot \lbrack \mathcal{F}_{s}^{-1}(f,q)]  \label{starpn} \\
&=&\cdot \lbrack \exp (-\frac{1}{2}i\hbar (\ ^{\shortmid }\mathbf{e}%
_{i_{s}}\otimes \ ^{\shortmid }e^{i_{s}}-\ ^{\shortmid }e^{i_{s}}\otimes \
^{\shortmid }\mathbf{e}_{i_{s}})+\frac{i\mathit{\ell }_{s}^{4}}{12\hbar }%
R^{i_{s}j_{s}a_{s}}(p_{a_{s}}\ ^{\shortmid }\mathbf{e}_{i_{s}}\otimes \
^{\shortmid }\mathbf{e}_{j_{a}}-\ ^{\shortmid }\mathbf{e}_{j_{s}}\otimes
p_{a_{s}}\ ^{\shortmid }\mathbf{e}_{i_{s}}))]f\otimes q  \notag \\
&=&f\cdot q-\frac{i}{2}\hbar \lbrack (\ ^{\shortmid }\mathbf{e}_{i_{s}}f)(\
^{\shortmid }e^{i_{s}}q)-(\ ^{\shortmid }e^{i_{s}}f)(\ ^{\shortmid }\mathbf{e%
}_{i_{s}}q)]+\frac{i\mathit{\ell }_{s}^{4}}{6\hbar }%
R^{i_{s}j_{s}a_{s}}p_{a_{s}}(\ ^{\shortmid }\mathbf{e}_{i_{s}}f)(\
^{\shortmid }\mathbf{e}_{j_{s}}q)+\ldots ..  \notag
\end{eqnarray}%
The antisymmetric coefficients $R^{i_{s}j_{s}a_{s}}$ define the
nonassociative part of the star product (in brief, we can write $\star $ or
correspondingly $\star _{N})$.  The string length constant $\mathit{\ell }$
characterizes the R-flux contributions from string theory. In formulas (\ref%
{starpn}), the tensor product $\otimes $ can be written also in an s-adapted
form $\otimes _{s}$. Explicit computations for R-flux deformations of
s-adapted geometric objects and (physical) equations, can be adapted and
classified with respect to decompositions on small parameters $\hbar $
(stating noncommutative properties) and $\kappa =\mathit{\ell }%
_{s}^{3}/6\hbar .$ For parametric decompositions, all tensor products turn
into usual multiplications which is very important for computing classical
and quantum effects in theories encoding nonassociative data. 

\vskip4pt For any phase space $\ _{s}^{\shortmid }\mathcal{M}$, we can lift
s-adapted geometric objects on the total space of a vector bundle $\
_{s}^{\shortmid }\mathcal{E}(\ _{s}^{\shortmid }\mathcal{M})$ and call $\
_{s}^{\shortmid }\mathcal{E}$ a s-vector bundle, see details and references
in \cite{partner07}. A star product (\ref{starpn}) deforms $\
_{s}^{\shortmid }\mathcal{E}(\ _{s}^{\shortmid }\mathcal{M})$ on $\
_{s}^{\shortmid }\mathcal{M}$ into respective nonassociative ones labeled by
a $\star $-symbol, $\ _{s}^{\shortmid }\mathcal{E}^{\star }$ on $\
_{s}^{\shortmid }\mathcal{M}^{\star }.$ Such geometric constructions were
considered in noncommutative form in \cite{svnc00,sv00} for the so-called
N-adapted Seiberg-Witten star product $\ast ,$ or $\ast _{N}$. Nevertheless,
to introduce nonassociative structures is not just a formal changing of, for
instance, $\ast _{N}$ into a $\star _{s}$ (\ref{starpn}) without R-flux
terms. For nonassociative models with R-flux $\star $-deformations, $\star :%
\mathbf{g\rightarrow g}^{\star }=(\mathbf{\breve{g}}^{\star },\mathbf{\check{%
g}}^{\star }),$ the $\star $-metrics contain nonassociative symmetric, $%
\breve{g}^{\star }$, and nonassociative nonsymmetric, $\mathbf{\check{g}}%
^{\star }$, components; the procedure of inverting metrics became nonlinear
and sophisticated. In general, all components of geometric s-objects depend
both on spacetime and momentum-like variables. Typically, the non-symmetry
of metrics is not considered in noncommutative $\ast $-theories. 

\vskip4pt Abstract geometric and tedious index/coordinate computations of
the fundamental geometric and physical objects on $\ _{s}^{\shortmid }%
\mathcal{M}^{\star }$ allow us to express all-important formulas for the
"star" d-metrics, d-connections, d-torsions, d-curvatures, etc.,  into certain 
$\hbar $ and $\kappa $-parametric forms which are provided in \cite%
{partner02,partner04}. Such computations can be considered for defining $%
\star $-versions of LC-connections, $\mathbf{\nabla \rightarrow \nabla }%
^{\star };$ or star product deformations of arbitrary d-connections, $\
^{\shortmid }\mathbf{D} \rightarrow \ ^{\shortmid }\mathbf{D}^{\star },$ or
canonical s-connections, $\ _{s}^{\shortmid }\widehat{\mathbf{D}}
\rightarrow \ _{s}^{\shortmid }\widehat{\mathbf{D}}^{\star },$ etc.
Correspondingly, we can compute the parametric and s-adapted forms for a
star product deformation of the Ricci tensor, or canonical s-tensor, $%
\mathcal{R}ic^{\star }[\mathbf{g}^{\star }, {\nabla }^{\star }]$ or $%
\widehat{\mathcal{R}}ic^{\star }[\mathbf{g}^{\star },\widehat{\mathbf{D}}%
^{\star }]$ etc. The $\hbar $- and $\kappa $-parametric terms determined by $%
\star $ deformations of pseudo-Riemannian metrics can be re-defined
equivalently as certain effective sources encoding nonassociative/
noncommutative data from string theory.

\subsection{Associative and commutative gauge gravity on phase spaces}

We outline necessary results on a model of gauge gravity theory on $\
^{\shortmid }\mathcal{M}$ when the gauge structure group is $\mathcal{G}%
r=\left( SO(4,1),SO(4,1\right) )$. In this theory, the de Sitter group $%
SO(4,1)$ may encode consequent nonlinear extensions of the affine structure
group $Af(4,1)$ and the Poincar\'{e} group $ISO(3,1)$, see details and
physical motivation in \cite{svnc00,sv00}, for commutative gauge gravity and
supergravity, and a recent paper \cite{partner07}, for nonassociative
generalizations. The commutative geometric parts of such phase space models
involve commutative nonholonomic (co) vector bundle spaces 
\begin{equation}
\ ^{\shortmid }\mathcal{E(\ ^{\shortmid }M)}:=\left( \ ^{\shortmid }\mathcal{%
E}=h\ ^{\shortmid }\mathcal{E}\oplus c\ ^{\shortmid }\mathcal{E},\
^{\shortmid }\mathcal{G}r=SO(4,1)\oplus \ ^{\shortmid }SO(4,1),\ ^{\shortmid
}\pi =(h\pi ,c\pi ),\mathcal{\ ^{\shortmid }M}\right) ,  \label{dvbundle}
\end{equation}%
which can be associated to respective tangent bundles $T\mathcal{(\
^{\shortmid }M)}$ and co-tangent bundles $T^{\ast }\mathcal{(\ ^{\shortmid}M)%
}.$ Such d- and s-vector bundles are enabled with N-adapted projections $\pi 
$ and $\ ^{\shortmid }\pi ,$ when brief notations like $\ ^{\shortmid }%
\mathcal{E}$ or $\ \mathcal{E}=\mathcal{E(M)}$ are used. In the above
formula, the group $\ ^{\shortmid }SO(4,1)$ is isomorphic to $SO(4,1)$ but
may have different parameterizations corresponding to different types of
spacetime coordinates and co-fiber momentum-like variables.

\vskip4pt A canonical de Sitter gauge gravitational connection on $\
^{\shortmid }\mathcal{E}$ is introduced as a 1-form 
\begin{equation}
\ ^{\shortmid }\widehat{\mathcal{A}}=\left[ 
\begin{array}{cc}
\ ^{\shortmid }\widehat{\mathcal{A}}_{\ \underline{\beta }}^{\underline{%
\alpha }} & l_{0}^{-1}\ ^{\shortmid }\chi ^{\underline{\alpha }} \\ 
l_{0}^{-1}\ ^{\shortmid }\chi _{\underline{\beta }} & 0%
\end{array}%
\right] ,  \label{cdSgp}
\end{equation}%
where $l_{0}$ is a dimensional constant which is used because of different
physical dimensions of $\ ^{\shortmid }\widehat{\mathcal{A}}$- and $%
^{\shortmid }\chi $-fields. In the nonholonomic gauge gravitational
potential (\ref{cdSgp}), $\ ^{\shortmid }\chi _{\ }^{\underline{\alpha }}= \
^{\shortmid }\chi _{\ \alpha }^{\underline{\alpha }}\ ^{\shortmid }\mathbf{e}%
^{\alpha }$ are for $8\times 8$ matrices $\ ^{\shortmid }\chi _{\ \alpha }^{%
\underline{\alpha }}(\ ^{\shortmid }u)\ $ subjected to the condition that $\
^{\shortmid }\mathbf{g}_{\alpha \beta }= \ ^{\shortmid }\chi _{\ \alpha }^{%
\underline{\alpha }}\ ^{\shortmid }\chi _{\ \beta }^{\underline{\beta }}\
^{\shortmid }\eta _{\underline{\alpha }\underline{\beta }},$ where the 8-d
dubbing of Minkovski metric can be written $\ ^{\shortmid }\eta _{\underline{%
\alpha }\underline{\beta }}=diag(1,1,1,-1,1,1,1,-1)$ in any point $\
^{\shortmid }u\in \ ^{\shortmid }\mathcal{M}.$ For $\ ^{\shortmid }\widehat{%
\mathcal{A}}_{\ \underline{\beta }}^{\underline{\alpha }}=\ ^{\shortmid }%
\widehat{\mathcal{A}}_{\ \underline{\beta }\gamma }^{\underline{\alpha }}\
^{\shortmid }\mathbf{e}^{\gamma }=\ ^{\shortmid }\widehat{\mathcal{A}}_{\ 
\underline{\beta }\gamma _{s}}^{\underline{\alpha }}\ ^{\shortmid }\mathbf{e}%
^{\gamma _{s}}$, with s-adapted $\ ^{\shortmid }\mathbf{e}^{\gamma _{s}}$ (%
\ref{nadapb}), the coefficients transform as $\ ^{\shortmid }\widehat{%
\mathcal{A}}_{\ \underline{\beta }\gamma }^{\underline{\alpha }}=\
^{\shortmid }\chi _{\ \alpha }^{\underline{\alpha }}\ ^{\shortmid }\chi _{%
\underline{\beta }\ }^{\ \beta }\ ^{\shortmid }\widehat{\Gamma }_{\ \beta
\gamma }^{\alpha }+\ ^{\shortmid }\chi _{\ \alpha }^{\underline{\alpha }}\ \
^{\shortmid }\mathbf{e}_{\gamma }(\ ^{\shortmid }\chi _{\underline{\beta }\
}^{\ \beta })$. Such formulas determined by the N-, or s-adapted coefficients of a
canonical d-/s-connection $\ _{s}^{\shortmid }\widehat{\mathbf{D}}= \{\
^{\shortmid }\widehat{\Gamma}_{\ \beta _{s}\gamma _{s}}^{\alpha _{s}}\}.$
Similar constructions can be done for an arbitrary linear connection $\
^{\shortmid }\Gamma _{\ \beta \gamma }^{\alpha },$ as in the metric-affine
geometry or a LC-connection $\ ^{\shortmid }\nabla $ for pseudo-Riemannian
models. Using the "hat-connection" $\ _{s}^{\shortmid }\widehat{\mathbf{D}}$, we can prove general decoupling and integration properties of various
modified Einstein equations in MGTs. 

\vskip4pt A d- / s-adapted metric structure $\ ^{\shortmid }\mathbf{g}%
_{\alpha \beta }\approx \quad ^{\shortmid }\mathbf{g}_{\alpha _{s}\beta
_{s}} $ allows to define respective Hodge d- /s-operators $\divideontimes
\approx \divideontimes _{N}\approx \divideontimes _{s}$ (we shall omit N- or 
$s$-labels for simplicity) and the absolute differential operator $\
^{\shortmid }d\ \approx \ _{s}^{\shortmid }d\ $ and skew product $\wedge $
on $\ _{s}^{\shortmid }\mathcal{M}.$ Defining in s-adapted geometric form
the curvature of (\ref{cdSgp}), $\ _{s}^{\shortmid }\widehat{\mathcal{F}}= \
_{s}^{\shortmid }d\ _{s}^{\shortmid }\widehat{\mathcal{A}}+\ _{s}^{\shortmid
}\widehat{\mathcal{A}} \wedge \ _{s}^{\shortmid }\widehat{\mathcal{A}},$ and
using $\divideontimes _{s},$ we derive in abstract geometrical form the
commutative gauge gravitational equations on $\ _{s}^{\shortmid }\mathcal{E}$, 
\begin{equation}
\ _{s}^{\shortmid }d(\divideontimes _{s}\ _{s}^{\shortmid }\widehat{\mathcal{%
F}})+\ _{s}^{\shortmid }\widehat{\mathcal{A}}\wedge (\divideontimes _{s}\
_{s}^{\shortmid }\widehat{\mathcal{F}})-(\divideontimes _{s}\
_{s}^{\shortmid }\widehat{\mathcal{F}})\wedge \ _{s}^{\shortmid }\widehat{%
\mathcal{A}}=-\lambda \ _{s}^{\shortmid }\widehat{\mathcal{J}}.
\label{ymgrcom}
\end{equation}%
The source in (\ref{ymgrcom}) is also parameterized in s-adapted form, 
\begin{equation}
\ _{s}^{\shortmid }\widehat{\mathcal{J}}=\left[ 
\begin{array}{cc}
\ _{s}^{\shortmid }\widehat{\mathcal{J}}_{\ \underline{\beta }}^{\underline{%
\alpha }} & -l_{0}\ _{s}^{\shortmid }t^{\underline{\alpha }} \\ 
-l_{0}\ _{s}^{\shortmid }t_{\underline{\beta }} & 0%
\end{array}%
\right] ,\mbox{ where }\ _{s}^{\shortmid }\widehat{\mathcal{J}}_{\ 
\underline{\beta }}^{\underline{\alpha }}=\ ^{\shortmid }\widehat{\mathcal{J}%
}_{\ \underline{\beta }\gamma _{s}}^{\underline{\alpha }}\ ^{\shortmid }%
\mathbf{e}^{\gamma _{s}}  \label{sourcgauge1}
\end{equation}%
is identified to zero for the model with LC-connection. It is induced
nonholonomically for the canonical d-connection but considered as a spin
density if elaborated on phase spaces  which are similar to  the
Riemann-Cartan theory. The 1-form $\ _{s}^{\shortmid }t^{\underline{\alpha }%
}= \ ^{\shortmid }t_{\ \alpha _{s}}^{\underline{\alpha }}\ ^{\shortmid }%
\mathbf{e}^{\alpha _{s}}$ is a phase space analogue of the energy-momentum
tensor for the matter. In above formulas, the constant $\lambda $ can be
related to the gravitational constant $l^{2}$ in 8-d extending by
analogy the 4-d formulas in GR. Other constants on the phase space (from
string gravity etc.) are introduced as in 4-d theories, for instance, we consider $l^{2}=2l_{0}^{2}%
\lambda ,\ \lambda _{1}=-3/l_{0}.$ Of course, we can re-define equivalently (%
\ref{ymgrcom}) in various other forms but the hat variables have the
priority to allow a general decoupling and integration of physically
important systems of nonlinear PDEs.  

\vskip4pt Using (\ref{cdSgp}) in s-adapted form, we can define a canonical
gauge s-operator $\ ^{\shortmid }\widehat{\mathcal{D}}_{\alpha _{s}}:= \
^{\shortmid }\widehat{\mathbf{D}}_{\alpha _{s}}+\ ^{\shortmid }\widehat{%
\mathcal{A}}_{\alpha _{s}}$ and write the nonassociative Yang-Mills
equations (\ref{ymgrcom}) in the form 
\begin{equation}
\ ^{\shortmid }\widehat{\mathcal{D}}_{\alpha _{s}}\ ^{\shortmid }\widehat{%
\mathcal{F}}^{\alpha _{s}\beta _{s}}=\ ^{\shortmid }\widehat{\mathbf{D}}%
_{\alpha _{s}}\ ^{\shortmid }\widehat{\mathcal{F}}^{\alpha _{s}\beta
_{s}}+[\ ^{\shortmid }\widehat{\mathcal{A}}_{\alpha _{s}},\ ^{\shortmid }%
\widehat{\mathcal{F}}^{\alpha _{s}\beta _{s}}]=-\lambda \ ^{\shortmid }%
\widehat{\mathcal{J}}^{\beta _{s}}.  \label{ymgrcom1}
\end{equation}%
In these formulas,  $[A,B]=AB-BA$ is the commutator on the Lie algebra of the chosen gauge
group $\ ^{\shortmid }\mathcal{G}r.$ The gauge gravitation fields on phase
space satisfy also the Bianchi identity (which is equivalent to the Jacobi 
identity):%
\begin{equation}
\lbrack \ ^{\shortmid }\widehat{\mathcal{D}}_{\mu _{s}}, [\ ^{\shortmid }%
\widehat{\mathcal{D}}_{\nu _{s}},\ ^{\shortmid }\widehat{\mathcal{D}}%
_{\alpha _{s}}]]+[\ ^{\shortmid }\widehat{\mathcal{D}}_{\alpha _{s}}, [\
^{\shortmid }\widehat{\mathcal{D}}_{\mu _{s}},\ ^{\shortmid }\widehat{%
\mathcal{D}}_{\nu _{s}}]]+ [\ ^{\shortmid }\widehat{\mathcal{D}}_{\nu _{s}},
[\ ^{\shortmid }\widehat{\mathcal{D}}_{\alpha _{s}}, \ ^{\shortmid }\widehat{%
\mathcal{D}}_{\mu _{s}}]]=0,  \label{bianchi}
\end{equation}%
considered for matrix operators with values in Lie algebra. 

We emphasize that the YM-like gravitational equations (\ref{ymgrcom}) can be postulated or derived in abstract geometric form as in \cite{misner} but using the canonical de Sitter gauge gravitational connection  (\ref{cdSgp}) instead of similar constructions involving the LC-connection and standard YM connections with the de Sitter structure group. Such abstract geometric formulations can be used for the non-variational models when the structure group is chosen to be of affine or Poincar\'{e} type. In the last cases, the Killing metric form is degenerated and the total bundle gauge theory is not variational. Nevertheless, we can elaborate a variational model by introducing an effective constant $a$ instead of the term  $l_{0}^{-1}\ ^{\shortmid }\chi _{\underline{\beta }}$ in  (\ref{cdSgp}). In such cases, an effective Lagrangian can be defined as in the YM theory which allows a variational proof of field equations of type (\ref{ymgrcom1}). Projecting such equations for the LC-connection, or the canonical d-connection, on a base spacetime Lorentz manifold, the constant $a$ disappears and the (affine) YM gauge equations transform into standard Einstein equations. Such a proof is provided for noncommutative generalizations  in \cite{svnc00} (see in that paper the references on previous works  with more details on N-adapted variational proofs etc.). For this work (involving nonholonomic generalizations of the BV method), it is enough to use  the abstract geometric definition of YM-like equations.  In the next subsection, we show that this approach can be extended for nonassociative gauge gravity theories  by using distortions of connections and star product deformations.

\subsection{Nonassociative star product deformation of de Sitter gauge gravity}

It is a tedious technical task to compute in explicit frame index or
coordinate forms of star product deformations of geometric and physical
objects on phase spaces \cite{aschieri17,partner02}. Non-trivial
N-connection structures make the formalism more sophisticated. The
Convention 2 from \cite{partner02} was formulated with the aim to compute
such $\star _{s}$-deformations in s-adapted form or using abstract geometric
methods. Applying on (co) vector/tangent bundles on corresponding phase
spaces and gauge geometric s-objects the twisted star product operator $%
\star _{s}$ (\ref{starpn}), we can define corresponding nonassociative
s-objects in gauge gravity theory. For instance, in a phase space, 
\begin{equation}
\star _{s}:\ \mathbf{g}_{s}\mathbf{\rightarrow g}_{s}^{\star }=(\breve{g}%
_{s}^{\star },\check{g}_{s}^{\star });\ \ _{s}^{\shortmid }\widehat{\mathbf{D%
}}=\ _{s}^{\shortmid }\nabla +\ _{s}^{\shortmid }\widehat{\mathbf{Z}}%
\rightarrow \ _{s}^{\shortmid }\widehat{\mathbf{D}}^{\star }=\
_{s}^{\shortmid }{\nabla }^{\star }+\ _{s}^{\shortmid }\widehat{\mathbf{Z}}%
^{\star };\divideontimes \rightarrow \breve{\divideontimes},
\label{canongeomobj}
\end{equation}%
when the Hodge operator \ $\breve{\divideontimes}$ on $\ _{s}^{\shortmid }%
\mathcal{M}^{^{\star }}$ and the parametric deformations can be defined
using the symmetric part $\breve{g}_{s}^{\star }$ of the star product
deformed s-metric. 

\vskip4pt The (co) vector bundles on $\ _{s}^{\shortmid }\mathcal{M}$, and
respective geometric s-objects subjected to $\star _{s}$-deformations define
certain nonassociative bundle spaces,  $\
_{s}^{\shortmid }\mathcal{E(}\ _{s}^{\shortmid }\mathcal{M})\rightarrow \ \
_{s}^{\shortmid }\mathcal{E}^{\star }(\ _{s}^{\shortmid }\mathcal{M}%
^{\star})= (\ _{s}^{\shortmid }\mathcal{E}^{\star },\ ^{\shortmid }\mathcal{G%
}r,\ ^{\shortmid }\pi , \ _{s}^{\shortmid }\mathcal{M}^{\star })$, for
double structure group $\ ^{\shortmid }\mathcal{G}r$ preserved as in (\ref%
{dvbundle}). Here we note that we can elaborate on more general classes of
nonassociative gauge models when, for instance, $\ ^{\shortmid }\mathcal{G}r$
transforms into some quantum groups (with group deforms which are additional
to $\star _{s}$-deformations). Such quantum group theories involve various
assumptions on algebraic structure and request new physical motivations
compared to the "nonassociative string theory R-flux deformation
philosophy". The procedure of general decoupling and integration for such
quantum gauge gravitational models is more sophisticated. In this work, we
 extend the research program outlined in \cite{partner02,partner04,partner06,partner07} for nonassociative
gravitational and matter field theories with gauge groups when the structure
of such groups and corresponding algebras are not subjected to quantum
deformations. For such theories, we can apply in direct form the AFCDM and
generate exact and parametric solutions of physically important systems of
nonlinear PDEs. 

\vskip4pt In abstract and nonholonomic s-adapted geometric forms, we define
and compute nonassociative deformations of type: $\star _{s}:\ ^{\shortmid }%
\widehat{\mathcal{A}}\rightarrow \ _{s}^{\shortmid }\widehat{\mathcal{A}}%
^{\star },$ with $\ ^{\shortmid }\mathbf{g}_{\alpha \beta }$ identified to $%
\breve{g}_{s}^{\star }$ if we consider zero powers of parameters $\hbar $
and $\kappa $ from (\ref{starpn}). We can prescribe also the nonholonomic
structure for the same $\ ^{\shortmid }\chi ^{\underline{\alpha }}$ and $\
^{\shortmid }\mathbf{e}^{\gamma _{s}}$ from (\ref{cdSgp}), with s-adapted
deformations $\star _{s}:\ \ ^{\shortmid }\widehat{\mathcal{A}}_{\ 
\underline{\beta }\gamma _{s}}^{\underline{\alpha }}\rightarrow  \
^{\shortmid} \widehat{\mathcal{A}}_{\ \underline{\beta }\gamma _{s}}^{\star%
\underline{\alpha }}.$ This allows us to compute nonassociative star product
deformations: 
\begin{equation}
\star _{s}:\ _{s}^{\shortmid }\widehat{\mathcal{F}}=\ _{s}^{\shortmid }d\
_{s}^{\shortmid }\widehat{\mathcal{A}}+\ _{s}^{\shortmid }\widehat{\mathcal{A%
}}\quad ^{s}\wedge ^{\star }\ _{s}^{\shortmid }\widehat{\mathcal{A}}%
\rightarrow \ _{s}^{\shortmid }\widehat{\mathcal{F}}^{\star }=\
_{s}^{\shortmid }d\ \ _{s}^{\shortmid }\widehat{\mathcal{A}}^{\star }+\
_{s}^{\shortmid }\widehat{\mathcal{A}}^{\star }\quad ^{s}\wedge ^{\star }\ 
\ _{s}^{\shortmid }\widehat{\mathcal{A}}^{\star },  \label{deformgauge}
\end{equation}%
with respective deformation of s-adapted anti-symmetric operator $\wedge
\rightarrow \ ^{s}\wedge ^{\star };$ $[A,B]\rightarrow \lbrack
A^{\star}, B^{\star }]^{\star }$. For generalized sources of the YM equations, $\star _{s}:\ ^{\shortmid }\widehat{\mathcal{J}}^{\beta _{s}}\rightarrow \ ^{\shortmid }\widehat{\mathcal{J}}^{\star \beta _{s}}$;
and  
\begin{equation*}
\ _{s}^{\shortmid }\widehat{\mathcal{D}}=\ _{s}^{\shortmid }\widehat{\mathbf{%
D}}+\ _{s}^{\shortmid }\widehat{\mathcal{A}}\rightarrow \ _{s}^{\shortmid }%
\widehat{\mathcal{D}}^{\star }=\ _{s}^{\shortmid }\widehat{\mathbf{D}}%
^{^{\star }}+\ _{s}^{\shortmid }\widehat{\mathcal{A}}^{^{\star }},\
_{s}^{\shortmid }\widehat{\mathcal{J}}\rightarrow \ _{s}^{\shortmid }%
\widehat{\mathcal{J}}^{\star }.
\end{equation*}%
This allows us to $\star _{s}$-deform the YM type equations for the de Sitter phase space gravity (\ref{ymgrcom}) and formulate their nonassociative versions, 
\begin{equation*}
\ ^{\shortmid }d(\breve{\divideontimes}\ ^{\shortmid }\widehat{\mathcal{F}}%
^{^{\star }})+\ ^{\shortmid }\widehat{\mathcal{A}}^{^{\star }}\wedge
(\divideontimes \ ^{\shortmid }\widehat{\mathcal{F}}^{^{\star
}})-(\divideontimes \ ^{\shortmid }\widehat{\mathcal{F}}^{^{\star }})\wedge
\ ^{\shortmid }\widehat{\mathcal{A}}^{^{\star }}=-\lambda \ ^{\shortmid }%
\widehat{\mathcal{J}}^{^{\star }}.
\end{equation*}%
This equation can be written also in nonholonomic s-adapted coefficient form
as a $\star _{s}$-deformation of (\ref{ymgrcom1}), 
\begin{equation}
\ ^{\shortmid }\widehat{\mathcal{D}}_{\alpha _{s}}^{^{\star }}\ ^{\shortmid }%
\widehat{\mathcal{F}}^{^{\star }\alpha _{s}\beta _{s}}=\ ^{\shortmid }%
\widehat{\mathbf{D}}_{\alpha _{s}}^{^{\star }}\ ^{\shortmid }\widehat{%
\mathcal{F}}^{^{\star }\alpha _{s}\beta _{s}}+[\ ^{\shortmid }\widehat{%
\mathcal{A}}_{\alpha _{s}}^{^{\star }},\ ^{\shortmid }\widehat{\mathcal{F}}%
^{^{\star }\alpha _{s}\beta _{s}}]^{\star }=-\lambda \ ^{\shortmid }\widehat{%
\mathcal{J}}^{^{\star }\beta _{s}}.  \label{nonassocymgreq1}
\end{equation}%
The above nonassociative YM equations (\ref{nonassocymgreq1}) are subjected
additionally to the conditions of $\star _{s}$-deformed Bianchi identities (%
\ref{bianchi}) involving a nonzero "Jacobiator", which is typical for
nonassociative theories. Such values can be computed as induced (effective)
ones by choosing respective s-adapted $\hbar $ and $\kappa $ parametric
decompositions. They reflect both the nonassociative and nonholonomic
structure of such phase space theories. We can consider certain analogies
with nonholonomic mechanics when the dynamical equations and conservation
laws are supplemented by additional non-integrable constraints, Lagrange
multiples, modified conservation laws etc. 

\vskip4pt We emphasize that the nonassociative and noncommutative theories determined by a general twisted product \cite{drinf89,blumenhagen16,aschieri17} are generic non-variational. So, the nonassociative gauge gravitational equations (\ref{nonassocymgreq1}) are, in general, non-variational and nonlocal.  Such issues related to nonassociative R-flux gravitational models are discussed in our partner works \cite{partner04,partner05,partner06,partner07,partner09}.  Here we note these important three points: 1) Effective variational theories encoding nonassociative data can be formulated after considering  $\hbar $ and $\kappa $ parametric decompositions of the geometric and physical s-objects in (\ref{nonassocymgreq1}).  2) Such physically important equations can be derived in low-energy limits of string theory (with corresponding nonholonomic re-parameterizations of terms with R-fluxes);  or as star product deformations of any variational or non-variational associative and commutative YM-like equations (\ref{ymgrcom}) or (\ref{ymgrcom1}). 3) To postulate or derive in abstract geometric form nonassociative gauge gravitational field equations using the nonassociative gauge gravitational potential $\ _{s}^{\shortmid }\widehat{\mathcal{A}}^{\star }$ (\ref{deformgauge})  is also possible. This approach is preferred for generic non-variational nonassociative theories. It allows an abstract nonassociative geometric generalization of the mathematical methods and results on BV and AQFT theory from \cite{rejzner20,brunetti22,brunetti09,rejzner16,fr12,rejzner14,fr13,df03,hr20,bd08,bf20}. Such methods are very powerful and important in elaborating new models of QG including nonassociative non-variational and nonlocal contributions from string and M-theory.  In this work, we study such possibilities for certain general and physically important classes of off-diagonal solutions of nonassociative gauge gravitational equations  (\ref{ymgrcom1}). This first step is important for elaborating physical  applications of the BV formalism to  generic nonlinear theories like GR and various modifications. Such results are important for elaborating QG models in the conditions a general nonlinear functional analysis theory does not exist and can't be formulated in a unique way for generic nonassociative non-variational theories.

\vskip4pt Finally, we note that an effective or matter field source $\
^{\shortmid }\widehat{\mathcal{J}}^{^{\star }\beta _{s}}$ (\ref{sourcgauge1}%
) can be correspondingly parameterized and physically motivated \cite%
{partner07} in some forms that the nonassociative YM equations (\ref%
{nonassocymgreq1}) present certain alternatives or gauge like
generalizations of the nonassociative star product deformed Einstein
equations. In nonassociative vacuum form, such models were considered in 
\cite{blumenhagen16,aschieri17} and, in s-adapted form, with extensions to
certain classes of nontrivial sources and off-diagonal solutions, in \cite%
{partner02,partner04,partner06}. 

\subsection{Quasi-stationary off-diagonal solutions in nonassociative gauge gravity}

To study the physical implications of nonassociativity in the framework of
gauge gravity theory it is important to derive certain physically important
solutions of the gravitational field equations (\ref{nonassocymgreq1}). This
is a very difficult technical problem because the nonassociative YM
equations consist of a strongly coupled system of nonlinear PDEs.
Surprisingly, the AFCDM allows to construction of such solutions (see the
first examples for cosmological solitonic hierarchies in \cite{partner07}).
This is possible if we parameterize and project the nonassociative YM
equations on the base phase space in a form which is equivalent to
nonassociative modified Einstein equations studied in our partner works. In
this subsection, we have shown how such nonholonomic projections can be defined in
certain forms the general decoupling and integration properties of
(nonassociative) gravitational equations which was proved in \cite{partner02}. This can  be
used for constructing physically important solutions. 

\subsubsection{Projections on phase spaces and effective parametric MGTs}

We can formulate a nonassociative gauge gravity model when the
nonassociative YM equations (\ref{nonassocymgreq1}) are equivalent to
certain nonassociative star product modifications of the Einstein equations
on the phase spaces. This is possible if we consider instead of the de
Sitter structure group $SO(4,1)$, the affine structure group $Af(4,1).$ Then
we construct a gauge potential which is similar to (\ref{cdSgp}). We consider some constants 
$\chi _{0}^{\underline{\alpha }}$ instead of $\ _{s}^{\shortmid }\chi ^{%
\underline{\alpha }}$ in the last line of the matrix $\ _{s}^{\shortmid }%
\widehat{\mathcal{A}}^{\star },$ which takes values into the double Poincar%
\'{e}-Lie algebra,%
\begin{equation}
\ _{s}^{\shortmid }\widehat{\mathcal{A}}^{\star }\rightarrow \
_{s}^{\shortmid }\widehat{\mathcal{A}}_{[P]}^{\star }=\left[ 
\begin{array}{cc}
\ _{s}^{\shortmid }\widehat{\mathcal{A}}_{\ \underline{\beta }}^{\star 
\underline{\alpha }} & l_{0}^{-1}\ _{s}^{\shortmid }\chi ^{\underline{\alpha 
}} \\ 
\chi _{0}^{\underline{\alpha }} & 0%
\end{array}%
\right] .  \label{affinpot}
\end{equation}%
The constants $\chi _{0}^{\underline{\alpha }}$ from (\ref{affinpot}) can be
fixed to be zero at the end of computations. Then, we can introduce such
constants in the source (\ref{sourcgauge1}) when further $\star _{s}-$%
transforms result into 
\begin{equation}
\ _{s}^{\shortmid }\widehat{\mathcal{J}}^{\star }=\left[ 
\begin{array}{cc}
\ _{s}^{\shortmid }\widehat{\mathcal{J}}_{\ \underline{\beta }}^{\star 
\underline{\alpha }}=0 & -l_{0}\ _{s}^{\shortmid }t^{\star \underline{\alpha 
}} \\ 
\chi _{0}^{\underline{\alpha }} & 0%
\end{array}%
\right] ,\mbox{ with } \ _{s}^{\shortmid }t^{\star \underline{\alpha }}=\chi
^{\underline{\alpha }\beta _{s}}\ \ ^{\shortmid }\mathcal{J}_{\alpha
_{s}\beta _{s}}^{\star }\ ^{\shortmid }\mathbf{e}^{\alpha _{s}}.
\label{sourcgauge1na}
\end{equation}%
In these formulas, $\ ^{\shortmid }\mathcal{J}_{\alpha _{s}\beta
_{s}}^{\star }$ is the star product deformation of the effective
energy-momentum tensor extended on phase space and written in s-adapted form 
$\ ^{\shortmid }\mathcal{J}_{\alpha _{s}\beta _{s}}$ on $\ _{s}^{\shortmid }%
\mathcal{M}^{\star }.$ Such nonholonomic nonassociative or commutative
sources are considered in nonassociative gravity and nonassociative
geometric flow theories, see details and references in \cite%
{partner02,partner04,partner06}. 

\vskip5pt For nonassociative gauge potentials and sources, respectively, of
type (\ref{affinpot}) and (\ref{sourcgauge1na}), the projections on $\
_{s}^{\shortmid}\mathcal{M}^{\star }$ of nonassociative YM equations (\ref%
{nonassocymgreq1}) transform into nonassociative s-adapted canonical
gravitational equations considered in our partner works, 
\begin{equation}
\ ^{\shortmid }\widehat{\mathbf{R}}ic_{\alpha _{s}\beta _{s}}^{\star }=\
^{\shortmid }\mathcal{J}_{\alpha _{s}\beta _{s}}^{\star }.
\label{nonassocaneinst}
\end{equation}%
In the vacuum case with $^{\shortmid }\mathcal{J}_{\alpha _{s}\beta
_{s}}^{\star }=0$ and $\ _{s}^{\shortmid }\widehat{\mathbf{D}}%
^{\star}\rightarrow \ _{s}^{\shortmid }{\nabla }^{\star },$ the system of
nonlinear PDEs (\ref{nonassocaneinst}) are just the vacuum gravitational
equations for nonassociative and noncommutative gravity studied in \cite%
{blumenhagen16,aschieri17}. More than that, both equations (\ref%
{nonassocymgreq1}) and (\ref{nonassocaneinst}) transform into standard
Einstein equations in GR if we construct the affine potential (\ref{affinpot}%
) for the standard LC-connection ${\nabla }$ on a pseudo-Riemannian
spacetime base. The above equations can be proven in s-adapted forms using a
tedious calculus considered in \cite{svnc00,sv00,partner07} for nonholonomic
commutative phase spaces and in noncommutative gauge gravity with
Seiberg-Witten product. 

\subsubsection{Off-diagonal parametric quasi-stationary solutions}

The system of nonlinear PDEs (\ref{nonassocaneinst}) can be decoupled and
integrated in certain very general forms when the coefficients off-diagonal
metrics and star-modified connections depend on all spacetime and
momentum-like coordinates \cite{partner02}. Such solutions can be generated
in a more simple form for  quasi-stationary
configurations with a Killing symmetry (this is enough for the purposes of
this work). For instance, we can use a time-like vector $\mathbf{e}%
_{4}=\partial _{4}=\partial _{t}$ (with $x^{4}=y^{4}=t$) and introduce such 
nonholonomic parameterizations of (effective) sources (\ref{sourcgauge1na}): 
\begin{equation}
\ ^{\shortmid }\mathcal{J}_{\ \ \beta _{s}}^{\star \alpha _{s}}(~^{\shortmid
}u)=diag[~_{1}^{\shortmid }\mathcal{J}^{\star }(x^{k_{1}})\delta
_{i_{1}}^{j_{1}},~_{2}^{\shortmid }\mathcal{J}^{\star
}(x^{k_{1}},x^{3})\delta _{b_{2}}^{a_{2}},~_{3}^{\shortmid }\mathcal{J}%
^{\star }(x^{k_{2}},p_{5})\delta _{a_{3}}^{b_{3}},~_{4}^{\shortmid }\mathcal{%
J}^{\star }(x^{k_{3}},p_{7})\delta _{a_{4}}^{b_{4}}].  \label{sourcparam}
\end{equation}%
In this formula, $k_{1}=1,2;k_{2}=1,2,3,4;k_{3}=1,2,...6;$ we can consider
variants when $x^{3}\rightarrow x^{4}$ (for locally anisotropic cosmological
configurations) or when $p_{5}\rightarrow p_{6},$ or when $p_{7}\rightarrow
p_{8},$ for generating other classes of off-diagonal solutions. In a more
general case, we can generate solutions for arbitrary shell parametrization $%
\ _{s}^{\shortmid }\mathcal{J}^{\star }(x^{k_{s-1}},p_{a_{s}}),$ for $s=3,4.$
Any class of locally anisotropic cosmological solutions, with Killing
symmetry $\mathbf{e}_{3}=\partial _{3},$ of nonassociative modified Einstein
equations (\ref{nonassocaneinst}) can be lifted on respective co-vector
bundles for generating solutions of nonassociative YM equations (\ref%
{nonassocymgreq1}), see details in \cite{partner07}. The goal of this
section is to show how the AFCDM can be used for generating quasi-stationary
solutions of the systems of nonlinear PDEs (\ref{nonassocaneinst}) and (\ref%
{nonassocymgreq1}). Here, we also note that the $\star $-labels of the shell
components in (\ref{sourcparam}) state that such effective sources encode in
parametric form\footnote{We consider the small parameters  $\hbar $ and $\kappa$;  we can use also other type physical constants like the gravitational one $G$, with extensions to higher dimensions, a BH mass, $%
M$, a cosmological constant, $\Lambda ,$ an electric charge, $e,$ etc.; for
simplicity, we shall write only $\ ^{\shortmid }\Im _{\ \ \beta _{s}}^{\star
\alpha _{s}}(\hbar ,\kappa ,\ ^{\shortmid }u)$ or $\ _{s}^{\shortmid }%
\mathbf{g}_{\ \ }^{\star }(\hbar ,\kappa ,\ ^{\shortmid }u)$ assuming that
(if necessary) we can introduce and emphasize another types of
physically important constants; this allows us to  define explicit classes of effective
sources and $s$--metrics encoding various types of parametrical
dependencies; using polarization functions, we can write $\ ^{\shortmid
}\eta _{\alpha _{s}}(\hbar ,\kappa ,\ ^{\shortmid }u)$ and $\ ^{\shortmid
}\eta _{i_{s-1}}^{a_{s}}(\hbar ,\kappa,\ ^{\shortmid }u)\ $} nonassociative
data for R-flux and matter field deformations as explained in details in 
\cite{partner02,partner04,partner06}. In this work, we shall consider that $%
\ _{s}^{\shortmid }\mathcal{J}^{\star }$ are certain generating sources
which can be prescribed following physical arguments, for instance, to model
nonassociative BH, or WH, or locally anisotropic cosmological
configurations. 

\vskip4pt Let us consider a prime $\ _{s}^{\shortmid }\mathbf{\mathring{g}}%
(\ ^{\shortmid }u)$ on $\ _{s}^{\shortmid }\mathcal{M}$. Geometrically, it
can be an arbitrary s-metric, or taken as an important solution of some
(modified) Einstein equations which allows applications in modern physics
and information theory. Our goal is to construct a family of target $\
_{s}^{\shortmid }\mathbf{g}(x^{k_{3}},p_{7})$ defining a solution of (\ref%
{nonassocymgreq1}), 
\begin{eqnarray}
\ _{s}^{\shortmid }\mathbf{\mathring{g}}(\ ^{\shortmid }u)\ &=&[\
^{\shortmid }\mathring{g}_{\alpha _{s}}(\ ^{\shortmid }u),\ ^{\shortmid }%
\mathring{N}_{i_{s-1}}^{a_{s}}(\ ^{\shortmid }u)]\rightarrow
\label{offdiagdefr} \\
\ _{s}^{\shortmid }\mathbf{g}^{\star }(x^{k_{3}},p_{7}) &=&[\ ^{\shortmid
}g_{\alpha _{s}}^{\star }(x^{k_{s-1}},p_{a_{s}})=\ ^{\shortmid }\eta
_{\alpha _{s}}(\ ^{\shortmid }u)\ ^{\shortmid }\mathring{g}_{\alpha _{s}}(\
^{\shortmid }u),\ ^{\shortmid }N_{i_{s-1}}^{a_{s}}(x^{k_{s-1}},p_{a_{s}})=\
^{\shortmid }\eta _{i_{s-1}}^{a_{s}}(\ ^{\shortmid }u)\ ^{\shortmid }%
\mathring{N}_{i_{s-1}}^{a_{s}}(\ ^{\shortmid }u)].  \notag
\end{eqnarray}%
In (\ref{offdiagdefr}), the $\eta $-polarization functions (i.e. phase space
gravitational polarization functions, which can be also considered as
generating functions) define star product deformations. To elaborate on  quantum models,
we can consider quantum perturbative or non-perturbative deformations, for
instance, of some solutions in GR or a MGTs extended on phase spaces. Here
we emphasize that parameterizations (\ref{sourcparam}) and (\ref{offdiagdefr}) used for (\ref{nadapb}) and (\ref{starpn}) allows to transform (\ref{nonassocaneinst}) into a system of nonlinear PDEs with general decoupling and integration properties \cite{partner02}. After a class of solutions is
defined in a certain general form by respective generating functions and
generating sources and integration functions, we can impose certain boundary
or Cauchy conditions. This allows us to elaborate on physical models with polarization of
physical constants or vacuum gravitational and phase space configurations,
deformation of horizons (for certain conditions) etc. 

\vskip4pt The ansatz for generating quasi-stationary solutions of
nonassociative gravitational YM equations (\ref{nonassocymgreq1}) with $\eta$%
-polarization functions and fixed energy parameter $p_{8}=E_{0}$ can be
parameterized (see details in \cite{partner02,partner04,partner06}) using
such quadratic line elements on phase space: 
\begin{eqnarray}
d\widehat{s}^{2} &=&g_{\alpha _{s}}(\hbar ,\kappa ,x^{i_{3}},p_{7})(\mathbf{e%
}^{a_{s}})^{2}=g_{i_{1}}(\hbar ,\kappa
,x^{k_{1}})(dx^{i_{1}})^{2}+g_{a_{2}}(\hbar ,\kappa ,x^{i_{1}},y^{3})[%
\mathbf{e}^{a_{2}}(\hbar ,\kappa ,x^{i_{1}},y^{3})]^{2}+  \label{ans1qs} \\
&&\ ^{\shortmid }g^{a_{3}}(\hbar ,\kappa ,x^{i_{2}},p_{6})[\ ^{\shortmid }%
\mathbf{e}_{a_{3}}(\hbar ,\kappa ,x^{i_{2}},p_{6})]^{2}+\ ^{\shortmid
}g^{a_{4}}(\hbar ,\kappa ,\ ^{\shortmid }x^{i_{3}},p_{7})[\ ^{\shortmid }%
\mathbf{e}_{a_{4}}(\hbar ,\kappa ,\ ^{\shortmid }x^{i_{3}},p_{7})]^{2} 
\notag \\
&=&\eta _{i_{1}}(\ ^{\shortmid }u)\mathring{g}_{i_{1}}(\ ^{\shortmid
}u)(dx^{i_{1}})^{2}+\eta _{a_{2}}(\ ^{\shortmid }u)\mathring{g}_{a_{2}}(\
^{\shortmid }u)(\mathbf{e}^{a_{2}})^{2}+\ ^{\shortmid }\eta ^{a_{3}}(\
^{\shortmid }u)\ ^{\shortmid }\mathring{g}^{a_{3}}(\ ^{\shortmid }u)(\
^{\shortmid }\mathbf{e}_{a_{3}})^{2}+\ ^{\shortmid }\eta ^{a_{4}}(\
^{\shortmid }u)\ ^{\shortmid }\mathring{g}^{a_{4}}(\ ^{\shortmid }u)(\
^{\shortmid }\mathbf{e}_{a_{4}})^{2},  \notag \\
&&\mbox{where }  \notag \\
\mathbf{e}^{a_{2}} &=&dy^{a_{2}}+N_{k_{1}}^{a_{2}}(\hbar ,\kappa
,x^{i_{1}},y^{3})dx^{k_{1}}=dy^{a_{2}}+\eta _{k_{1}}^{a_{2}}(\ ^{\shortmid
}u)\mathring{N}_{k_{1}}^{a_{2}}(\ ^{\shortmid }u)dx^{k_{1}},  \notag \\
\ ^{\shortmid }\mathbf{e}_{a_{3}} &=&dp_{a_{3}}+\ ^{\shortmid
}N_{a_{3}k_{2}}(\hbar ,\kappa ,,x^{i_{2}},p_{5})dx^{k_{2}}=dp_{a_{3}}+\
^{\shortmid }\eta _{a_{3}k_{2}}(\ ^{\shortmid }u)\ ^{\shortmid }\mathring{N}%
_{a_{3}k_{2}}(\ ^{\shortmid }u)dx^{k_{2}},  \notag \\
\ ^{\shortmid }\mathbf{e}_{a_{4}} &=&dp_{a_{4}}+\ ^{\shortmid
}N_{a_{4}k_{3}}(\hbar ,\kappa ,\ ^{\shortmid }x^{i_{3}},p_{7})d\ ^{\shortmid
}x^{k_{3}}\ =dp_{a_{4}}+\ ^{\shortmid }\eta _{a_{4}k_{3}}(\ ^{\shortmid }u)\
^{\shortmid }\mathring{N}_{a_{4}k_{3}}(\ ^{\shortmid }u)dx^{k_{3}}.  \notag
\end{eqnarray}%
In Appendix, we provide explicit formulas for off-diagonal solutions with s-coefficients depending on generating and integration functions  and on generating sources $\ _{s}^{\shortmid }\mathcal{J}^{\star }$(\ref{sourcparam}). Such values  can be chosen in different forms allowing to construct physically important exact or parametric solutions. 

\vskip4pt Any quasi-stationary solution (\ref{ans1qs}) can be written in
some off-diagonal functional forms with labels stating certain basic
properties defined by corresponding classes of generation functions and
(effective) sources and integration functions/ constants, 
\begin{eqnarray}
\ ^{\shortmid }\mathbf{g}^{\star }[\mathfrak{Cqs}] &=&\ _{s}^{\shortmid }%
\mathbf{g}^{\star }[\mathit{%
solit,horiz,polarizc,sg,regular,geomflow,therm,kinetic,ramif,filament,anhol,...%
}]  \notag \\
&=&\ ^{\shortmid }g_{\alpha \beta }^{\star }(\hbar ,\kappa
;x^{i},x^{3}p_{a};\partial _{4},\partial _{8};\ _{s}^{\shortmid }\mathcal{J}%
^{\star },\ _{s}^{\shortmid }\Lambda ;\ ^{\shortmid }\mathring{g}_{\alpha
\beta },\ _{s}^{\shortmid }\eta \sim \ ^{\shortmid }\zeta _{\alpha
_{s}}(1+\kappa \ ^{\shortmid }\chi _{\alpha _{s}}),..)d\ ^{\shortmid
}u^{\alpha }\otimes d\ ^{\shortmid }u^{\beta },\mbox{ or }  \notag \\
\ ^{\shortmid }\mathbf{g}^{\star }[\mathfrak{Clacs}] &=&\ ^{\shortmid
}g_{\alpha \beta }^{\star }(\hbar ,\kappa ;x^{i},t,p_{a};\partial
_{3},\partial _{8};\ _{s}^{\shortmid }\mathcal{J}^{\star },\ _{s}^{\shortmid
}\Lambda ;\ ^{\shortmid }\mathring{g}_{\alpha \beta },\ _{s}^{\shortmid
}\eta \sim \ ^{\shortmid }\zeta _{\alpha _{s}}(1+\kappa \ ^{\shortmid }\chi
_{\alpha _{s}}),...)d\ ^{\shortmid }u^{\alpha }\otimes d\ ^{\shortmid
}u^{\beta }  \label{baspropsol}
\end{eqnarray}%
In (\ref{baspropsol}), we consider such abstract labels for quasi-stationary
configurations in $\ ^{\shortmid }\mathbf{g}^{\star }[\mathfrak{Cqs}]:$
vacuum polarizations, for instance, as solitonic hierarchies, $\mathit{solit;%
}$ deformations of horizons (creation and disappearance), $\mathit{horiz;}$
polarization of constants, $\mathit{polarizc}$; singularities, $\mathit{sg}$%
; and regularizations, $\mathit{regular;}$ geometric flows evolution, $%
\mathit{geomflow}$, thermodynamic properties, $\mathit{therm}$; kinetic
properties, $\mathit{kinetic}$; ramification, $\mathit{ramif}$; \ filaments, 
$\mathit{filament}$; nonholonomic constraints $\mathit{anhol}$; and other
possible variants which are determined by generating and integration
functions and generating sources. Corresponding nonlinear symmetries
relating effective generating sources to shell effective cosmological
constants; and when Horava-Lifshitz and Finsler-Lagrange-Hamilton structures 
\cite{hlfin} can be modelled on (nonassociative) phase spaces, in GR, or
other type MGTs. 

\vskip4pt In explicit form, nonassociative off-diagonal solutions $\
^{\shortmid }\mathbf{g}^{\star }[\mathfrak{Cqs}]$ can be generated by
prescribing a corresponding class of integration functions (\ref%
{integrfunctrf}) \ for s-metrics (\ref{offdiagpolfr}). We also need
additional assumptions on the type polarization functions and nonlinear
symmetries (\ref{nonlinsym}). For more special conditions on parametric
decompositions, we can generate regular BH configurations on phase space as
in (\ref{sol4of}). Here we note that physical properties of nonassociative
phase space WH and black ellipsoid, BE, and BH solutions with singularities
were studied in \cite{partner02,partner04,partner05,partner06}. The
(nonassociative) vacuum structure in gauge gravity and projections on phase
spaces may encode solitonic hierarchies branching of solutions and filament
structures as we explained in \cite{partner07} for locally anisotropic
cosmological solutions of type $\ ^{\shortmid }\mathbf{g}^{\star }[\mathfrak{%
Clacs}]$ (\ref{baspropsol}). To generate such nonassociative accelerating
cosmological models we can use the time-like duality symmetries as we
explain in (\ref{cosmvariables}). 

\vskip4pt We conclude this sections with such remarks:
\begin{enumerate}
\item The modified YM equations for nonassociative gauge gravity (\ref{nonassocymgreq1}) and, for phase space projections, (\ref{nonassocaneinst}) are characterized by generic nonlinear solutions 
    $\ ^{\shortmid }\mathbf{g}^{\star }[\mathfrak{Cqs}]$ or 
    $\ ^{\shortmid }\mathbf{g}^{\star }[\mathfrak{Clacs}]$ as in (\ref{baspropsol}) defining a very rich geometric structure. Any solution is characterized by a nonassociative geometric flow thermodynamics, and respective nonlinear symmetries. Such nonassociative theories with general twisted star products are not variational, see details in \cite{partner04,partner05,partner06}. To elaborate on \ new methods of
quantization and constructing respective QG gravity models and formulating a generalized nonassociative BV formalism, we have to apply advanced geometric methods of the nonassociative geometric flow theory in
certain forms correlated to the AFCDM.

\item For parametric decompositions, the corresponding gauge gravity and
phase gravity equations and their generic off-diagonal solutions $\
^{\shortmid }\mathbf{g}^{\star }[\mathfrak{Cqs}]$ or $\ ^{\shortmid }\mathbf{%
g}^{\star }[\mathfrak{Clacs}]$ can be formulated in the framework of some
effective variational theories on phase spaces with respective effective
Lagrange or Hamilton densities. In this work, such densities  encode nonassociative data.
This approach allows us to develop in nonholonomic form (for nonassociative gauge
gravity theories) the BV formalism using rigorous mathematical results from
pAQFT \cite{bv81,rejzner20,brunetti22}.
\end{enumerate}


\section{The BV formalism for the nonassociative classical gauge gravity theory}

\label{sec03} In this section, we consider phase spaces modelled by
quasi-stationary off-diagonal solutions $\ ^{\shortmid }\mathbf{g}^{\star }[%
\mathfrak{Cqs}]$ (\ref{baspropsol}) of the YM equations (\ref%
{nonassocymgreq1}) for a classical nonassociative gauge gravity theory with
star product. The main goals are to generalize for such configurations the
principles of locality and homology  \cite{rejzner20,brunetti22} to construct nonassociative gauge
models, which will then be used to quantize using nonassociative and
nonholonomic deformation principles. 

\subsection{Geometric preliminaries and kinematic structure}

We choose a region $\ ^{\shortmid }\mathcal{U}\subset \ _{s}^{\shortmid }%
\mathcal{M}$ which is time-oriented and globally hyperbolic both for
projections on the Lorentz spacetime manifold and for typical cofibers. This
means that both h- and c-components are Cauchy surfaces. In brief, we write $%
\ ^{\shortmid }\mathcal{U}_{\mathfrak{Cqs}}^{\star }$ if such a phase space
region is defined by a $\ ^{\shortmid }\mathbf{g}^{\star }[\mathfrak{Cqs}]$
with corresponding parametric dependencies. Respectively, we write $\
^{\shortmid }\mathcal{E}_{\mathfrak{Cqs}}^{\star }= \ _{s}^{\shortmid }%
\mathcal{E}^{\star }(\ ^{\shortmid }\mathcal{U}_{\mathfrak{Cqs}}^{\star })$ (%
\ref{dvbundle}) and omit star-labels for configurations with $[00]=[\hbar
^{0},\kappa ^{0}],$ when $\ ^{\shortmid }\mathcal{U}_{\mathfrak{Cqs}}$ and $%
\ ^{\shortmid }\mathcal{E}_{\mathfrak{Cqs}}$ are determined by a $\
^{\shortmid }\mathbf{g}[\mathfrak{Cqs}]$ which, for instance, contains
dependencies on parameters depending BHs, WHs or other types physically
important solutions in an associative and commutative gravity theory. By
choosing an explicit quasi-stationary solution (\ref{offdiagpolfr}) or (\ref%
{sol4of}), we decide what kind of s-objects define our (nonassociative)
model, for instance, using certain classes of s-tensors, s-connections,
scalar fields etc.\footnote{\label{fnconv}We shall use also such
conventions: $\ ^{\shortmid }\mathcal{E}_{\mathfrak{Cqs}}^{C}$ denote the
space of smooth compactly supported sections of $\ ^{\shortmid }\mathcal{E}_{%
\mathfrak{Cqs}};$ for complexifications of topological duals of $\
^{\shortmid }\mathcal{E}_{\mathfrak{Cqs}}$ and $\ ^{\shortmid }\mathcal{E}_{%
\mathfrak{Cqs}}^{C},$ which are equipped with the strong topology, we write
respectively $\ ^{\shortmid }\mathcal{E}_{\mathfrak{Cqs}}^{\prime }$ and $\
^{\shortmid }\mathcal{E}_{\mathfrak{Cqs}}^{\prime C};$ we denote by 
$sec(\ ^{\shortmid }\mathcal{E}_{\mathfrak{Cqs}}^{\star })^{\ast }$ the space of
smooth section of the dual bundle $(\ ^{\shortmid }\mathcal{E}_{\mathfrak{Cqs%
}}^{\star })^{\ast };$ then, $(\ ^{\shortmid }\mathcal{E}_{\mathfrak{Cqs}%
}^{\star })^{!}$ denotes the complexification of the space of sections of $%
(\ ^{\shortmid }\mathcal{E}_{\mathfrak{Cqs}}^{\star })^{\ast }$ tensored
with the bundle of densities over $\ ^{\shortmid }\mathcal{U}_{\mathfrak{Cqs}%
}^{\star };$ and we write $(\ ^{\shortmid }\mathcal{E}_{\mathfrak{Cqs}%
}^{\star }(\ ^{\shortmid}\mathcal{U}_{\mathfrak{Cqs}}^{\star })^{n})^{!}$
for the complexified space of sections of the $n$-fold exterior tensor
product of a star product deformed bundle, seen as a s-vector bundle over $%
(\ ^{\shortmid }\mathcal{U}_{\mathfrak{Cqs}}^{\star })^{n}.$ We have to
introduce an "abuse of notations" comparing to \cite%
{misner,rejzner20,partner02,partner07} because the procedure of quantization
of nonlinear systems encoding nonassociative data depends in an explicit
form on the type of quasi-stationary configurations $\mathfrak{Cqs.}$ To
make the notation system more simple we shall denote (if not ambiguous),
for instance, the elements of $\ ^{\shortmid }\mathcal{E}_{\mathfrak{Cqs}%
}^{\star }$ by $\ ^{\shortmid }\mathcal{\varphi }^{\star }$ even such
elements carry s-indices (which can be invoked when it becomes necessary).} 

\vskip4pt For scalar fields, the configuration space is just $\ ^{\shortmid }%
\mathcal{E}_{\mathfrak{Cqs}}= \mathcal{C}^{\infty }(\ ^{\shortmid }\mathcal{U%
}_{\mathfrak{Cqs}},\mathbb{R}).$ Considering YM fields with $\ ^{\shortmid }%
\mathcal{G}r=SO(4,1)\oplus \ ^{\shortmid}SO(4,1) $ we can define off-shell
configuration space as $\ ^{\shortmid }\mathcal{E}:= \Omega ^{1}(\
^{\shortmid }\mathcal{U}_{\mathfrak{Cqs}},\ ^{\shortmid }\mathcal{G}r)$, see
details and references in \cite{rejzner20}. If structure groups are not
semi-simple and compact, for instance, $\ ^{\shortmid }\mathcal{G}%
r=Af(4,1)\oplus \ ^{\shortmid}Af(4,1) $, we can introduce some auxiliary
constants which transform to zero after projections on the base spacetime
but allow to define nondegenerate total metric structures, and take come
compact regions. In this work, we shall consider nontrivial bundles which is
enough for the perturbative treatment of (nonassociative) gauge models. Our
nonholonomic approach also includes a non-perturbative treatment because we
use geometric s-objects on $\ ^{\shortmid }\mathcal{U}_{\mathfrak{Cqs}}.$
Classical observables are functionals on $\ ^{\shortmid }\mathcal{E}_{\mathfrak{Cqs}}^{\star }$. Such a set can be  equipped with the natural Frech\'{e}t topology if the space of smooth functionals 
$\mathcal{C}^{\infty }(\ ^{\shortmid }\mathcal{E}_{\mathfrak{Cqs}},\mathbb{R}) $ is considered. This means that an observable assigns to a given field
configuration a number (which is the value of a measurement at a given point
in a spacetme). Smoothness is used to introduce well-defined algebraic
structures. 

\vskip4pt Next, we extend to nonassociative phase spaces the important
notion of spacetime support of a functional (which typically encodes \textit{%
localization} properties of observables and \textit{additivity}). For the
goals of this work (working with quasi-stationary phase space configurations
subjected to star product deformations, for simplicity, in linear parametric
form up to $[\hbar ^{1},\kappa ^{1}]$), we define the \textit{phase space
support of a functional} as 
\begin{eqnarray*}
\sup p\ ^{\shortmid }F_{\mathfrak{Cqs}}^{\star } &=&\{\ ^{\shortmid }u\in \
^{\shortmid }\mathcal{U}_{\mathfrak{Cqs}}\mid \forall 
\mbox{ neighborhoods
of }\ ^{\shortmid }u\exists \ ^{\shortmid }\mathcal{\varphi }_{1}^{\star },\
^{\shortmid }\mathcal{\varphi }_{2}^{\star }\in \ ^{\shortmid }\mathcal{E}_{%
\mathfrak{Cqs}},\sup p\ ^{\shortmid }\mathcal{\varphi }_{2}^{\star }\subset
\ ^{\shortmid }\mathcal{U}_{\mathfrak{Cqs}} \\
\mbox{ such that } &&\ ^{\shortmid }F_{\mathfrak{Cqs}}^{\star }(\
^{\shortmid }\mathcal{\varphi }_{1}^{\star }+\ ^{\shortmid }\mathcal{\varphi 
}_{2}^{\star })\neq \ ^{\shortmid }F_{\mathfrak{Cqs}}^{\star }(\ ^{\shortmid
}\mathcal{\varphi }_{1}^{\star })\}.
\end{eqnarray*}%
A functional $\ ^{\shortmid }F_{\mathfrak{Cqs}}^{\star }$ is \textit{additive%
} on phase space configuration $\mathfrak{Cqs}$ if 
\begin{eqnarray*}
\ ^{\shortmid }F_{\mathfrak{Cqs}}^{\star }(\ ^{\shortmid }\mathcal{\varphi }%
_{1}^{\star }+\ ^{\shortmid }\mathcal{\varphi }_{2}^{\star }+\ ^{\shortmid }%
\mathcal{\varphi }_{3}^{\star }) &=&\ ^{\shortmid }F_{\mathfrak{Cqs}}^{\star
}(\ ^{\shortmid }\mathcal{\varphi }_{1}^{\star }+\ ^{\shortmid }\mathcal{%
\varphi }_{2}^{\star })-\ ^{\shortmid }F_{\mathfrak{Cqs}}^{\star }(\
^{\shortmid }\mathcal{\varphi }_{3}^{\star })+\ ^{\shortmid }F_{\mathfrak{Cqs%
}}^{\star }(\ ^{\shortmid }\mathcal{\varphi }_{2}^{\star }+\ ^{\shortmid }%
\mathcal{\varphi }_{3}^{\star }), \\
\mbox{ for }\ ^{\shortmid }\mathcal{\varphi }_{1}^{\star }+\ ^{\shortmid }%
\mathcal{\varphi }_{2}^{\star }+\ ^{\shortmid }\mathcal{\varphi }_{3}^{\star
} &\in &\ ^{\shortmid }\mathcal{E}_{\mathfrak{Cqs}}\mbox{ and }\sup p\
^{\shortmid }\mathcal{\varphi }_{1}^{\star }\cap \sup p\ ^{\shortmid }%
\mathcal{\varphi }_{3}^{\star }=\varnothing .
\end{eqnarray*}%
Then, a functional $\ ^{\shortmid }F_{\mathfrak{Cqs}}^{\star }$ is \textit{%
local} if it can be written in the form 
\begin{equation}
\ ^{\shortmid }F_{\mathfrak{Cqs}}^{\star }(\ ^{\shortmid }\mathcal{\varphi }%
^{\star })=\int_{\ ^{\shortmid }\mathcal{U}_{\mathfrak{Cqs}}}\ ^{\shortmid
}\omega ^{\star }(j_{u}^{k}(\ ^{\shortmid }\mathcal{\varphi }^{\star }))\
^{\shortmid }\delta \ ^{\shortmid }\mu _{\mathfrak{Cqs}}(\ ^{\shortmid }u),
\label{measure1}
\end{equation}%
where $\ ^{\shortmid }\omega ^{\star }$ is a function on the jet bundle over 
$\ ^{\shortmid }\mathcal{U}_{\mathfrak{Cqs}},$ subjected to star product
deformations up to $[\hbar ^{1},\kappa ^{1}],$ and $j_{u}^{k}(\ ^{\shortmid }%
\mathcal{\varphi }^{\star })=(\ ^{\shortmid }u,\ ^{\shortmid }\mathcal{%
\varphi }^{\star }(\ ^{\shortmid }u),\ ^{\shortmid }\partial \ ^{\shortmid }%
\mathcal{\varphi }^{\star }(\ ^{\shortmid }u),...),$ with derivatives up to
order $k,$ is the $k$-jet of $\ ^{\shortmid }\mathcal{\varphi }^{\star }$at
the point $\ ^{\shortmid }u,$ up to $[\hbar ^{1},\kappa ^{1}].$ Here we note
that our approach involves quasi-stationary off-diagonal configurations
defined as parametric solutions for $\mathfrak{Cqs.}$ 

\vskip4pt We denote by $\ _{loc}^{\shortmid }\mathcal{F}_{\mathfrak{Cqs}%
}^{\star }$ the spaces of compactly supported smooth local functions on $\
^{\shortmid }\mathcal{E}_{\mathfrak{Cqs}}^{\star }.$ Respectively, the
commutative algebra $\ ^{\shortmid }\mathcal{F}_{\mathfrak{Cqs}}$ of
multilocal functionals is defined as the completion of $\ _{loc}^{\shortmid }%
\mathcal{F}_{\mathfrak{Cqs}}$ with respect ot the point wise product $\
^{\shortmid }F_{\mathfrak{Cqs}}\cdot \ ^{\shortmid }G_{\mathfrak{Cqs}}(\
^{\shortmid }\mathcal{\varphi })=\ ^{\shortmid }F_{\mathfrak{Cqs}}(\
^{\shortmid }\mathcal{\varphi })\ ^{\shortmid }G_{\mathfrak{Cqs}}(\
^{\shortmid }\mathcal{\varphi }).$ Star product deformations result in a
nonassociative $\ _{loc}^{\shortmid }\mathcal{F}_{\mathfrak{Cqs}}^{\star }.$
We can introduce also regular functionals $\ _{reg}^{\shortmid }\mathcal{F}_{%
\mathfrak{Cqs}}$ and say that $\ ^{\shortmid }F_{\mathfrak{Cqs}}\in \
_{reg}^{\shortmid }\mathcal{F}_{\mathfrak{Cqs}}$ if all the derivatives $\
^{\shortmid }F_{\mathfrak{Cqs}}^{(n)}(\ ^{\shortmid }\mathcal{\varphi })$
are smooth, when for all $^{\shortmid }\mathcal{\varphi }\subset \
^{\shortmid }\mathcal{E}_{\mathfrak{Cqs}},n\in \mathbb{N}$ we have $\
^{\shortmid }F_{\mathfrak{Cqs}}^{(n)}(\ ^{\shortmid }\mathcal{\varphi })\in
(\ ^{\shortmid }\mathcal{E}_{\mathfrak{Cqs}}(\ ^{\shortmid }\mathcal{U}_{%
\mathfrak{Cqs}})^{n})^{!\text{ }}.$ Similarly, star product deformations
define $\ _{reg}^{\shortmid }\mathcal{F}_{\mathfrak{Cqs}}^{\star }.$

\subsection{Nonassociative dynamics and nonlinear and linear symmetries}

For a general nonassociative star product (\ref{starpn}), we are not able to
formulate a unique variational derivation of physically important systems of
PDEs in nonassociative geometric flow and gravitational theories, see
details in \cite{partner04,partner05,partner06}. Such equations can be
postulated in abstract geometric form and then we can consider nonholonomic
parametric deformations for a class of solutions and construct some
nonholonomic Lagrangians or Hamiltonians \cite{brunetti09}. In $\mathfrak{%
Cqs,}$ various types of non-trivial compact solutions can be found. To get
around such obstructions with can consider an effective Lagrangian density for
a $\ ^{\shortmid}\mathcal{U}_{\mathfrak{Cqs}}$ with a cutoff function $\
^{\shortmid }f\in \ ^{\shortmid }\mathcal{\check{U}}_{\mathfrak{Cqs}}:=%
\mathcal{C}^{\infty }(\ ^{\shortmid }\mathcal{U}_{\mathfrak{Cqs}},\mathbb{R}%
).$ This way, we can define all relevant s-objects and, for instance,
nonassociative deformations of the Euler-Lagrange derivative in a way which
is independent of $\ ^{\shortmid}f.$ 

\subsubsection{Effective $\star $-deformed Lagrangians and Euler-Lagrange
s-operators}

\label{ss32}A nonassociative star product (\ref{starpn}) deformation of a
generalized Lagrangian $\ ^{\shortmid }L$ on a fixed configuration $\
^{\shortmid }\mathcal{U}_{\mathfrak{Cqs}}\subset \ _{s}^{\shortmid }\mathcal{%
M},$ i.e. $\star :\ ^{\shortmid }L\rightarrow \ ^{\shortmid }L^{\star },$
for a map $\ ^{\shortmid }L:\ ^{\shortmid }\mathcal{U}_{\mathfrak{Cqs}%
}\rightarrow \ _{loc}^{\shortmid }\mathcal{F}_{\mathfrak{Cqs}}$ defined by
three properties:

\begin{enumerate}
\item \textit{additivity}, when $\ ^{\shortmid }L^{\star }(\ ^{\shortmid }%
\mathcal{\varphi }_{1}^{\star }+\ ^{\shortmid }\mathcal{\varphi }_{2}^{\star
}+\ ^{\shortmid }\mathcal{\varphi }_{3}^{\star })=\ ^{\shortmid }L^{\star
}(\ ^{\shortmid }\mathcal{\varphi }_{1}^{\star }+\ ^{\shortmid }\mathcal{%
\varphi }_{2}^{\star })-\ ^{\shortmid }L^{\star }(\ ^{\shortmid }\mathcal{%
\varphi }_{2}^{\star })+\ ^{\shortmid }L^{\star }(\ ^{\shortmid }\mathcal{%
\varphi }_{2}^{\star }+\ ^{\shortmid }\mathcal{\varphi }_{3}^{\star })$ for $%
\ ^{\shortmid }\mathcal{\varphi }_{1}^{\star },\ ^{\shortmid }\mathcal{%
\varphi }_{2}^{\star },\ ^{\shortmid }\mathcal{\varphi }_{3}^{\star }\in \
^{\shortmid }\mathcal{\check{U}}_{\mathfrak{Cqs}}^{\star },$ for $\
^{\shortmid }\mathcal{\check{U}}_{\mathfrak{Cqs}}\rightarrow \ ^{\shortmid }%
\mathcal{\check{U}}_{\mathfrak{Cqs}}^{\star },$ with $\sup p\ ^{\shortmid }%
\mathcal{\varphi }_{1}^{\star }\cap \sup p\ ^{\shortmid}\mathcal{\varphi }%
_{3}^{\star }=\varnothing ;$

\item \textit{support,} when $\sup p(\ ^{\shortmid }L^{\star }(\ ^{\shortmid}%
\mathcal{\varphi }^{\star }))\subseteq \sup p(\ ^{\shortmid }\mathcal{%
\varphi }^{\star })$ is defined by star product deformation of $\sup p(\
^{\shortmid }L(\ ^{\shortmid }\mathcal{\varphi }))\subseteq \sup p(\
^{\shortmid }\mathcal{\varphi });$

\item \textit{covariance}, when for a local Minkowski spacetime and then for
a cofiber, we consider an isometry group $\ ^{\shortmid }\mathcal{P}
=(P_{+}^{\uparrow },\ ^{\shortmid }P_{+}^{\uparrow }),$ where $%
P_{+}^{\uparrow }$ is the proper orthocronous Poincar\'{e} group, we require
the property $\ ^{\shortmid }L^{\star }(\ ^{\shortmid }f)(\rho^{\ast }\
^{\shortmid }\mathcal{\varphi }^{\star })=\ ^{\shortmid }L^{\star }(\rho
_{\ast }\ ^{\shortmid }f)(\ ^{\shortmid }\mathcal{\varphi }^{\star }),$ for
every $\rho \in \ ^{\shortmid }\mathcal{P}.$
\end{enumerate}

The abstract geometric formalism \cite{misner,partner02} can be extended to
the spaces of all generalized Lagrangians $\ ^{\shortmid }\mathcal{L}=\{\
^{\shortmid }L\},$ when we assume that the Lagrangians satisfy the condition 
$\ ^{\shortmid }L^{\ast }=\ ^{\shortmid }L,$ where $\ast $ is an involution
which is different from the star product. Such an involution is not just the
complex conjugation, but for graded geometries, it also swaps the order of
factors. Then all geometric s-objects are subjected to star product
deformations, $\star :\ ^{\shortmid }\mathcal{L}\rightarrow \ ^{\shortmid }%
\mathcal{L}^{\star },$ which allow to work with nonassociative generalized
Lagrangians $\ ^{\shortmid }\mathcal{L}^{\star }.$ 

\vskip4pt Even in nonassociative geometry with general twist products, we
are not able to formulate a general and uniquely defined variational
calculus, the abstract geometric formalism allows us to perform BV classical
and quantum constructions for any class of quasi-stationary configurations $%
\mathfrak{Cqs.}$ The equivalence classes of  $\ ^{\shortmid}f $  on the space of 
generating and integration functions (\ref{integrfunctrf})  are related via
nonlinear symmetries (\ref{nonlinsym}). In certain effective parametric
forms, we can always define corresponding actions $\ ^{\shortmid }S^{\star
}(\ ^{\shortmid }L^{\star })$ considering equivalence classes of Lagrangians 
$\ ^{\shortmid }L_{1}^{\star } $ and $\ ^{\shortmid }L_{2}^{\star },$ which
are equivalent, $\ ^{\shortmid }L_{1}^{\star }\sim \ ^{\shortmid
}L_{2}^{\star}, $ if $\sup p(\ ^{\shortmid }L_{1}- \ ^{\shortmid }L_{2})(\
^{\shortmid}f)\subset \sup p(d\ ^{\shortmid }f),\forall \ ^{\shortmid }f\in
\ ^{\shortmid }\mathcal{\check{U}}_{\mathfrak{Cqs}}.$ In brief, we shall
write $\ ^{\shortmid }S^{\star }$ instead of $\ ^{\shortmid }S_{\mathfrak{Cqs%
}}^{\star },$ or $\ ^{\shortmid}S^{\star }(\ ^{\shortmid }L_{\mathfrak{Cqs}%
}^{\star })$, if such simplifications do not result in ambiguities. 

\vskip4pt Using canonical nonholonomic geometric variables (\ref%
{canongeomobj}) and (\ref{deformgauge}), two important examples of
nonassociative generalized Lagrangians are written in the form:%
\begin{eqnarray}
\ _{\varphi }^{\shortmid }L^{\star }(\ ^{\shortmid }f)[\ ^{\shortmid }%
\mathcal{\varphi }^{\star }] &=&\frac{1}{2}\int_{\ ^{\shortmid }\mathcal{U}_{%
\mathfrak{Cqs}}}\left( \ ^{\shortmid }\widehat{\mathbf{D}}_{\alpha
_{s}}^{^{\star }}\ ^{\shortmid }\mathcal{\varphi }^{\star }\ ^{\shortmid }%
\widehat{\mathbf{D}}^{^{\star }\alpha _{s}}\ ^{\shortmid }\mathcal{\varphi }%
^{\star }-m^{2}(\ ^{\shortmid }\mathcal{\varphi }^{\star })^{2}\right) \
^{\shortmid }f\ \delta ^{8}\mu ,\mbox{ free scalar field };  \notag \\
\ _{gr}^{\shortmid }L^{\star }(\ ^{\shortmid }f)[\ \ _{s}^{\shortmid }%
\widehat{\mathcal{A}}^{\star }] &=&-\frac{1}{2}\int_{\ ^{\shortmid }\mathcal{%
U}_{\mathfrak{Cqs}}}\ ^{\shortmid }f\quad tr(\ ^{\shortmid }\widehat{%
\mathcal{F}}^{^{\star }}\wedge (\divideontimes \ ^{\shortmid }\widehat{%
\mathcal{F}}^{^{\star }})),\mbox{ nonassociative gauge gravitational field },
\label{effectgauge}
\end{eqnarray}%
where trace, $tr,$ is in the Killing metric over the Lie algebra of the gauge group; the effective mass $m$ includes distortions of connections; and the measure $\delta ^{8}\mu $ is defined by a chosen 
$\mathfrak{Cqs.}$ 

\vskip4pt For any $\ ^{\shortmid }L^{\star }\in \ ^{\shortmid }\mathcal{L}%
^{\star },\ ^{\shortmid }\mathcal{\varphi }^{\star }\in \ ^{\shortmid }%
\mathcal{E}_{\mathfrak{Cqs}}^{\star },$ we define a functional%
\begin{equation*}
\ ^{\shortmid }\delta \ ^{\shortmid }L^{\star }(\ ^{\shortmid }\mathcal{%
\varphi }_{1}^{\star })[\ ^{\shortmid }\mathcal{\varphi }^{\star }]:=\
^{\shortmid }L^{\star }(\ ^{\shortmid }f)[\ ^{\shortmid }\mathcal{\varphi }%
^{\star }+\ ^{\shortmid }\mathcal{\varphi }_{1}^{\star }]-\ ^{\shortmid
}L^{\star }(\ ^{\shortmid }f)[\ ^{\shortmid }\mathcal{\varphi }^{\star }],
\end{equation*}%
when $\ ^{\shortmid }\delta \ ^{\shortmid }L^{\star }$ acts on $\
^{\shortmid }\mathcal{U}_{\mathfrak{Cqs}}\times \ ^{\shortmid }\mathcal{E}_{%
\mathfrak{Cqs}}^{\star }$ and $\ ^{\shortmid }f\equiv 1$ on $\sup p\
^{\shortmid }\mathcal{\varphi }_{1}^{\star }$ and such a map does not depend
on choice of $\ ^{\shortmid }f.$ This functional allows to introduce in
effective form the \textit{Euler-Lagrange derivative} of $\ ^{\shortmid
}S^{\star },$%
\begin{eqnarray}
&&\ ^{\shortmid }d\ ^{\shortmid }S^{\star }:\ ^{\shortmid }\mathcal{E}_{%
\mathfrak{Cqs}}^{\star }\rightarrow \ ^{\shortmid }\mathcal{E}_{\mathfrak{Cqs%
}}^{^{\prime }C}\mbox{ defined by }  \label{effectelag} \\
&&<\ ^{\shortmid }d\ ^{\shortmid }S^{\star }(\ ^{\shortmid }\mathcal{\varphi 
}^{\star }),\ ^{\shortmid }\mathcal{\varphi }_{1}^{\star }>:=\lim_{\varsigma
\rightarrow 0}\frac{1}{\varsigma }\ ^{\shortmid }\delta \ ^{\shortmid
}L^{\star }(\varsigma \ ^{\shortmid }\mathcal{\varphi }_{1}^{\star })[\
^{\shortmid }\mathcal{\varphi }^{\star }]=\int \frac{\ ^{\shortmid }\delta \
^{\shortmid }L^{\star }(\ ^{\shortmid }f)}{\ ^{\shortmid }\delta \
^{\shortmid }\mathcal{\varphi }^{\star }}\ ^{\shortmid }\mathcal{\varphi }%
_{1}^{\star }(\ ^{\shortmid }u),\mbox{ for }  \notag \\
&&\ ^{\shortmid }\mathcal{\varphi }_{1}^{\star }\in \ ^{\shortmid }\mathcal{E%
}_{\mathfrak{Cqs}}^{C},\frac{\ ^{\shortmid }\delta \ ^{\shortmid }L^{\star
}(\ ^{\shortmid }f)}{\ ^{\shortmid }\delta \ ^{\shortmid }\mathcal{\varphi }%
^{\star }}\in (\ ^{\shortmid }\mathcal{E}_{\mathfrak{Cqs}}^{\star })\subset
\ ^{\shortmid }\mathcal{E}_{\mathfrak{Cqs}}^{^{\prime }C};  \notag \\
&&\ ^{\shortmid }d\ ^{\shortmid }S^{\star }(\ ^{\shortmid }\mathcal{\varphi }%
^{\star })\equiv 0\mbox{ defines the effective field equations };  \notag \\
&&\ _{Sol}^{\shortmid }\mathcal{E}_{\mathfrak{Cqs}}^{\star }%
\mbox{ denotes
the spaces of solutions defined as the zero locus of the 1-form }d\
^{\shortmid }S^{\star },\ _{Sol}^{\shortmid }\mathcal{E}_{\mathfrak{Cqs}%
}^{\star }\subset \ ^{\shortmid }\mathcal{E}_{\mathfrak{Cqs}}^{\star }; 
\notag \\
&&\ _{Sol}^{\shortmid }\mathcal{F}_{\mathfrak{Cqs}}%
\mbox{ denotes the space
of off-shell functionals in the space of functionals on  }\
_{Sol}^{\shortmid }\mathcal{E}_{\mathfrak{Cqs}}^{\star }.  \notag
\end{eqnarray}

In above formulas, we can identify the space of solutions of $\ ^{\shortmid
}S^{\star }(\ ^{\shortmid }\mathcal{\varphi }^{\star })\equiv 0$\ to be
equivalent with the class of parametric solutions for $\ ^{\shortmid }%
\mathcal{E}_{\mathfrak{Cqs}}^{\star }=\ _{s}^{\shortmid }\mathcal{E}^{\star
}(\ ^{\shortmid }\mathcal{U}_{\mathfrak{Cqs}}^{\star })$ (\ref{dvbundle}).
Here we also note that in all formulas involving $\ ^{\shortmid }\mathcal{%
\varphi }^{\star }$ or $\ ^{\shortmid }\mathcal{\varphi }$, we can introduce
abstract indices $\check{\alpha},$ with inverse hat, for (nonassociative)
fields labeling degree of freedom of corresponding scalar fields, any type of
(gravitational) gauge fields etc. So, we can write $\ ^{\shortmid }\mathcal{%
\varphi }^{\star }=\{\ ^{\shortmid }\mathcal{\varphi }_{\check{\alpha}%
}^{\star }\}$ and use notations like $\frac{\ ^{\shortmid }\delta S^{\star }%
}{\ ^{\shortmid }\delta \ ^{\shortmid }\mathcal{\varphi }_{\check{\alpha}%
}^{\star }(\ ^{\shortmid }u)}$ for $\frac{\ ^{\shortmid }\delta L^{\star }(\
^{\shortmid }f)}{\ ^{\shortmid }\delta \ ^{\shortmid }\mathcal{\varphi }_{%
\check{\alpha}}^{\star }(\ ^{\shortmid }u)}$ evaluated at $\ ^{\shortmid
}f\equiv 1.$

\subsubsection{Nonassociative configuration phase spaces' symmetries and
Noether theorem}

A specific type of nonassociative nonlinear symmetries (\ref{nonlinsym})
characterize the quasi-stationary off-diagonal solutions $\ ^{\shortmid }%
\mathbf{g}^{\star }[\mathfrak{Cqs}]$ (\ref{offdiagpolfr}). Introducing
effective Lagrangians and actions as in (\ref{effectelag}), we impose other
types of symmetries for nonassociative parametric gravitational and matter
field interactions. Let us analyse how additional symmetries characterize
respective effective systems. 

\vskip4pt Geometrically, we study additional symmetries defined as a vector
field $\ ^{\shortmid }\mathcal{X}^{\star }(\ ^{\shortmid }u)$ on $\
^{\shortmid} \mathcal{E}_{\mathfrak{Cqs}}^{\star }$ such that 
\begin{equation*}
\ ^{\shortmid }\partial _{\ ^{\shortmid }\mathcal{X}}^{\star }\ ^{\shortmid
}S^{\star }\equiv 0,\mbox{ where }\ ^{\shortmid }\partial _{\ ^{\shortmid }%
\mathcal{X}}^{\star }\ ^{\shortmid }S^{\star }:=\int \frac{\ ^{\shortmid
}\delta L^{\star }(\ ^{\shortmid }f)}{\ ^{\shortmid }\delta \ ^{\shortmid }%
\mathcal{\varphi }^{\star }}\ ^{\shortmid }\mathcal{X}^{\star },\
^{\shortmid }f\equiv 1\mbox{ on }\sup p\ \ ^{\shortmid }\mathcal{X}^{\star }.
\end{equation*}%
In abstract geometric form, 
\begin{equation}
\ ^{\shortmid }\mathcal{X}^{\star }=\int \ ^{\shortmid }\mathcal{X}^{\star
}(\ ^{\shortmid }u)\frac{\ ^{\shortmid }\delta }{\ ^{\shortmid }\delta \
^{\shortmid }\mathcal{\varphi }^{\star }}\in \Gamma (T\ ^{\shortmid }%
\mathcal{E}_{\mathfrak{Cqs}}^{\star })  \label{aux2}
\end{equation}%
is identified with a map from $\ ^{\shortmid }\mathcal{E}_{\mathfrak{Cqs}%
}^{\star }$ to $\ ^{\shortmid }\mathcal{E}_{\mathfrak{Cqs}}^{\star } \
^{\shortmid }\mathcal{E}_{\mathfrak{Cqs}}^{\star C\mathbb{C}}$ for sections $%
\Gamma ,$ where $\mathbb{C}$ is used for complexification. We can introduce
the anti-fields $\frac{\ ^{\shortmid }\delta }{\ ^{\shortmid }\delta \
^{\shortmid }\mathcal{\varphi }^{\star }}\equiv \ _{\ddag }^{\shortmid }%
\mathcal{\varphi }^{\star }$ identified as a basis on the fiber $T_{\varphi
}\ ^{\shortmid }\mathcal{E}_{\mathfrak{Cqs}}^{\star }$ and consider $\
^{\shortmid }\mathcal{X}^{\star }$ and $\ _{\ddag }^{\shortmid }\mathcal{%
\varphi }^{\star }$ as s-vectors for configurations $\ ^{\shortmid }\mathbf{g%
}^{\star }[\mathfrak{Cqs}]$ involving s-adapted frames. The value $\
^{\shortmid }\partial _{\ ^{\shortmid }\mathcal{X}}^{\star }\ ^{\shortmid
}S^{\star }$ is just the insertion of 1-form $\ ^{\shortmid }d\ ^{\shortmid
}S^{\star }$ into a vector field $\ ^{\shortmid }\mathcal{X}^{\star }.$

For our research, we can focus on local symmetries of a nonassociative
system which can be expressed as $\ ^{\shortmid }\mathcal{X}^{\star }=I+\
^{\shortmid }\omega ^{\star }\ ^{\shortmid }\rho ^{\star }(\ ^{\shortmid}\xi
),$ where $I$ is a symmetry that vanishes identically on $\
_{Sol}^{\shortmid }\mathcal{E}_{\mathfrak{Cqs}}^{\star }$ (\ref{effectelag}%
), $\ ^{\shortmid }\omega ^{\star }\ $\ is a local function defied as a map $%
\ ^{\shortmid }\mathcal{E}_{\mathfrak{Cqs}}^{\star }\rightarrow \
^{\shortmid }\mathcal{U}_{\mathfrak{Cqs}}$ and $\ ^{\shortmid }\rho ^{\star} 
$ determine infinitesimal symmetries, when $\ ^{\shortmid }\xi \in \
^{\shortmid }\mathfrak{g}_{c}.$ Applying such formulas, the multiplication
with an element of $\Gamma (T\ ^{\shortmid }\mathcal{E}_{\mathfrak{Cqs}%
}^{\star }),$ and $\ ^{\shortmid }\rho ^{\star }: \ ^{\shortmid }\mathfrak{g}
_{c} \rightarrow \Gamma (T\ ^{\shortmid }\mathcal{E}_{\mathfrak{Cqs}}^{\star
})$ is a double Lie-algebra morphism defined by a given local action $\
^{\shortmid }\sigma ^{\star }$ of Lie d-algebra $\ ^{\shortmid }\mathfrak{g}%
_{c}$ on $\ _{Sol}^{\shortmid }\mathcal{E}_{\mathfrak{Cqs}}^{\star }$ when 
\begin{equation}
\ ^{\shortmid }\rho ^{\star }(\ ^{\shortmid }\xi )\ ^{\shortmid }F_{%
\mathfrak{Cqs}}[\ ^{\shortmid }\mathcal{\varphi }^{\star }]:=<\ ^{\shortmid
}F_{\mathfrak{Cqs}}^{(1)}(\ ^{\shortmid }\mathcal{\varphi }^{\star }),\
^{\shortmid }\sigma ^{\star }(\ ^{\shortmid }\xi )\ ^{\shortmid }\mathcal{%
\varphi }^{\star }>\equiv \int_{\ ^{\shortmid }\mathcal{U}_{\mathfrak{Cqs}}}%
\frac{\ ^{\shortmid }\delta F^{\star }(\ ^{\shortmid }f)}{\ ^{\shortmid
}\delta \ ^{\shortmid }\mathcal{\varphi }^{\star }(\ ^{\shortmid }u)}\
^{\shortmid }\sigma ^{\star }(\ ^{\shortmid }\xi )\ ^{\shortmid }\mathcal{%
\varphi }^{\star }(\ ^{\shortmid }u).  \label{aux01}
\end{equation}%
In these formulas, we use $\ ^{\shortmid }\mathfrak{g}_{c}$ with a subscript
"c" stating that the maps are on corresponding spaces of smooth compactly
supported sections over nonassociative deformed vector bundles over $\
^{\shortmid }\mathcal{U}_{\mathfrak{Cqs}}$ and when the action $\
^{\shortmid }\sigma ^{\star }$ on $\ _{Sol}^{\shortmid }\mathcal{E}_{%
\mathfrak{Cqs}}^{\star }$ is defined to be local. For linear parametric
deformations, we can consider nonassociative gauge gravity models with $\
^{\shortmid }\rho ^{\star }\simeq \rho ^{\star }$ when nonassociativity is
encoded into functionals like $F^{\star },$ or $\ ^{\shortmid }\mathcal{%
\varphi }^{\star }$ and $\ ^{\shortmid }\sigma ^{\star }.$ 

\vskip4pt For any nonassociative gauge model defined by a quasi-stationary
solution of (\ref{nonassocymgreq1}), the presence of local symmetries
implies that the effective equations of motion $\ _{Sol}^{\shortmid }%
\mathcal{E}_{\mathfrak{Cqs}}^{\star }$ (\ref{effectelag}) have orbits of the
action $\ ^{\shortmid}\sigma ^{\star }$ (i.e. have redundancies). This is
formulated mathematically as the second Noether theorem: 
\begin{equation}
\int \frac{\ ^{\shortmid }\delta ^{\shortmid }S^{\star }}{\ ^{\shortmid
}\delta \ ^{\shortmid }\mathcal{\varphi }_{\check{\alpha}}^{\star }\ }\
^{\shortmid }\mathcal{X}_{\check{\alpha}}^{\star }(\ ^{\shortmid }u)~\
^{\shortmid }\delta ^{8}\ ^{\shortmid }\mu (\ ^{\shortmid }u)=\int \
^{\shortmid }\mathcal{\varphi }_{\check{\beta}}^{\star }(Q_{~\check{\alpha}%
}^{\check{\beta}})^{\ast }\frac{\ ^{\shortmid }\delta ^{\shortmid }S^{\star }%
}{\ ^{\shortmid }\delta \ ^{\shortmid }\mathcal{\varphi }_{\check{\alpha}%
}^{\star }\ }\ ^{\shortmid }\delta ^{8}\ ^{\shortmid }\mu =0,
\label{noetht2}
\end{equation}%
for $\ast $ denoting the formal adjoint of a differential operator obtained
using integration by parts. Formulas (\ref{noetht2}) state the condition to
be a symmetry for any local and compactly supported s-vector can be
expressed as a differential s-operator%
\begin{equation*}
\ ^{\shortmid }\mathcal{X}_{\check{\alpha}}^{\star }(\ ^{\shortmid }u)[\
^{\shortmid }\mathcal{\varphi }^{\star }]=Q_{~\check{\alpha}}^{\check{\beta}%
}(\ ^{\shortmid }\mathcal{\varphi }^{\star })\ ^{\shortmid }\mathcal{\varphi 
}_{\check{\beta}}^{\star }(\ ^{\shortmid }u)=a(\ ^{\shortmid }u)[\
^{\shortmid }\mathcal{\varphi }^{\star }]+b^{\alpha }(\ ^{\shortmid }u)[\
^{\shortmid }\mathcal{\varphi }^{\star }]\ ^{\shortmid }\widehat{\mathbf{D}}%
_{\alpha }^{\star }\ ^{\shortmid }\mathcal{\varphi }^{\star }(\
^{\shortmid}u)+...
\end{equation*}%
The effective  equations of motion with $\frac{\ ^{\shortmid }\delta ^{\shortmid
}S^{\star }}{\ ^{\shortmid }\delta \ ^{\shortmid }\mathcal{\varphi }_{\check{%
\alpha}}^{\star }\ }$ and encoding nonassociative data are not all
independent and related both to linear and nonlinear symmetries (\ref%
{nonlinsym}) stated for quasi-stationary solutions $\ ^{\shortmid }\mathbf{g}%
^{\star }[\mathfrak{Cqs}].$ For a general nonassociative gauge gravity
theory with twisted star product, it is not possible to define a unique
variational calculus and prove a general form of second Noether theorem (\ref%
{noetht2}) as in \cite{fulp03}. Such nonholonomic constructions can be
performed in abstract geometric form with further parametric decompositions.
This allows to define the space $\ _{inv}^{\shortmid }\mathcal{F}_{\mathfrak{%
Cqs}}^{\star }$ of functionals on the solution space $\ _{Sol}^{\shortmid }%
\mathcal{E}_{\mathfrak{Cqs}}^{\star }$ (\ref{effectelag}) that are invariant
on the actions of symmetries $\ ^{\shortmid }\rho ^{\star }$ encoding in
parametric form off-diagonal solutions of gauge gravitational equations with
nonassociative data.

\subsection{Ghosts and the BV complex of nonassociative quasi-stationary solutions}

Any class of quasi-stationary or locally anisotropic solutions $\
^{\shortmid }\mathbf{g}^{\star }[\mathfrak{Clacs}]$ or $\ ^{\shortmid }%
\mathbf{g}^{\star }[\mathfrak{Cqs}]$ (\ref{baspropsol}) may involve
singularities, nonassociative data, gravitational polarizations, solitonic
hierarchies etc. In the presence of local symmetries, the equations of
motion (\ref{effectelag}) have redundancies when the Cauchy problem is not
well posed. In certain cases, a redundancy can be removed by taking the
quotient by the action of infinitesimal symmetries (\ref{noetht2}), which
does not solve the issues related to the existence of nonlinear symmetries (%
\ref{nonlinsym}) and the non-variational properties of general twisted star
products. We can approach such problems following the guiding idea of
homology when (instead of taking quotients) we go to a large class of
nonholonomic geometries with distortion of connections. For certain classes
of nonassociative physically important systems of PDEs, we can prove general
decoupling and integration properties and select equivalent classes of
solutions with physically important properties, which are better behaved and
can be used as a first step towards quantization. The goal of this
subsection is to characterize the space $\ _{inv}^{\shortmid }\mathcal{F}_{%
\mathfrak{Cqs}}^{\star }$ of functionals on $\ _{Sol}^{\shortmid }\mathcal{E}%
_{\mathfrak{Cqs}}^{\star }$ in a way that will facilitate quantization. Here
we note that nonholonomic methods for deformation quantization of GR and
Finsler-like MGT on phase spaces was elaborated in \cite{vacaru07,vacaru13}.
In this work, those nonholonomic geometric constructions are extended for
nonassociative theories with twisted star product. 

\subsubsection{Nonassociative local symmetries and ghosts}

The homological interpretation of the associative and commutative parts of
the space $\ ^{\shortmid }\mathcal{F}_{\mathfrak{Cqs}}^{\star }$ (of
functionals on the solution space $\ _{Sol}^{\shortmid }\mathcal{E}_{%
\mathfrak{Cqs}}^{\star }$ (\ref{effectelag})) denoted respectively $\
^{\shortmid }\mathcal{F}_{\mathfrak{Cqs}}$ and $\ ^{\shortmid }\mathcal{E}_{%
\mathfrak{Cqs}}$, is similar to that reviewed in section 2.3 of \cite%
{rejzner20} and formulated in rigorous mathematical form in \cite{rejzner16}%
. Those proofs can not be extended for general twist products when a unique
variational formulation of physical models is not possible. Nevertheless, we
can study possible physical implications of nonlinear and linear symmetries
of respective classes of solutions that can be modified for parametric
deformations on phase spaces to encode nonassociative geometric data using a
respective abstract and nonholonomic geometric formalism and R-flux
deformations. 

\vskip4pt We consider the Lie d-algebra $\ ^{\shortmid }\mathfrak{g}_{c}$ on 
$\ _{Sol}^{\shortmid }\mathcal{E}_{\mathfrak{Cqs}}^{\star }$ characterizing
the infinitesimal local symmetries as we explained in (\ref{aux01}).
Considering such symmetries as s-adapted derivations on functionals that are
themselves compactly supported and considered for respective classes of
solutions, we can work directly with $\ ^{\shortmid }\mathfrak{g}$ (dropping
the condition of compact support for symmetries on phase space and further
deformations). For our purposes, we shall work with the space of
symmetry-invariant variables defined by certain functionals $\ ^{\shortmid }%
\mathcal{B}^{\star} $ such that 
 $\ ^{\shortmid }\partial _{\ ^{\shortmid }\rho ^{\star }(\ ^{\shortmid }\xi
)\ }\ ^{\shortmid }\mathcal{B}^{\star }=0;\ \forall \ ^{\shortmid }\xi \in \
^{\shortmid }\mathfrak{g}$. For physical applications, such conditions can be satisfied in linear
parametric form. Here we note that the spaces of invariants under the action
of a Lie algebra as in the above formula and respective homological algebra
and homology groups constructions can be characterized using the
Chevalley-Eilenberg complex. We cite \cite{rejzner16,rejzner20} for precise
definitions and emphasize that in this paper we work on phase spaces with d-algebras, i.e.
couples of algebras corresponding to h- and c-spaces, which are star-product
deformed in parametric form. Advanced topological homologic methods which
are very sophisticate for researchers in theoretical physics are not
considered in this work. 

\vskip4pt For further geometric constructions, we use a s-vector co-bundle $%
\ _{s}^{\shortmid }\mathcal{E}(\ _{s}^{\shortmid }\mathcal{M})$ on
respective phase space $\ _{s}^{\shortmid}\mathcal{M}$, when a star product (%
\ref{starpn}) deforms such spaces into respective nonassociative ones
labelled by a $\star $-symbol, $\ _{s}^{\shortmid }\mathcal{E}^{\star }$ on $%
\ _{s}^{\shortmid }\mathcal{M}^{\star },$ see details in \cite{partner07}. A 
\textit{graded s-adapted phase s-vector co-bundle} is by definition $\
_{s}^{\shortmid }\overline{\mathcal{E}}:= \ _{s}^{\shortmid }\mathcal{E}%
\oplus \ ^{\shortmid }\mathfrak{g[1],}$ and, in $\star $-deformed form, $\
_{s}^{\shortmid }\overline{\mathcal{E}}^{\star }:=\ _{s}^{\shortmid }%
\mathcal{E}^{\star }\oplus \ ^{\shortmid }\mathfrak{g[1],}$ when the
functionals on $\ ^{\shortmid }\mathfrak{g[1]}$ are identified with the
exterior d-algebra over the duals $\ ^{\shortmid }\mathfrak{g}^{\prime },$
i.e. $\wedge \ ^{\shortmid }\mathfrak{g}^{\prime},$ for $\ ^{\shortmid }%
\mathfrak{g}=(h\ ^{\shortmid }\mathfrak{g},c\ \mathfrak{^{\shortmid }%
\mathfrak{g})}$. The ghost phase space fields are introduced as evaluation
functionals 
\begin{equation}
\ ^{\shortmid }\xi ^{I}(\ ^{\shortmid }u)=\ ^{\shortmid }c^{I}(\ ^{\shortmid
}u)[\ ^{\shortmid }\xi ],  \label{ghosts}
\end{equation}%
where an abstract index $I$ is used. Such ghosts are related to the graded
d-algebra of the Chevalley-Eilenberg complex 
\begin{eqnarray}
\mathcal{C}\ _{s}^{\shortmid }\mathcal{E}&:= &\mathcal{C}_{[ml]}^{\infty }(\
_{s}^{\shortmid }\overline{\mathcal{E}},\mathbb{C})=(\wedge \ ^{\shortmid }%
\mathfrak{g}^{\prime }\widehat{\otimes }\ ^{\shortmid }\mathcal{F}_{%
\mathfrak{Cqs}},\ _{ce}^{\shortmid }\gamma );%
\mbox{ and, for star product
deformations,}  \label{cec} \\
\mathcal{C}\ _{s}^{\shortmid }\mathcal{E}^{\star }&:= &\mathcal{C}%
_{[ml]}^{\infty }(\ _{s}^{\shortmid }\overline{\mathcal{E}}^{\star },\mathbb{%
C})=(\wedge \ ^{\shortmid }\mathfrak{g}^{\prime }\widehat{\otimes }\
^{\shortmid }\mathcal{F}_{\mathfrak{Cqs}}^{\star },\ _{ce}^{\shortmid
}\gamma ),  \notag
\end{eqnarray}%
where $[ml]\,\,$ refers to the space of multilocal functionals on $\
_{s}^{\shortmid }\overline{\mathcal{E}}.$ In (\ref{cec}), $\widehat{\otimes}$
is the appropriately completed tensor product and the grading of $\mathcal{C}%
\ _{s}^{\shortmid }\mathcal{E}$ is called the pure ghost number $%
\#pg=(\#_{h}+\#_{c})pg,$ defined as a sum of h- and c-ghosts. 

\vskip4pt Using the complex (\ref{cec}), we introduce the nonassociative
Chevalley-Eilenberg differential $\ _{ce}^{\shortmid }\gamma ,$ defined by 
\begin{eqnarray*}
(\ _{ce}^{\shortmid }\gamma \ ^{\shortmid }\mathcal{B}^{\star })(\
^{\shortmid }\mathcal{\varphi }^{\star },\ ^{\shortmid }\xi ) &:=&\partial
_{\ ^{\shortmid }\rho (\ ^{\shortmid }\xi )\ }\ ^{\shortmid }\mathcal{B}%
^{\star }(\ ^{\shortmid }\mathcal{\varphi }^{\star }),\mbox{ for }\
^{\shortmid }\xi \in \ ^{\shortmid }\mathfrak{g}^{\prime },\mbox{ or } \\
\ _{ce}^{\shortmid }\gamma \ ^{\shortmid }\mathcal{B}^{\star } &=&\partial
_{\ ^{\shortmid }\rho (\ ^{\shortmid }c)\ }\ ^{\shortmid }\mathcal{B}^{\star
},\mbox{ in terms of evaluation funtionals, i.e. ghosts}.
\end{eqnarray*}%
In these formulas, $\ ^{\shortmid }\mathcal{B}^{\star }\in \ ^{\shortmid }%
\mathcal{F}_{\mathfrak{Cqs}}^{\star }$ and $\ _{ce}^{\shortmid }\gamma \
^{\shortmid }\mathcal{B}^{\star }\in \mathcal{C}_{[ml]}^{\infty }(\
_{s}^{\shortmid }\mathcal{E}^{\star },\ ^{\shortmid }\mathfrak{g});$ and \ $%
\ _{ce}^{\shortmid }\gamma $ encodes the action $^{\shortmid }\rho $ of the
gauges d-algebra $\ ^{\shortmid }\mathfrak{g}$ on $\ ^{\shortmid }\mathcal{F}%
_{\mathfrak{Cqs}}^{\star }.$ We express for the ghosts field (\ref{ghosts}) 
\begin{equation*}
\ _{ce}^{\shortmid }\gamma \ ^{\shortmid }c=-\frac{1}{2}[\ ^{\shortmid }c,\
^{\shortmid }c]\mbox{ and }\ _{ce}^{\shortmid }\gamma \ ^{\shortmid }\omega
^{\star }(\ ^{\shortmid }\xi _{1},\ ^{\shortmid }\xi _{2})=\ ^{\shortmid
}\omega ^{\star }([\ ^{\shortmid }\xi _{1},\ ^{\shortmid }\xi _{2}])\in
\wedge ^{2}\ ^{\shortmid }\mathfrak{g}^{\prime }
\end{equation*}%
for any form $\ ^{\shortmid }\omega ^{\star }\in \ ^{\shortmid }\mathfrak{g}%
^{\prime }.$ Because $\ _{ce}^{\shortmid }\gamma \ ^{\shortmid }\mathcal{B}%
^{\star }\equiv 0$ if $\ ^{\shortmid }\mathcal{B}^{\star }\in \
_{inv}^{\shortmid }\mathcal{F}_{\mathfrak{Cqs}}^{\star },$ we can consider
that the zero holonomy group $H^{0}(\mathcal{C}\ _{s}^{\shortmid }\mathcal{E}%
^{\star })$ characterizes the nonassociative gauge invariant functionals.
Such formulas generalize on (nonassociative) phase space the constructions
from section 2.3.2 in \cite{rejzner20}. 

\subsubsection{Nonassociative classical BV complex}

Let us characterize the nonassociative geometric properties of the space $\
_{inv}^{\shortmid }\mathcal{F}_{\mathfrak{Cqs}}^{\star }.$ We consider
s-vector fields on $\ _{s}^{\shortmid }\overline{\mathcal{E}}^{\star }:=\
_{s}^{\shortmid }\mathcal{E}^{\star }\oplus \ ^{\shortmid }\mathfrak{g[1]}$
instead of s-vector fields on $\ _{s}^{\shortmid }\mathcal{E}^{\star }.$ We
call $\ _{s}^{\shortmid }\overline{\mathcal{E}}^{\star }$ an extended
configuration phase space (equivalently, graded phase space) involving a
nonholonomic splitting (with corresponding nonholonomic h- and c-splitting).
For quasi-stationary configurations, we take $\ _{s}^{\shortmid }\mathcal{E}%
^{\star }$ as $\ _{Sol}^{\shortmid }\mathcal{E}_{\mathfrak{Cqs}}^{\star }$
and write $\ _{Sol}^{\shortmid }\overline{\mathcal{E}}^{\star }.$ Now, we
are able to construct respective nonholonomic s-adapted, $\ ^{\shortmid }%
\mathcal{BV},$ and nonassociative, $\ ^{\shortmid }\mathcal{BV}^{\star },$
BV complexes when $\star :\ ^{\shortmid }\mathcal{BV\rightarrow }\
^{\shortmid }\mathcal{BV}^{\star }.$ We may distinguish such complexes by
respective labels like $\ _{Sol}^{\shortmid }\mathcal{BV}^{\star }$ or $\
^{\shortmid }\mathcal{BV}_{\mathfrak{Cqs}}^{\star }.$ In abstract geometric
form, a nonassociative BV complex is defined by an underlying d-algebra of
multilocal polyvector fields (which can be s-adapted) on $\ _{s}^{\shortmid }%
\overline{\mathcal{E}}^{\star }$. This space is modelled as the space of multilocal
compactly supported functional on the graded nonholonomic s-adapted
manifold (generalized cotangent s-vector bundle)%
\begin{equation}
T^{\ast }[-1]\ _{s}^{\shortmid }\overline{\mathcal{E}}^{\star }\equiv \
_{s}^{\shortmid }\mathcal{E}^{\star }[0]\oplus \ ^{\shortmid }\mathfrak{%
g[1]\oplus (\ _{s}^{\shortmid }\mathcal{E}^{\star })}^{!}\mathfrak{%
[-1]\oplus \ ^{\shortmid }\mathfrak{g}}^{!}\mathfrak{\mathfrak{[-2]},}
\label{extendsp}
\end{equation}%
where labels (with ! and possible additionally ones with $Sol,$ or $%
\mathfrak{Cqs}$) are explained in footnote \ref{fnconv}. In explicit form,
the elements of $\ ^{\shortmid }\mathcal{BV}^{\star }$ are multilocal
functionals of the nonassociative field multiplets $\ ^{\shortmid }\mathcal{%
\varphi }^{\star }=\{\ ^{\shortmid }\mathcal{\varphi }_{\check{\alpha}%
}^{\star }\}$ and of corresponding antifields $\ _{\ddag }^{\shortmid } 
\mathcal{\varphi }^{\star }=\{\ \ _{\ddag }^{\shortmid }\mathcal{\varphi }_{%
\check{\alpha}}^{\star }\}$ as defined for formulas (\ref{aux2}). In a more
general context, an index $\check{\alpha}$ runs through all the physical and
ghost indices on nonassociative phase spaces, with possible nonholonomic
dyadic or h- and c-splitting of phase space. We use the convention that the
phase anti-field are right derivatives $\frac{_{r}^{\shortmid }\delta }{%
\delta \ ^{\shortmid }\mathcal{\varphi }_{\check{\alpha}}},$ with for graded
manifolds are different from the left derivatives $\frac{_{l}^{\shortmid
}\delta }{\delta \ ^{\shortmid }\mathcal{\varphi }_{\check{\alpha}}}.$ For
phase and spacetime indices, such derivatives can be defined in N-adapted or
s-adapted forms and then subjected to star product deformations.\footnote{%
Both in h- and c-forms, the d-algebra has two gradings: the ghost numbers $%
\#gh=(\#_{h}+\#_{c})gh,$ for the main phase space gradings, and the
antifield numbers $\#af=(\#_{h}+\#_{c})af.$ In this work, we consider a
gauge  gravity theory on phase space with a structure d-group $\ ^{\shortmid }%
\mathfrak{g}$ when the R-flux deformations result in effective variational
model with distorted connections and sources but preserving the ghost and
antighost number and $\ ^{\shortmid }\mathfrak{g.}$ For such conditions, the
functionals of physical fields on (nonassociative) phase spaces have both
numbers equal to 0; and, for functionals of ghosts, $\#af=0$ and $\#gh=\#pg, 
$ when, for the "pure" ghost grading, a phase ghost $\ ^{\shortmid }c$ has $%
\#pg=1.$ All s-vector fields have a non-zero antifield number computed $%
\#af(\ \ _{\ddag }^{\shortmid }\mathcal{\varphi }_{\check{\alpha}}^{\star
})=1+$ $\#pg(\ ^{\shortmid }\mathcal{\varphi }_{\check{\alpha}}^{\star }),$
and $\#gh=-\#af.$} 

\vskip4pt The complex $\ ^{\shortmid }\mathcal{BV}^{\star }$ can be
considered as the space of graded multivector fields quipped with a
generalized Schouten bracket (which is an antibracket), 
\begin{equation*}
\star :\{\ ^{\shortmid }\mathcal{X},\ ^{\shortmid }\mathcal{Y}\}\rightarrow
\{\ ^{\shortmid }\mathcal{X}^{\star },\ ^{\shortmid }\mathcal{Y}^{\star
}\}:=\sum\nolimits_{\check{\alpha}}\left( \langle \frac{_{r}^{\shortmid
}\delta \ ^{\shortmid }\mathcal{X}^{\star }}{\delta \ ^{\shortmid }\mathcal{%
\varphi }_{\check{\alpha}}},\frac{_{l}^{\shortmid }\delta \ ^{\shortmid }%
\mathcal{Y}^{\star }}{\delta \ _{\ddag }^{\shortmid }\mathcal{\varphi }_{%
\check{\alpha}}^{\star }}\rangle +\langle \frac{_{r}^{\shortmid }\delta \
^{\shortmid }\mathcal{X}^{\star }}{\delta \ \ _{\ddag }^{\shortmid }\mathcal{%
\varphi }_{\check{\alpha}}^{\star }},\frac{_{l}^{\shortmid }\delta \
^{\shortmid }\mathcal{Y}^{\star }}{\delta \ ^{\shortmid }\mathcal{\varphi }_{%
\check{\alpha}}}\rangle \right) .
\end{equation*}%
To define variational configurations, we use linear parametric deformations.
Using such an antibracket, we can introduce locally a right derivation $\
^{\shortmid }\delta _{S}\rightarrow ~^{\shortmid }\delta _{S}^{\star },$
when 
\begin{eqnarray*}
~^{\shortmid }\delta _{S}^{\star }\ ^{\shortmid }\mathcal{X}^{\star }
&=&\{^{\shortmid }\mathcal{X}^{\star },\ ^{\shortmid }L^{\star }(\
^{\shortmid }f)\},\ ^{\shortmid }f\equiv 1\mbox{ on }\sup p\ ^{\shortmid }%
\mathcal{X},\ ^{\shortmid }\mathcal{X\in \ ^{\shortmid }V}^{\star }; \\
&=&\{^{\shortmid }\mathcal{X}^{\star },\ ^{\shortmid }S^{\star }(\
^{\shortmid }f)\}.
\end{eqnarray*}

Because we work with linear parametric decompositions, for any $\
^{\shortmid }\gamma \ ^{\shortmid }\mathcal{X}^{\star },$ we can find an
effective action $\ ^{\shortmid }\Theta ^{\star }$ that $\ ^{\shortmid
}\gamma \ ^{\shortmid }\mathcal{X}^{\star }=\{\ ^{\shortmid }\mathcal{X}%
^{\star },\ ^{\shortmid }\Theta ^{\star }\}.$ Using such effective
decompositions, we define the \textit{nonassociative classical BV
differential }as 
\begin{equation}
\ ^{\shortmid }\mathcal{\aleph }^{\star }=\{\bullet ,\ ^{\shortmid }S^{\star
}+\ ^{\shortmid }\Theta ^{\star }\}:=\{\bullet ,\ ^{\shortmid
}S_{ext}^{\star }\},\mbox{ with extended action }\ ^{\shortmid
}S_{ext}^{\star }.  \label{nBVdif}
\end{equation}%
The nilpotent property, $(\ ^{\shortmid }\mathcal{\aleph }^{\star })^{2}=0,$
results in the \textit{nonassociative classical master equation} (nCME),
which modulo terms vanishing in the limit of a constant $\ ^{\shortmid }f$ $%
\ $is written in the form 
\begin{equation}
\{\ ^{\shortmid }L_{ext}^{\star }(\ ^{\shortmid }f),\ ^{\shortmid
}L_{ext}^{\star }(\ ^{\shortmid }f)\}=0,  \label{nonassocme}
\end{equation}%
for an effective $\ ^{\shortmid }L_{ext}^{\star }$ used for constructing $\
^{\shortmid }S_{ext}^{\star }.$\footnote{%
For associative and commutative configurations on a Lorentz manifold, the
formula (\ref{nonassocme}) transforms into the CME (11) in \cite{rejzner20},
where another system of notations is used. In this work, we have to consider
a star product deformed classical BV differential (\ref{nBVdif}) with an
"abuse" of notations, like $\ ^{\shortmid }\mathcal{\aleph }^{\star }$ and $%
\ ^{\shortmid }\Theta ^{\star },$ because we elaborate our theory on
nonassociative phase spaces which requires  more sophisticated geometric and index-type notations.} Here we note that the operator $\ ^{\shortmid }\mathcal{\aleph }^{\star }$ increases the ghost
number by one (it is of order 1 in $\ \#gh$) and can be expressed as $\
^{\shortmid }\mathcal{\aleph }^{\star }=\ ^{\shortmid }\delta ^{\star }+\
^{\shortmid }\gamma ^{\star }$, where the extension of $\ ^{\shortmid
}\delta _{S}^{\star }$ is denoted $\ ^{\shortmid }\delta ^{\star }$ (an
operator of order -1 in $\#af$) and the extension of $\ _{ce}^{\shortmid
}\gamma $ is denoted $\ ^{\shortmid }\gamma ^{\star }$ (an operator of order
0). 

\vskip4pt Above formulas (\ref{nBVdif}) and (\ref{nonassocme}) define a
nonassociative variant of the Koszul-Tate complex $(\ ^{\shortmid }\mathcal{%
BV}_{\mathfrak{Cqs}}^{\star },\ ^{\shortmid }\delta ^{\star })$ which is a
resolution for nonlinear and linear symmetries considered in our
nonassociative gauge gravity theory. It would be noted a resolution if the
(non) linear symmetries were not independent. $\ ^{\shortmid }\mathcal{BV}_{%
\mathfrak{Cqs}}^{\star}$ has a simpler algebraic structure when the
quotients/ spaces of orbits and nonlinear symmetries are resolved than in
the case of $_{inv}^{\shortmid }\mathcal{F}_{\mathfrak{Cqs}}^{\star }.$ 

\vskip4pt To introduce the gravitational gauge fixing we use a star product
automorphism acting on s-adapted generators in the form%
\begin{equation*}
~^{\shortmid }\alpha _{\psi }^{\star }(\ \ _{\ddag }^{\shortmid }\mathcal{%
\phi }_{\check{\alpha}}^{\star }(~^{\shortmid }u)):=\frac{~^{\shortmid
}\delta ^{\star }~^{\shortmid }\psi ^{\star }(\ ^{\shortmid }f)}{%
~^{\shortmid }\delta \ ^{\shortmid }\mathcal{\varphi }_{\check{\alpha}%
}(~^{\shortmid }u)}\mbox{ and }~^{\shortmid }\alpha _{\psi }^{\star }(\
^{\shortmid }\phi ^{I}(\ ^{\shortmid }u))=\ ^{\shortmid }\phi ^{I}(\
^{\shortmid }u),
\end{equation*}%
where $\ ^{\shortmid }f(\ ^{\shortmid }u)=1,$ see formulas (\ref{ghosts}),
and a gauge fixing phase space fermion $^{\shortmid }\psi _{M}^{\star }(\
^{\shortmid }f)$ is stated as a fixed generalized Lagrangian of ghost number
-1. Similarly to \cite{fr12,rejzner20}, we can choose $\ ^{\shortmid }\alpha
_{\psi }^{\star }$ such that $\#af=0$, and in a form that $\ ^{\shortmid }\alpha
_{\psi }^{\star }$ lives the star product antibracket invariant when the
R-flux deformed action results into hyperbolic equations. Here we note that
a Lorenz-like gauge for nonassociative gauge gravity theory (\ref%
{nonassocymgreq1}) we need to extend the BV complex with antighosts $\
^{\shortmid }\overline{C}_{\star }=\{\ ^{\shortmid }\overline{C}_{\star
}^{I}\}$, of degree -1, and so-called Nakanishi-Lautrup field $\
^{\shortmid}B_{\star }=\{\ ^{\shortmid }B_{\star }^{I}\}$, degree 0, which
form a trivial pair when $\ ^{\shortmid }\mathcal{\aleph }^{\star }\
^{\shortmid }\overline{C}_{\star }^{I}=i\ ^{\shortmid }B_{\star }^{I}$ and $%
\ ^{\shortmid }\mathcal{\aleph }^{\star }\ ^{\shortmid }B_{\star }^{I}=0$.
Such operators act on an extended nonassociative configuration space $\
_{s}^{\shortmid }\overline{\mathcal{E}}^{\star }\equiv \ _{s}^{\shortmid }%
\mathcal{E}^{\star }[0]\oplus \ ^{\shortmid }\mathfrak{g[1]\oplus \
^{\shortmid }\mathfrak{g[0]}\oplus \ ^{\shortmid }\mathfrak{g[-1],}}$ when
the gauge fixing fermion defined and computed%
\begin{equation*}
~^{\shortmid }\psi ^{\star }(\ ^{\shortmid }f)=i\int_{M}\left( \frac{\alpha 
}{2}\kappa (\ ^{\shortmid }\overline{C}_{\star },\ ^{\shortmid }B_{\star
})+<\ ^{\shortmid }\overline{C}_{\star },\breve{\divideontimes}\ ^{\shortmid
}d(\breve{\divideontimes}\ ^{\shortmid }\widehat{\mathcal{A}}^{^{\star }})>\
^{\shortmid }\delta \ ^{\shortmid }\mu _{\mathfrak{Cqs}}(\ ^{\shortmid
}u),\right)
\end{equation*}%
with a measure as in (\ref{measure1}) and $\ ^{\shortmid }\widehat{\mathcal{A%
}}^{^{\star }}$ (\ref{affinpot}), see details in section 2.3 of \cite%
{rejzner20}. 

\vskip4pt We conclude that for gauge-fixed theories, gradings are convenient
to be redefined considering that $\#ta$ is the total antifield number (1 for
the antifield generators and 0 for fields). Using the decomposition $\
^{\shortmid }\mathcal{\aleph }^{\star }=\ ^{\shortmid }\delta ^{\star }+\
^{\shortmid }\gamma ^{\star }$\ and for a total action $\ ^{\shortmid
}S_{ext}^{\star },$ where $\ ^{\shortmid }S^{\star }$ denotes the $\#ta=0$
term, and when $\ ^{\shortmid }\Theta ^{\star }:=\ ^{\shortmid
}S_{ext}^{\star }-\ ^{\shortmid }S^{\star }$ as in formulas (\ref{nBVdif}).
Expressing 
\begin{equation*}
~^{\shortmid }\delta ^{\star }\ =\{\cdot ,\ ^{\shortmid }S^{\star }\}%
\mbox{
and }\ ^{\shortmid }\gamma ^{\star }=\{\cdot ,\ ^{\shortmid }\Theta ^{\star
}\},
\end{equation*}%
where the differential $\ ^{\shortmid }\delta ^{\star }$ acts trivially both
on fields and antifields (giving $\ ^{\shortmid }\delta ^{\star }\ _{\ddag
}^{\shortmid }\mathcal{\varphi }_{\check{\alpha}}^{\star }=$ $\frac{%
^{\shortmid }\delta \ ^{\shortmid }S^{\star }}{\delta \ _{\ddag }^{\shortmid
}\mathcal{\varphi }_{\check{\alpha}}^{\star }}$), we derive nonassociative
gauge-fixed euqations of motion which are hyperbolic equations of motion of $%
\ ^{\shortmid }S^{\star }.$ This is true for any solution of type $\
_{Sol}^{\shortmid }\mathcal{E}_{\mathfrak{Cqs}}^{\star }$ and respective
nonlinear symmetries. 

\subsection{Linearization and nonassociative classical BV operator and the M%
\o ller maps}

On nonassociative phase spaces with parametric decompositions, we can split
and linearize respectively the extended action $\ ^{\shortmid
}S_{ext}^{\star }$ in a form that%
\begin{eqnarray}
\ ^{\shortmid }S_{0}^{\star } &=&\ ^{\shortmid }S_{00}^{\star }+\
^{\shortmid }\Theta _{0}^{\star },\mbox{ the quadratic term in (anti) fields}%
,\#ta(\ ^{\shortmid }S_{00}^{\star })=0,\#ta(\ ^{\shortmid }\Theta
_{0}^{\star })=1;  \notag \\
\ ^{\shortmid }V^{\star } &=&\ ^{\shortmid }V_{0}^{\star }+\ ^{\shortmid
}\Theta ^{\star },\mbox{ the interacton term };  \label{interact1} \\
\ ^{\shortmid }S^{\star } &=&\ ^{\shortmid }S_{00}^{\star }+\ ^{\shortmid
}V_{0}^{\star },\mbox{ the total antifield independent part of the action}. 
\notag
\end{eqnarray}%
Using above formulas we can defined the linearized nonassociative
differentials (respective BRST and BV type):%
\begin{equation}
\ ^{\shortmid }\gamma ^{\star }\ ^{\shortmid }\mathcal{F}:=\{\ ^{\shortmid }%
\mathcal{F},\ ^{\shortmid }\Theta _{0}^{\star }\}\mbox{ and }\ ^{\shortmid }%
\mathcal{\aleph }_{0}^{\star }=~^{\shortmid }\delta _{0}^{\star }+\
^{\shortmid }\gamma _{0}^{\star },  \label{aux02c}
\end{equation}%
where $~^{\shortmid }\delta _{0}^{\star }(\ _{\ddag }^{\shortmid }\mathcal{%
\varphi }_{\star }^{\check{\alpha}})=-$ $\frac{^{\shortmid }\delta \
^{\shortmid }S_{00}^{\star }}{\delta \ ^{\shortmid }\mathcal{\varphi }_{%
\check{\alpha}}^{\star }}.$ Such operators allow us to introduce two
important nonassociative differential operators, $^{\shortmid }P_{\star }^{%
\check{\alpha}\breve{\beta}}(\ ^{\shortmid }u)$ and $\ ^{\shortmid }K_{\star 
\breve{\beta}}^{\check{\alpha}},$ when 
\begin{equation*}
\frac{_{l}^{\shortmid }\delta \ ^{\shortmid }S_{00}^{\star }}{\delta \
^{\shortmid }\mathcal{\varphi }_{\check{\alpha}}^{\star }}(\ ^{\shortmid }%
\mathcal{\varphi }_{\check{\alpha}}^{\star }):=\ ^{\shortmid }P_{\star }^{%
\check{\alpha}\breve{\beta}}(\ ^{\shortmid }u)(\ ^{\shortmid }\mathcal{%
\varphi }_{\breve{\beta}}^{\star }),\mbox{ in brief },=\ ^{\shortmid
}P_{\star }\ ^{\shortmid }\mathcal{\varphi }^{\star };\mbox{ and }\ \frac{%
_{r}^{\shortmid }\delta _{l}^{\shortmid }\delta \ \ ^{\shortmid }\Theta
_{0}^{\star }}{\delta \ ^{\shortmid }\mathcal{\varphi }_{\check{\alpha}%
}^{\star }(\ ^{\shortmid }u_{1})\delta \ _{\ddag }^{\shortmid }\mathcal{%
\varphi }_{\star }^{\breve{\beta}}(\ ^{\shortmid }u_{1})}(\ ^{\shortmid }%
\mathcal{\varphi }_{\check{\alpha}}^{\star }):=\ ^{\shortmid }K_{\star 
\breve{\beta}}^{\check{\alpha}}(\ ^{\shortmid }u)(\ ^{\shortmid }\mathcal{%
\varphi }_{\breve{\beta}}^{\star }).
\end{equation*}

On nonassociative phase space, we can assume a gauge fixing in such a way
that $\ ^{\shortmid }P_{\star }$ is Green hyperbolic for any $h$- and $c$%
-component as for (associative and commutative) gauge and gravity theories
was shown in \cite{fr12}. We introduce a double Green function of motion
operator $\ ^{\shortmid }P_{\star }$ in canonical form (with hat
d-operators) by $\ ^{\shortmid }\mathbf{g}^{\star }[\mathfrak{Clacs}],$ when 
$\ ^{\shortmid }\widehat{\triangle }_{\star }^{A/R}=(h\triangle _{\star
}^{A/R},c\triangle _{\star }^{A/R}),$ where $A$ and $B$ mean respectively
"advanced" and "retarded". This allows to define a nonassociative
Pauli-Jordan function%
\begin{equation}
\ ^{\shortmid }\widehat{\triangle }_{\star }=\ ^{\shortmid }\widehat{%
\triangle }_{\star }^{R}-\ ^{\shortmid }\widehat{\triangle }_{\star }^{A}
\label{nonassocpj}
\end{equation}%
and prove in abstract geometric form such properties (nonassociative
generalizations of \cite{fr12,hol08,rejzner14}): For any $\ ^{\shortmid }%
\widehat{\triangle }_{\star }$ being a retarded, advanced or causal
propagator corresponding to $\ ^{\shortmid }P_{\star }\ ^{\shortmid }%
\mathcal{\varphi }^{\star }=0,$ there are satisfied the consistency
conditions 
\begin{equation}
\sum_{\breve{\beta}}\left[ (-1)^{\shortmid \check{\alpha}\shortmid }\
^{\shortmid }K_{\star \breve{\beta}}^{\check{\alpha}}(\ ^{\shortmid
}u^{\prime })\ ^{\shortmid }\widehat{\triangle }_{\star }^{\ast \breve{\beta}%
\breve{\gamma}}(\ ^{\shortmid }u^{\prime },\ ^{\shortmid }u)+\ ^{\shortmid
}K_{\star \breve{\beta}}^{\breve{\gamma}}(\ ^{\shortmid }u)\ ^{\shortmid }%
\widehat{\triangle }_{\star }^{\ast \check{\alpha}\breve{\beta}}(\
^{\shortmid }u^{\prime },\ ^{\shortmid }u)\right] =0,  \label{consistcond1}
\end{equation}%
which are determined for a $\ ^{\shortmid }S_{00}^{\star }$ being invariant
under $\ ^{\shortmid }\gamma _{0}^{\star },$ and when the nCME (\ref%
{nonassocme}) are satisfied. 

\vskip4pt For associative and commutative configurations, we can define an
interacting quantum BV operator \cite{fr13} by taking a free one and
twisting it with quantum M\o ller map \cite{df03,hr20}. Then, it is possible
to prove that the resulting method is local. Nonassociative theories with
twister star product (\ref{starpn}) introduce a generic nonlocal structure
both for classical and quantum models. The advantage of the cited
mathematical formulas is that it works for nonlocal operators on phase
spaces. In this subsection, we consider such constructions for
nonassociative classical gauge gravity models which in parametric form can
be defined by an effective action $^{\shortmid }S^{\star }=\ ^{\shortmid
}S_{0}^{\star }+\ ^{\shortmid }V^{\star },$. Here, the interaction term $\
^{\shortmid }V^{\star }$ is a star product deformation of local compactly
supported functions. 

\vskip4pt We assume that $^{\shortmid }S^{\star }$ and necessary functionals 
$\ ^{\shortmid }\mathcal{F}_{\mathfrak{Cqs}}^{\star }$ are determined on the
solution space $\ _{Sol}^{\shortmid }\mathcal{E}_{\mathfrak{Cqs}}^{\star }$ (%
\ref{effectelag})), with respective non-trivial (non) linear symmetries. We
define a nonassociative generalization of the maps from \cite{df03,hr20} in
the form: 
\begin{align}
\ ^{\shortmid }r_{\lambda V}^{\star -1}(\ ^{\shortmid }\mathcal{F}_{%
\mathfrak{Cqs}}^{\star })(\ ^{\shortmid }\mathcal{\varphi }^{\star }):=& \
^{\shortmid }\mathcal{F}_{\mathfrak{Cqs}}^{\star }(\ ^{\shortmid }r_{\lambda
V}^{\star -1}(\ ^{\shortmid }\mathcal{\varphi }^{\star })),\mbox{ where }
\label{mmap} \\
\ ^{\shortmid }r_{\lambda V}^{\star -1}(\ ^{\shortmid }\mathcal{\varphi }%
^{\star })& =\ ^{\shortmid }\mathcal{\varphi }^{\star }+\ ^{\shortmid
}\lambda \ ^{\shortmid }\widehat{\triangle }_{\star }^{R}\ ^{\shortmid
}V_{(1)}^{\star }(\ ^{\shortmid }\mathcal{\varphi }^{\star })%
\mbox{ is the
inverse nonassociative M\o ller map};  \notag \\
\ ^{\shortmid }r_{\lambda V}^{\star }(\ ^{\shortmid }\mathcal{\varphi }%
^{\star })& =\ ^{\shortmid }\mathcal{\varphi }^{\star }-\ ^{\shortmid
}\lambda \ ^{\shortmid }\widehat{\triangle }_{\star \ ^{\shortmid
}S_{0}^{\star }}^{R}\ ^{\shortmid }V_{(1)}^{\star }(\ ^{\shortmid
}r_{\lambda V}^{\star }(\ ^{\shortmid }\mathcal{\varphi }^{\star }))%
\mbox{
is the classical nonassociative M\o ller map}.  \notag
\end{align}%
In above formulas, the maps are inverted as formal power series on small $\
^{\shortmid }\lambda $ when $\ ^{\shortmid }r_{\lambda V}^{\star }$ goes
from certain nonassociative interactions to a fee theory. One holds the
so-called intertwining relation:%
\begin{equation*}
\lbrack \ ^{\shortmid }r_{\lambda V}^{\star }(\ ^{\shortmid }\mathcal{F}_{%
\mathfrak{Cqs}}^{\star }),\ ^{\shortmid }r_{\lambda V}^{\star }(\
^{\shortmid }\mathcal{G}_{\mathfrak{Cqs}}^{\star })]=\ ^{\shortmid
}r_{\lambda V}^{\star }[\ ^{\shortmid }\mathcal{F}_{\mathfrak{Cqs}}^{\star
},\ ^{\shortmid }\mathcal{G}_{\mathfrak{Cqs}}^{\star }]_{\ ^{\shortmid
}r_{\lambda V}^{\star }},
\end{equation*}%
where, respectively, $[.,.]$ and $[.,.]_{V}$ are the free and the
interacting Poisson brackets (as for the Pielers bracket). For instance, we have 
\begin{equation}
\lbrack \ ^{\shortmid }\mathcal{F}^{\star },\ ^{\shortmid }\mathcal{G}%
^{\star }]=\sum_{\check{\alpha},\breve{\beta}}\langle \frac{_{r}^{\shortmid
}\delta \ ^{\shortmid }\mathcal{F}^{\star }}{\delta \ ^{\shortmid }\mathcal{%
\varphi }_{\check{\alpha}}^{\star }},\ ^{\shortmid }\widehat{\triangle }%
_{\star \check{\alpha}\breve{\beta}}\frac{_{l}^{\shortmid }\delta \
^{\shortmid }\mathcal{G}^{\star }}{\delta \ ^{\shortmid }\mathcal{\varphi }_{%
\breve{\beta}}^{\star }}\rangle ,  \label{pielersb}
\end{equation}%
for $\ ^{\shortmid }\mathcal{F}^{\star },\ ^{\shortmid }\mathcal{G}%
^{\star}\in \ ^{\shortmid }\mathcal{BV}_{\mathfrak{Cqs}}^{\star }.$ To close 
$\ ^{\shortmid}\mathcal{BV}_{\mathfrak{Cqs}}^{\star }$ under such brackets
we have to extend the constructions to a large space, for instance,
considering microcausal functions on $T^{\ast }[-1]\ _{s}^{\shortmid }%
\overline{\mathcal{E}}^{\star }$ (\ref{extendsp}). This allows us to work
with functionals that are smooth and compactly supported satisfying
conditions similar to (14) - (16) in \cite{rejzner20}. The effective
potential $\ ^{\shortmid }V_{(1)}^{\star }(\ ^{\shortmid }\mathcal{\varphi }%
^{\star })$ in (\ref{mmap}) is defined in a form that $\ ^{\shortmid
}r_{\lambda V}^{\star -1}$ maps the solutions encoded into free equations of
motion into the solutions generated by the interacting equations of motion,
i.e.:%
\begin{equation*}
\ ^{\shortmid }r_{\lambda V}^{\star -1}\frac{^{\shortmid }\delta \
^{\shortmid }S_{0}^{\star }}{\delta \ ^{\shortmid }\mathcal{\varphi }^{\star
}}=\ ^{\shortmid }r_{\lambda V}^{\star -1}(\ ^{\shortmid }P_{\star }\
^{\shortmid }\mathcal{\varphi }^{\star })=\ ^{\shortmid }P_{\star }\
^{\shortmid }\mathcal{\varphi }^{\star }+\ ^{\shortmid }\lambda \
^{\shortmid }P_{\star }\circ \ ^{\shortmid }\widehat{\triangle }_{\star
}^{R}\ ^{\shortmid }V_{(1)}^{\star }(\ ^{\shortmid }\mathcal{\varphi }%
^{\star })=\ ^{\shortmid }P_{\star }\ ^{\shortmid }\mathcal{\varphi }^{\star
}+\ ^{\shortmid }\lambda \ ^{\shortmid }V_{(1)}^{\star }(\ ^{\shortmid }%
\mathcal{\varphi }^{\star }).
\end{equation*}

The formulas (\ref{mmap}) can be generalized to the case when nonassociative
gauge symmetries are present. In such cases, $\ ^{\shortmid }S_{0}^{\star }$
has two terms then the first one ($\ ^{\shortmid }S_{00}^{\star },$ it does
not depend on antifields) defines $\ ^{\shortmid }P_{\star }$ and $\
^{\shortmid }\widehat{\triangle }_{\star }^{R}$. We can consider 
\begin{equation*}
\ ^{\shortmid }r_{\lambda V}^{\star -1}(\ ^{\shortmid }\mathcal{\varphi }_{%
\check{\alpha}}^{\star })=\ ^{\shortmid }\mathcal{\varphi }_{\check{\alpha}%
}^{\star }+\ ^{\shortmid }\widehat{\triangle }_{\star \check{\alpha}\breve{%
\beta}}\frac{_{l}^{\shortmid }\delta \ ^{\shortmid }V^{\star }}{\delta \
^{\shortmid }\mathcal{\varphi }_{\breve{\beta}}^{\star }}(\ ^{\shortmid }%
\mathcal{\varphi }^{\star })\mbox{ and }\ ^{\shortmid }r_{\lambda V}^{\star
}(\ ^{\shortmid }\mathcal{\varphi }_{\check{\alpha}}^{\star })=\ ^{\shortmid
}\mathcal{\varphi }_{\check{\alpha}}^{\star }-\ ^{\shortmid }\widehat{%
\triangle }_{\star \check{\alpha}\breve{\beta}}\frac{_{l}^{\shortmid }\delta
\ ^{\shortmid }V^{\star }}{\delta \ ^{\shortmid }\mathcal{\varphi }_{\breve{%
\beta}}^{\star }}(\ ^{\shortmid }r_{\lambda V}^{\star }(\ ^{\shortmid }%
\mathcal{\varphi }^{\star })).
\end{equation*}%
Using such operators, we can formulate and prove such results for $nCME(\
^{\shortmid }S^{\star }),$ i.e. the classical master equations for the
nonassociative gauge gravity with star product deformation (\ref{starpn}):%
\begin{eqnarray}
\ ^{\shortmid }r_{\lambda V}^{\star -1}(\{\ ^{\shortmid }\mathcal{X}^{\star
},\ ^{\shortmid }S_{0}^{\star }\}) &=&\{\ ^{\shortmid }r_{\lambda V}^{\star
-1}(\ ^{\shortmid }\mathcal{X}^{\star }),\ ^{\shortmid }S_{0}^{\star }+\
^{\shortmid }V^{\star }\}-  \label{aux02a} \\
&&\int \frac{_{r}^{\shortmid }\delta \ ^{\shortmid }\mathcal{X}^{\star }}{%
\delta \ ^{\shortmid }\mathcal{\varphi }_{\check{\alpha}}^{\star }(\
^{\shortmid }u^{\prime })}(\ ^{\shortmid }r_{\lambda V}^{\star -1}(\
^{\shortmid }\mathcal{\varphi }^{\star }))\ ^{\shortmid }\widehat{\triangle }%
_{\star \check{\alpha}\breve{\beta}}(\ ^{\shortmid }u^{\prime },\
^{\shortmid }u)\frac{_{l}^{\shortmid }\delta (nCME(\ ^{\shortmid }S^{\star
}))}{\delta \ ^{\shortmid }\mathcal{\varphi }_{\breve{\beta}}^{\star }(\
^{\shortmid }u)},  \notag
\end{eqnarray}%
which allows to write the classical BV operator (\ref{nBVdif}) of the
nonassociative gauge gravity theory as%
\begin{equation}
\ ^{\shortmid }\mathcal{\aleph }^{\star }=\ ^{\shortmid }r_{\lambda
V}^{\star -1}\circ \ ^{\shortmid }\mathcal{\aleph }_{0}^{\star }~\circ \
^{\shortmid }r_{\lambda V}^{\star }.  \label{aux02b}
\end{equation}%
Details on proofs of formulas (\ref{aux02a}) \ and (\ref{aux02b}) in
associative and commutative forms are provided in \cite{rejzner20} (for the
Theorem 2.15 and Corollary 2.16 and appendix of that work) and \cite{fr13}.
Nonassociative generalizations can be obtained using the abstract geometric
formalism for nonholonomic parametric deformations on phase spaces. 

\section{BV-scheme and quantization of nonassociative gauge de Sitter gravity%
}

\label{sec04}The goal of this section is to show how the BV formalism \cite%
{brst75,brst76,barv18,bv81} can be applied for quantizing quasi-parametric
off-diagonal solutions. We cite also recent mathematical developments \cite%
{rejzner20,rejzner16,rejzner14,brunetti22,hol08,fr13,df03,hr20} using certain methods for quantizing the classical nonassociative gauge gravity theories from the previous two sections. The twisted star product (\ref{starpn}) can be written in the form $\star =\star _{\hbar }+\star _{\kappa }$, where $\star _{\hbar }$
includes terms with $\hbar ^{0}$ and $\hbar ^{1}$ and $\star _{\kappa }$ encodes terms with are proportional to $\kappa $ with possible mutiplications on terms which are also $\hbar ^{0}$- and 
$\hbar ^{1}$-parametric. The quantization scheme is elaborated using the Planck constant $\hbar$ and when the string parameter $\kappa $  defines nonassociative R-flux deformations. For simplicity, we consider only the $\kappa $-linear terms when the higher order contributions can be computed recurrently. 

\subsection{Free nonassociative gauge gravitational fields}

We can construct a quantized algebra of free nonassociative filed using
deformation quantization, DQ, encoding also $\kappa $-parametric R-flux
deformation of the classical s-algebras $(\ _{\mu c}^{\shortmid }\mathcal{BV}%
_{\mathfrak{Cqs}}^{\star },\lfloor \cdot ,\cdot \rfloor)$. Such
noncommutative gauge gravity theories were elaborated in \cite{svnc00,sv00}
when the DQ formalism for GR and MGTs was studied in N-adapted form in \cite%
{vacaru07,vacaru13}. In our approach, we equip the nonassociative phase
space of formal power series $\ _{\mu c}^{\shortmid }\mathcal{BV}_{\mathfrak{%
Cqs}}^{\star }[[\hbar ,\kappa ]]$ with a nonassociative star product $\star
=\star _{\hbar }+\star _{\kappa }$ (\ref{starpn}), when the noncommutative
component $\star _{\hbar }$ is used for defining the quantum s-operator
product of quantum observables as in \cite{hr20,rejzner20,df03,fr13}. Here
we note that the meaning of the string constant $\kappa $ differs from that
of $\hbar $ used for quantization. 

\vskip4pt For any s-operators $\ ^{\shortmid }\mathcal{F}^{\star }$ and $\
^{\shortmid }\mathcal{G}^{\star },$ we can consider such a deformation of
the point wise product,%
\begin{equation*}
\ ^{\shortmid }\mathcal{F}^{\star }\star _{s}\ ^{\shortmid }\mathcal{G}%
^{\star }:=\ ^{\shortmid }m\circ \exp (i\hbar \ ^{\shortmid }D_{W}^{\star
})(~^{\shortmid }\mathcal{F}^{\star }\otimes _{s}\ ^{\shortmid }\mathcal{G}%
^{\star }),
\end{equation*}%
where the multiplication s-operator is considered on associative and commutative phase space as 
\begin{equation*}
\ ^{\shortmid }m(~^{\shortmid }\mathcal{F}^{\star }\otimes _{s}\ ^{\shortmid
}\mathcal{G}^{\star })(\ ^{\shortmid }\mathcal{\varphi }^{\star
})=~^{\shortmid }\mathcal{F}^{\star }(\ ^{\shortmid }\mathcal{\varphi }%
^{\star })\cdot _{s}\ ^{\shortmid }\mathcal{G}^{\star }(\ ^{\shortmid }%
\mathcal{\varphi }^{\star }),
\end{equation*}%
and a functional differential s-operator is defined 
 $\ ^{\shortmid }D_{W}^{\star }:=\frac{1}{2}\sum_{\check{\alpha},\breve{\beta}%
}\langle \ ^{\shortmid }W_{\check{\alpha}\breve{\beta}}^{\star },\frac{%
_{l}^{\shortmid }\delta }{\delta \ ^{\shortmid }\mathcal{\varphi }_{\check{%
\alpha}}^{\star }}\otimes _{s}\frac{_{r}^{\shortmid }\delta }{\delta \
^{\shortmid }\mathcal{\varphi }_{\breve{\beta}}^{\star }}\rangle $. 
In this formula, we use $\ ^{\shortmid }W_{\check{\alpha}\breve{\beta}%
}^{\star }$ as the phase space R-flux deformation of so called 2-point
function of a Hardmard state. In \cite{rad96,hol08}, similar details are provided   on how to
chose such an $W$-operator to be positive definite, satisfy the appropriate
wave conditions and when 
\begin{equation*}
\ ^{\shortmid }W_{\check{\alpha}\breve{\beta}}^{\star }=\frac{i}{2}\
^{\shortmid }\widehat{\triangle }_{\star \check{\alpha}\breve{\beta}}+\
^{\shortmid }H_{\check{\alpha}\breve{\beta}}^{\star }.
\end{equation*}%
In this formula, $\ ^{\shortmid }H_{\check{\alpha}\breve{\beta}}^{\star }$ is a
symmetric bi-solution for $\ ^{\shortmid }P_{\star }.$ The nonholonomic
s-adapted structure can be prescribed in such a form that additionally to
the standard properties there satisfied also the consistency condition (see (%
\ref{consistcond1})) for the symmetric part:%
\begin{equation}
\sum_{\breve{\beta}}\left[ (-1)^{\shortmid \ ^{\shortmid }\mathcal{\varphi }%
_{\check{\alpha}}^{\star }\shortmid }\ ^{\shortmid }K_{\star \breve{\beta}}^{%
\check{\alpha}}(\ ^{\shortmid }u^{\prime })\ ^{\shortmid }H_{\star }^{\ast 
\breve{\beta}\breve{\gamma}}(\ ^{\shortmid }u^{\prime },\ ^{\shortmid }u)+\
^{\shortmid }K_{\star \breve{\beta}}^{\breve{\gamma}}(\ ^{\shortmid }u)\
^{\shortmid }H_{\star }^{\ast \check{\alpha}\breve{\beta}}(\ ^{\shortmid
}u^{\prime },\ ^{\shortmid }u)\right] =0.  \label{consistcond2}
\end{equation}%
So, $\ ^{\shortmid }\gamma _{0}^{\star }$ is a right derivation with respect
to the star product if the conditions (\ref{consistcond2}) are satisfied.
Here we note that $~^{\shortmid }\delta _{0}^{\star }$ is also a right
derivation with respect to $\star $ since $\ ^{\shortmid }W_{\check{\alpha}%
\breve{\beta}}^{\star }$ is a solution for the linearized parametric
nonassociative phase space equations of motion s-operator $\ ^{\shortmid
}P_{\star }$.

\subsection{Interacting nonassociative gauge gravitational fields}

In this subsection, we consider the space $\ _{reg}^{\shortmid }\mathcal{BV}%
_{\mathfrak{Cqs}}^{\star }$ of regular functional for which the derivatives
at every point are smooth compactly supported functions on $T^{\ast }[-1]\
_{s}^{\shortmid }\overline{\mathcal{E}}^{\star }$ (\ref{extendsp}). 

\subsubsection{Time-ordered products and Peierls bracket}

For any $\ ^{\shortmid }\mathcal{F}^{\star },$ the time-ordering operator, $%
\ ^{\shortmid }\mathcal{T}^{\star },$ is defined using the formula for an
internal kernel, $\ ^{\shortmid }K_{\check{\alpha}\breve{\beta}}^{\star },$ 
\begin{equation}
\ ^{\shortmid }\mathcal{T}^{\star }\ ^{\shortmid }\mathcal{F}^{\star }(\
^{\shortmid }\mathcal{\varphi }^{\star }):=e^{\frac{\hbar }{2}\ ^{\shortmid }%
\mathcal{D}_{\ ^{\ ^{\shortmid }\widehat{\triangle }_{\star }^{F}}}^{\star
_{\kappa }}},\mbox{ for }\ ^{\shortmid }\mathcal{D}_{K}^{\star _{\kappa
}}:=\sum_{\check{\alpha},\breve{\beta}}\langle \ ^{\shortmid }K_{\check{%
\alpha}\breve{\beta}}^{\star },\frac{_{l}^{\shortmid }\delta }{\delta \
^{\shortmid }\mathcal{\varphi }_{\check{\alpha}}^{\star }}\otimes _{s}\frac{%
_{r}^{\shortmid }\delta }{\delta \ ^{\shortmid }\mathcal{\varphi }_{\breve{%
\beta}}^{\star }}\rangle ,  \label{aux02d}
\end{equation}%
is defined to include parametric $\kappa $-terms from $\star _{\kappa }$ and 
$\ ^{\shortmid }\widehat{\triangle }_{\star }^{F}= \frac{i}{2}\left( \
^{\shortmid }\widehat{\triangle }_{\star }^{R}+ \ ^{\shortmid }\widehat{%
\triangle }_{\star }^{A}\right) +\ ^{\shortmid }H_{\star }.$ We state that
formally $\ ^{\shortmid }\mathcal{T}^{\star }$ is a s-operator of
convolution corresponding to oscillating Gaussian measure with covariance $%
i\hbar \ ^{\shortmid }\widehat{\triangle }_{\star }^{F}$ when,%
\begin{equation*}
\ ^{\shortmid }\mathcal{T}^{\star }\ ^{\shortmid }\mathcal{F}^{\star }(\
^{\shortmid }\mathcal{\varphi }^{\star })=\int \ ^{\shortmid }\mathcal{F}%
^{\star }(\ ^{\shortmid }\mathcal{\varphi }^{\star }-\ ^{\shortmid }\mathcal{%
\varphi }_{1}^{\star })\ ^{\shortmid }\delta \ ^{\shortmid }\mu _{i\hbar \
^{\shortmid }\widehat{\triangle }_{\star }^{F}}(\ ^{\shortmid }\mathcal{%
\varphi }_{1}^{\star }).
\end{equation*}%
In such a formula, $\ ^{\shortmid }\delta \ ^{\shortmid }\mu _{i\hbar \ ^{\shortmid }%
\widehat{\triangle }_{\star }^{F}}(\ ^{\shortmid }\mathcal{\varphi }%
_{1}^{\star })$ is chosen to correspond to some $\ ^{\shortmid }\delta \
^{\shortmid }\mu _{\mathfrak{Cqs}}(\ ^{\shortmid }u).$ Using such formulas,
we define the time-ordered product $\cdot _{\mathcal{T}^{\star }}$ on $\
_{reg}^{\shortmid }\mathcal{BV}_{\mathfrak{Cqs}}^{\star }[[\hbar ,\kappa ]]$  as 
\begin{equation*}
\ ^{\shortmid }\mathcal{F}_{1}^{\star }\cdot _{\mathcal{T}^{\star }}\
^{\shortmid }\mathcal{F}_{2}^{\star }:=\ ^{\shortmid }\mathcal{T}^{\star }(\
^{\shortmid }\mathcal{T}^{\star -1}\ ^{\shortmid }\mathcal{F}_{1}^{\star
}\cdot \ ^{\shortmid }\mathcal{T}^{\star -1}\ ^{\shortmid }\mathcal{F}%
_{2}^{\star }).
\end{equation*}%
We note that $\cdot _{\mathcal{T}^{\star }}$ consists a time-ordered version
of $\star =\star _{\hbar }+\star _{\kappa }$ (\ref{starpn}) when 
\begin{eqnarray}
\ ^{\shortmid }\mathcal{F}_{1}^{\star }\cdot _{\mathcal{T}^{\star }}\
^{\shortmid }\mathcal{F}_{2}^{\star } &=&\ ^{\shortmid }\mathcal{F}%
_{1}^{\star }\star \ ^{\shortmid }\mathcal{F}_{2}^{\star }%
\mbox{ if the
support of }\ ^{\shortmid }\mathcal{F}_{1}^{\star }%
\mbox{ is later than the
support of }\ ^{\shortmid }\mathcal{F}_{2}^{\star };  \label{ppropa} \\
\ ^{\shortmid }\mathcal{F}_{1}^{\star }\cdot _{\mathcal{T}^{\star }}\
^{\shortmid }\mathcal{F}_{2}^{\star } &=&\ ^{\shortmid }\mathcal{F}%
_{2}^{\star }\star \ ^{\shortmid }\mathcal{F}_{1}^{\star }%
\mbox{ if the
support of }\ ^{\shortmid }\mathcal{F}_{2}^{\star }%
\mbox{ is later than the
support of }\ ^{\shortmid }\mathcal{F}_{1}^{\star }.  \notag
\end{eqnarray}

Let us consider as in AQFT a QFT model of assigning s-algebras of observables%
\begin{equation*}
\ ^{\shortmid }\mathcal{U}^{\star }(\ ^{\shortmid }\mathcal{O}^{\star })=(\
_{\mu c}^{\shortmid }\mathcal{BV}_{\mathfrak{Cqs}}^{\star }[[\hbar ,\kappa
]], \star _{\hbar }+\star _{\kappa })\subset \ _{s}^{\shortmid }\mathcal{M}%
^{\star }
\end{equation*}%
on a given nonassociative phase space. In parametric form, we approximate $\
^{\shortmid }\mathcal{U}^{\star } (\ ^{\shortmid }\mathcal{O}%
^{\star})\approx \ ^{\shortmid }\mathcal{U}_{\mathfrak{Cqs}}^{\star } (\
^{\shortmid }\mathcal{O}^{\star })$ for a compact phase space region $\
^{\shortmid }\mathcal{O}^{\star }=\ _{s}^{\shortmid }\mathcal{O}^{\star
}=(h\ ^{\shortmid }\mathcal{O}^{\star },c\ ^{\shortmid }\mathcal{O}^{\star}) 
$ defined by a $\ ^{\shortmid }\mathbf{g}^{\star }[\mathfrak{Cqs}]$ (\ref%
{baspropsol}), when a phase space differential $\ ^{\shortmid }d^{\star} $
and star product $\star $ are assigned of such chain of complexes in
s-algebras of regions. Here we note that the decoupling properties of
nonassociative gauge gravity equations does not depend on the existence, or
not, of certain Cauchy hypersurfaces in $\ _{s}^{\shortmid }\mathcal{M}%
^{\star }.$ We can consider such a hypersurfrace in the form that every
in-extendible causal curve intersects it exactly once (with a conventional
splitting on $h$- and $v$-parts) in any neighborhood $\ _{s}^{\shortmid }%
\mathcal{N}^{\star }\subset \ _{s}^{\shortmid }\mathcal{M}^{\star }$. Above
stated conditions allow us to extend on nonassociative phase spaces two
important axioms considered of associative and commutative spacetimes in
section 3.2.2 of \cite{rejzner20}: 
\begin{itemize}
\item \textbf{Axiom 1 on weak causality on nonassociative phase space } $\
_{s}^{\shortmid }\mathcal{M}^{\star }:$ For any h- and c-components which
are of spacelike signature, the commutator 
 $\lbrack \ \ ^{\shortmid }\mathcal{U}^{\star }(\ ^{\shortmid }\mathcal{O}%
_{1}^{\star }),\ \ ^{\shortmid }\mathcal{U}^{\star }(\ ^{\shortmid }\mathcal{%
O}_{1}^{\star })]=\ ^{\shortmid }d^{\star }\ ^{\shortmid }\mathcal{X}^{\star}$ 
for some $\ ^{\shortmid }\mathcal{X}^{\star }\in \ \ ^{\shortmid }\mathcal{U}%
^{\star }(\ ^{\shortmid }\mathcal{O}^{\star })$ when $\ ^{\shortmid }%
\mathcal{O}^{\star }$ contains both $\ ^{\shortmid }\mathcal{O}_{1}^{\star }$
and $\ ^{\shortmid }\mathcal{O}_{2}^{\star }.$

\item \textbf{\ Axiom 2 on time-slicing of nonassociative phase space} $\
_{s}^{\shortmid }\mathcal{M}^{\star }$: For any $\ _{s}^{\shortmid }\mathcal{%
N}^{\star }$ of a Cauchy hypersurface in the region $\ ^{\shortmid }\mathcal{%
O}^{\star }\subset \ _{s}^{\shortmid }\mathcal{M}^{\star },$ the map $\
^{\shortmid }\mathcal{U}^{\star }(\ ^{\shortmid }\mathcal{N}^{\star })$ and $%
\ ^{\shortmid }\mathcal{U}^{\star }(\ ^{\shortmid }\mathcal{O}^{\star })$
are quasi-isomorphic (i.e. are isomorphic on the level of cohomology groups
of corresponding $h$- and $c$-components).
\end{itemize}

Let us take $\ ^{\shortmid }\mathcal{F}_{1}^{\star },\ ^{\shortmid }\mathcal{F}%
_{2}^{\star }\in \ ^{\shortmid }\mathcal{U}^{\star }(\ ^{\shortmid }\mathcal{%
O}^{\star })$ and denote $\ _{s}^{\shortmid }\mathcal{N}_{+}^{\star }$ and $%
\ _{s}^{\shortmid }\mathcal{N}_{-}^{\star }$ the respective Cauchy surfaces
to the future and past of $h$- and $c$-components of $\ ^{\shortmid }%
\mathcal{O}^{\star }.$ The Axiom 2 implies that there are localization maps $\ ^{\shortmid }\beta _{\pm }^{\star }(\ ^{\shortmid }\mathcal{F}_{1}^{\star })$ and a $\ ^{\shortmid }\mathcal{F}_{3}^{\star }$
that $\ ^{\shortmid }\beta _{-}^{\star }(\ ^{\shortmid }\mathcal{F}%
_{1}^{\star }) - \ ^{\shortmid }\beta _{+}^{\star }(\ ^{\shortmid }\mathcal{F%
}_{1}^{\star })= \ ^{\shortmid }\mathcal{\aleph }_{0}^{\star }\ ^{\shortmid }%
\mathcal{F}_{3}^{\star },$ see formulas (\ref{aux02b}). Here we note that on
base spacetime with fixed (co) fiber coordinates $\ ^{\shortmid }\mathcal{%
\aleph }_{0}^{\star }$ transforms into BV-operator $s_{0}$ from \cite%
{fr13,rejzner20}. Using Axiom 1 and the properties (\ref{ppropa}), we can
define and compute the nonassociative $\star $-star commutator $\
^{\shortmid }\mathcal{F}_{1}^{\star }$ and$\ ^{\shortmid }\mathcal{F}%
_{2}^{\star }$ in the form%
\begin{eqnarray*}
i\hbar \lbrack \ ^{\shortmid }\mathcal{F}_{2}^{\star },\ ^{\shortmid }%
\mathcal{F}_{1}^{\star }]_{\star } &=&\ ^{\shortmid }\mathcal{F}_{2}^{\star
}\star \ ^{\shortmid }\mathcal{F}_{1}^{\star }-\ ^{\shortmid }\mathcal{F}%
_{1}^{\star }\star \ ^{\shortmid }\mathcal{F}_{2}^{\star } = \ ^{\shortmid }%
\mathcal{F}_{2}^{\star }\star \ ^{\shortmid }\beta _{+}^{\star }(\
^{\shortmid }\mathcal{F}_{1}^{\star })-\ ^{\shortmid }\beta _{-}^{\star }(\
^{\shortmid }\mathcal{F}_{1}^{\star })\star \ ^{\shortmid }\mathcal{F}%
_{2}^{\star } \\
&=&\ ^{\shortmid }\mathcal{F}_{2}^{\star }\cdot _{\mathcal{T}^{\star }}\
^{\shortmid }\beta _{+}^{\star }(\ ^{\shortmid }\mathcal{F}_{1}^{\star })-\
^{\shortmid }\beta _{-}^{\star }(\ ^{\shortmid }\mathcal{F}_{1}^{\star
})\cdot _{\mathcal{T}^{\star }}\ ^{\shortmid }\mathcal{F}_{2}^{\star }~%
\mbox{mod}~\mbox{Im}\ ^{\shortmid }d^{\star },
\end{eqnarray*}%
considering modulo the image $od$ of $\ ^{\shortmid }d^{\star }.$ This
allows us to express a quantum nonassociative version of the Peierls bracket
(\ref{pielersb}), when 
\begin{equation*}
i\hbar \lbrack \ ^{\shortmid }\mathcal{F}_{2}^{\star },\ ^{\shortmid }%
\mathcal{F}_{1}^{\star }]_{\star }=\ ^{\shortmid }\mathcal{F}_{2}^{\star
}\cdot _{\mathcal{T}^{\star }}\ ^{\shortmid }\mathcal{\aleph }_{0}^{\star }\
^{\shortmid }\mathcal{F}_{3}^{\star }~\mbox{mod}~\hbar ^{2},\mbox{Im}\
^{\shortmid }d^{\star }.
\end{equation*}%
For $\ ^{\shortmid }d^{\star }\ ^{\shortmid }\mathcal{F}_{2}^{\star }=0,$ we
can introduce an antibracket in the right side, 
\begin{equation*}
i\hbar \lbrack \ ^{\shortmid }\mathcal{F}_{2}^{\star },\ ^{\shortmid }%
\mathcal{F}_{1}^{\star }]=\ ^{\shortmid }\mathcal{\aleph }_{0}^{\star }(\
^{\shortmid }\mathcal{F}_{2}^{\star }\cdot _{\mathcal{T}^{\star }}\
^{\shortmid }\mathcal{F}_{3}^{\star })+i\hbar \{\ ^{\shortmid }\mathcal{F}%
_{2}^{\star },\ ^{\shortmid }\mathcal{F}_{3}^{\star }\}~\mbox{mod}~\hbar
^{2},\mbox{Im}\ ^{\shortmid }d^{\star }.
\end{equation*}%
So, we can consider 
\begin{equation*}
\lbrack \ ^{\shortmid }\mathcal{F}_{2}^{\star },\ ^{\shortmid }\mathcal{F}%
_{1}^{\star }]=\{\ ^{\shortmid }\mathcal{F}_{2}^{\star },\ ^{\shortmid }%
\mathcal{F}_{3}^{\star }\}~\mbox{mod}~\hbar ,\mbox{Im}\ ^{\shortmid}d^{\star
}
\end{equation*}%
as the intrinsic definition of the Peierls bracket for a given antibracket
and the time-ordered product in a nonassociative phase space theory
satisfying the Axiom 2.

\subsubsection{Interactions and the renormalization problem}

We model nonassociative phase space interactions (\ref{interact1}) as in the
classical master equations (\ref{aux02a}) assuming that $\ ^{\shortmid
}V^{\star }\in \ _{reg}^{\shortmid }\mathcal{BV}_{\mathfrak{Cqs}}^{\star }.$
The quantum observable of the free theory associated with this
nonassociative phase space potential as $\ ^{\shortmid }\mathcal{T}^{\star
}\ ^{\shortmid}V^{\star }.$ This is a quantization map (in our approach, it
is nonholonomic s-adapted and determined by a configuration $\mathfrak{Cqs}$
and respective parametric deformations) which involves also a normal
ordering $\ ^{\shortmid }\mathcal{T}^{\star }\ ^{\shortmid }V^{\star }\equiv
:\ ^{\shortmid }V^{\star }:.$ 

\vskip4pt Considering formal power series on a small $\ ^{\shortmid }\lambda 
$ as in (\ref{mmap}), we define the formal S-matrix 
\begin{equation*}
\ ^{\shortmid }\mathcal{S}^{\star }(\ ^{\shortmid }\lambda \ ^{\shortmid
}V^{\star }):=e^{i\ ^{\shortmid }\lambda :\ ^{\shortmid }V^{\star }:/~\hbar
}=\ ^{\shortmid }\mathcal{T}^{\star }(e^{i\ ^{\shortmid }\lambda \
^{\shortmid }V^{\star }/~\hbar }),
\end{equation*}%
where $\ ^{\shortmid }\mathcal{S}^{\star }(\ ^{\shortmid }\lambda :\
^{\shortmid }V^{\star }:)\in \ _{reg}^{\shortmid }\mathcal{BV}_{\mathfrak{Cqs%
}}^{\star }((\hbar ))[[\ ^{\shortmid }\lambda ,\kappa ]]$, explicit
computations involve $\kappa $-linear terms. The formulas  can be recurrently
extended as power series on $\kappa $ if we consider such string R-flux
deformations. Interacting nonholonomic s-fields are considered as elements
of $\ _{reg}^{\shortmid }\mathcal{BV}_{\mathfrak{Cqs}}^{\star }[[\hbar ,\
^{\shortmid }\lambda ,\kappa ]]$ \ given and computed as 
\begin{eqnarray}
\ ^{\shortmid }r_{\lambda V}^{\star }(\ ^{\shortmid }\mathcal{F}^{\star })
&:=&[(e^{i\ ^{\shortmid }\lambda :\ ^{\shortmid }V^{\star }:/~\hbar
})^{\star }]^{-1}\star (e_{\ ^{\shortmid }\mathcal{T}}^{i\ ^{\shortmid
}\lambda :\ ^{\shortmid }V^{\star }:/~\hbar }\cdot _{\mathcal{T}^{\star }}:\
^{\shortmid }\mathcal{F}^{\star }:)  \label{roper} \\
&=&-i\hbar \frac{d}{d\ ^{\shortmid }\mu }\ ^{\shortmid }\mathcal{S}^{\star
}(\ ^{\shortmid }\lambda \ ^{\shortmid }V^{\star })^{-1}\ ^{\shortmid }%
\mathcal{S}^{\star }(\ ^{\shortmid }\lambda \ ^{\shortmid }V^{\star }+\
^{\shortmid }\mu \ ^{\shortmid }\mathcal{F}^{\star })_{\mid ^{\shortmid }\mu
=0},  \notag
\end{eqnarray}%
when $\ ^{\shortmid }r_{0}^{\star }(\ ^{\shortmid }\mathcal{F}^{\star })=: \
^{\shortmid }\mathcal{F}^{\star }:,$ for $\ ^{\shortmid }\lambda =0.$ Using
the s-operator (\ref{roper}), we can define the interacting nonassociative
star product 
\begin{equation*}
\ ^{\shortmid }\mathcal{F}_{1}^{\star }\star _{s}^{int}\ ^{\shortmid }%
\mathcal{F}_{2}^{\star }:=\ ^{\shortmid }r_{V}^{\star -1}(\ ^{\shortmid
}r_{V}^{\star }(\ ^{\shortmid }\mathcal{F}_{1}^{\star })\star _{s}\
^{\shortmid }r_{V}^{\star }(\ ^{\shortmid }\mathcal{F}_{2}^{\star })),
\end{equation*}%
which encode in parametric form nonassociative quantum regularizations of
the constants in (\ref{starpn}). 

\vskip4pt On nonassociative phase spaces, we face the same problem of
quantization as on typical Lorentz manifolds when quantum interactions and
observables are local but not regular. For details on the corresponding
renormalization problem, we cite \cite{rejzner20}, section 3.2.4. Here we
also note that our goal is to elaborate on models of physical interactions
which are usually local and encoding in effective parametric forms certain
nonassociative data. Both on base spacetime Lorentz manifold and for co-fiber constructions the time-order product $\cdot _{\mathcal{T}^{\star }}$ is not well defined on local and nonlinear functionals because of singularities of $\ ^{\shortmid }\widehat{\triangle}_{\star }^{F}$. So, the
renormalization problem of nonassociative gauge gravitational fields (in our
gauge models) is then to extend the effective parametric functional $\
^{\shortmid }\mathcal{S}^{\star }$ to local arguments by using time-ordering
products on nonassociative phase spaces,%
\begin{eqnarray*}
\ ^{\shortmid }\mathcal{S}^{\star } &=&\sum\nolimits_{n=0}^{\infty }\frac{1}{%
n!}\ ^{\shortmid }\mathcal{F}_{n}^{\star }(\ ^{\shortmid }V^{\star },...,\
^{\shortmid }V^{\star }), \mbox{ for } \\
&& \ ^{\shortmid }\mathcal{F}_{n}^{\star }(\ ^{\shortmid }F_{1}^{\star
},...,\ ^{\shortmid }F_{n}^{\star }) :=\ ^{\shortmid }F_{1}^{\star }\cdot _{%
\mathcal{T}^{\star }}...\cdot _{\mathcal{T}^{\star }}\ ^{\shortmid
}F_{n}^{\star }.
\end{eqnarray*}%
In these formulas, the s-adapted time-ordered product of $n$ local functionals is well defined by choosing pairwise disjointed supports respectively defined on base and typical fibers. To construct a causal  perturbation theory on phase space we can extend $\ ^{\shortmid }\mathcal{F}_{n}^{\star }$ to arbitrary
local functional following the Epstein and Glaser process. In this case,  the causal factorization property is stated as  
\begin{equation*}
\ ^{\shortmid }\mathcal{F}_{n}^{\star }(\ ^{\shortmid }F_{1}^{\star }\otimes
...\otimes \ ^{\shortmid }F_{n}^{\star })=\ ^{\shortmid }\mathcal{F}%
_{n}^{\star }(\ ^{\shortmid }F_{1}^{\star }\otimes ...\otimes \ ^{\shortmid
}F_{k}^{\star })\star _{s}\ ^{\shortmid }\mathcal{F}_{n-j}^{\star }(\
^{\shortmid }F_{k+1}^{\star }\otimes ...\otimes \ ^{\shortmid }F_{n}^{\star
})
\end{equation*}%
if the supports of $\ ^{\shortmid }F_{1}^{\star },...,\ ^{\shortmid
}F_{k}^{\star }$ are later than the supports of $\ ^{\shortmid
}F_{k+1}^{\star },...,\ ^{\shortmid }F_{n}^{\star }.$ 

\subsubsection{Renormalized nonassociative QME and quantum BV operator}

\label{ss423}We can apply the BV-formalism elaborated for the pAQFT \cite%
{fr13,rejzner20} by extending on nonassociative phase spaces the conditions
that the free classical nonassociative BV s-operator $\ ^{\shortmid }%
\mathcal{\aleph }_{0}^{\star }$ from (see formulas (\ref{nBVdif}) and (\ref%
{aux02b})) is defined 
\begin{eqnarray}
\ ^{\shortmid }\mathcal{\aleph }_{0}^{\star }(e_{\ ^{\shortmid }\mathcal{T}%
}^{i\ :\ ^{\shortmid }V^{\star }:/~\hbar }) &=&\ ^{\shortmid }\mathcal{%
T^{\star }(}e_{\ }^{i\ \ ^{\shortmid }V^{\star }/~\hbar }-i\hbar \
^{\shortmid }\widehat{\triangle }_{\star }e_{\ }^{i\ \ ^{\shortmid }V^{\star
}/~\hbar })  \label{nBVdif1} \\
&=&\ ^{\shortmid }\mathcal{T^{\star }}(e_{\ }^{i\ \ ^{\shortmid }V^{\star
}/~\hbar }(\frac{i}{\hbar }\{\ ^{\shortmid }V^{\star },\ \ ^{\shortmid
}S_{0}^{\star }\}+\frac{i}{2\hbar }\{\ ^{\shortmid }V^{\star },\ \
^{\shortmid }V^{\star }\}+\ ^{\shortmid }\widehat{\triangle }_{\star }(\
^{\shortmid }V^{\star })))=0.  \notag
\end{eqnarray}%
Such computation is performed using the identity satisfied by $\
^{\shortmid }\mathcal{T^{\star }}$ (\ref{aux02d}),%
 $\ ^{\shortmid }\delta _{0}^{\star }(\ ^{\shortmid }\mathcal{T^{\star }}\
^{\shortmid }F^{\star })=\ ^{\shortmid }\mathcal{T^{\star }}(~^{\shortmid
}\delta _{0}^{\star }\ ^{\shortmid }F^{\star }-i\hbar \ ^{\shortmid }%
\widehat{\triangle }_{\star }\ ^{\shortmid }F^{\star })$. 
This follows from the consistency conditions (\ref{consistcond2}) which for
the decomposition $\ ^{\shortmid }\mathcal{\aleph }_{0}^{\star }=\
^{\shortmid }\delta _{0}^{\star }+\ ^{\shortmid }\gamma _{0}^{\star }$ (\ref%
{aux02c}), result in $\ ^{\shortmid }\mathcal{T^{\star }}$ $\circ \ \
^{\shortmid }\gamma _{0}^{\star }=\ ^{\shortmid }\gamma _{0}^{\star }\circ \
^{\shortmid }\mathcal{T^{\star }}.$ In above formulas, the nonassociative BV
Laplacian is defined and computed as%
\begin{equation*}
\ ^{\shortmid }\widehat{\triangle }_{\star }\ ^{\shortmid }\mathcal{X}%
^{\star }=(-1)^{(1+\#gh(\ ^{\shortmid }\mathcal{X}^{\star }))}\sum_{\check{%
\alpha}}\int \frac{_{r}^{\shortmid }\delta ^{2}\ ^{\shortmid }\mathcal{X}%
^{\star }}{\delta \ ^{\shortmid }\mathcal{\varphi }_{\check{\alpha}}^{\star
}\delta \ \ _{\ddag }^{\shortmid }\mathcal{\varphi }_{\check{\alpha}}^{\star
}}\ ^{\shortmid }\delta ^{8}\ ^{\shortmid }\mu .
\end{equation*}

Using the classical master equations nCME (\ref{nonassocme}) and, for
symmetry reasons, setting $\ ^{\shortmid }\widehat{\triangle }_{\star }\ \
^{\shortmid }S_{0}^{\star }=0,$ when 
\begin{equation*}
\ ^{\shortmid }\mathcal{\aleph }_{0}^{\star }(e_{\ ^{\shortmid }\mathcal{T}%
}^{i\ ^{\shortmid }:\ ^{\shortmid }V^{\star }:/~\hbar })=\frac{i}{\hbar }%
e_{\ ^{\shortmid }\mathcal{T}}^{i\ ^{\shortmid }:\ ^{\shortmid }V^{\star
}:/~\hbar }\cdot _{\mathcal{T}^{\star }}\ ^{\shortmid }\mathcal{T^{\star }}%
\left( \frac{1}{2}\{\ \ ^{\shortmid }S_{0}^{\star }+\ ^{\shortmid }V^{\star
},\ \ ^{\shortmid }S_{0}^{\star }+\ ^{\shortmid }V^{\star }\}-i\hbar \
^{\shortmid }\widehat{\triangle }_{\star }(\ \ ^{\shortmid }S_{0}^{\star }+\
^{\shortmid }V^{\star })\right) ,
\end{equation*}%
we write (\ref{nBVdif1}) as nonassociative quantum master equation, nQME:%
\begin{equation}
\frac{1}{2}\{\ \ ^{\shortmid }S_{0}^{\star }+\ ^{\shortmid }V^{\star },\ \
^{\shortmid }S_{0}^{\star }+\ ^{\shortmid }V^{\star }\}=i\hbar \ ^{\shortmid
}\widehat{\triangle }_{\star }(\ \ ^{\shortmid }S_{0}^{\star }+\ ^{\shortmid
}V^{\star }).  \label{nqme}
\end{equation}%
This equation can be considered as a condition on $\ ^{\shortmid }V^{\star }$
stating the locality of the\textit{\ nonassociative quantum BV s-operator} $%
\ ^{\shortmid }\widehat{\mathcal{\aleph }}_{0}^{\star }$ (in general, a
nonassociative star product defines a nonlocal structure but it can
distinguished in parametric form as local ones on base and co-fiber
spaces). In the free theories,  we define and compute parametrically%
\begin{equation}
\ ^{\shortmid }\widehat{\mathcal{\aleph }}_{0}^{\star }:=(\ ^{\shortmid }%
\mathcal{T^{\star }})^{-1}\circ \ ^{\shortmid }\mathcal{\aleph }_{0}^{\star
}\circ \ ^{\shortmid }\mathcal{T^{\star }=}\ ^{\shortmid }\mathcal{\aleph }%
_{0}^{\star }-i\hbar \ ^{\shortmid }\widehat{\triangle }_{\star }
\label{nqbvsop}
\end{equation}%
which follows from (\ref{nBVdif1}) and (\ref{nqme}). We omit here the
s-labels for the geometric s-objects which have to be introduced on a
nonholonomic phase space determined by s-adapted off-diagonal solutions in
nonassociative gauge gravity. 

\vskip4pt The nonassociative quantum BV s-operator (\ref{nqbvsop}) can be
generalized on regular functionals for the interacting nonassociative gauge
fields (see formulas (\ref{roper}) \ and (\ref{aux02b})) 
\begin{equation*}
\ ^{\shortmid }\widehat{\mathcal{\aleph }}^{\star }=\ ^{\shortmid
}r_{\lambda V}^{\star -1}\circ \ ^{\shortmid }\widehat{\mathcal{\aleph }}%
_{0}^{\star }~\circ \ ^{\shortmid }r_{\lambda V}^{\star }.
\end{equation*}%
This is a nonassociative quantum twist of the free classical BV s-operator
by a non-local map involving both the free and the quantum interacting 
theories. In the classical limit, we obtain the formulas (\ref{aux02a}).
Nevertheless, the operator $\ ^{\shortmid }\widehat{\mathcal{\aleph }}%
^{\star }$ is local and characterizes nonassociative quantum gauge invariant
observables. This follows form the property that assuming nQME, we can
compute 
\begin{eqnarray}
\ ^{\shortmid }\widehat{\mathcal{\aleph }}^{\star }\ ^{\shortmid }F^{\star }
&=&e_{\ ^{\shortmid }\mathcal{T}}^{i\ :\ ^{\shortmid }V^{\star }:/~\hbar
}\cdot _{\mathcal{T}^{\star }}\ ^{\shortmid }\mathcal{\aleph }_{0}^{\star
}(e_{\ ^{\shortmid }\mathcal{T}}^{i~:\ ^{\shortmid }V^{\star }:/~\hbar
}\cdot _{\mathcal{T}^{\star }}:\ ^{\shortmid }F^{\star }:)  \label{aux03d} \\
&=&\{\ \ ^{\shortmid }F^{\star },\ \ ^{\shortmid }S_{0}^{\star }+\
^{\shortmid }V^{\star }\}-i\hbar \ ^{\shortmid }\widehat{\triangle }_{\star
}(\ \ ^{\shortmid }F^{\star })=\ ^{\shortmid }\mathcal{\aleph }_{0}^{\star
}-i\hbar \ ^{\shortmid }\widehat{\triangle }_{\star }(\ \ ^{\shortmid
}F^{\star }).  \notag
\end{eqnarray}%
This operator is also nilpotent by definition. 

\vskip4pt We can extend the nQME and $\ ^{\shortmid }\widehat{\mathcal{%
\aleph }}^{\star }$ to local s-observables by replacing $\cdot _{\mathcal{T}%
^{\star }}$ with the renormalized time-ordered product on nonasociative
phase spaces by generalizing in abstract nonholonomic geometric form the
results stated by Theorem 3.4 in \cite{fr13}. In parametric effective form,
the associative product $\cdot _{r}$on $\ ^{\shortmid }\mathcal{T}_{r}(\
^{\shortmid }\mathcal{F}^{\star })$ is given by 
\begin{eqnarray}
\ ^{\shortmid }\mathcal{F}_{1}^{\star }\cdot _{\mathcal{T}_{r}^{\star }}\
^{\shortmid }\mathcal{F}_{2}^{\star } &:=&\ ^{\shortmid }\mathcal{T}_{r}(\
^{\shortmid }\mathcal{T}_{r}^{-1}\ \ ^{\shortmid }\mathcal{F}_{1}^{\star
}\cdot \ ^{\shortmid }\mathcal{T}_{r}^{1}\ \ ^{\shortmid }\mathcal{F}%
_{2}^{\star }),\mbox{ for }  \label{aux03a} \\
&&\ ^{\shortmid }\mathcal{T}_{r}:\ \ ^{\shortmid }\mathcal{F}^{\star
}[[\hbar\ ]] \rightarrow \ ^{\shortmid }\mathcal{T}_{r}(\ ^{\shortmid }%
\mathcal{F}^{\star })[[\hbar \ ,\kappa ]]  \notag
\end{eqnarray}%
defined as $\ ^{\shortmid }\mathcal{T}_{r}=(\oplus _{n}\ ^{\shortmid }%
\mathcal{T}_{r}^{n})\circ \ \ ^{\shortmid }\beta ^{\star },$ where $\ ^{\shortmid }\beta ^{\star }$ is the inversion of multiplication  for 
$\ ^{\shortmid }\mathcal{T}_{r}|_{\ ^{\shortmid }\mathcal{F}_{loc}^{\star}}=id, $ 
so $:\ ^{\shortmid }V^{\star }:=\ ^{\shortmid }V^{\star }.$ 

\vskip4pt The $\cdot _{\mathcal{T}_{r}^{\star }}$defined by (\ref{aux03a})
is an associative and commutative product and we can use it in place of $%
\cdot _{\mathcal{T}^{\star }}$ and define the renormalized parametric nQME
and the quantum nonassiative BV s-operators using formulas (\ref{nBVdif1})
and (\ref{nqbvsop}). For applications, we can consider the terms
proportional to $\kappa ^{0}$ and $\kappa ^{1}$ (the linear ones on $\kappa $
encoding nonassociative data in effective quantum form). We can simplify
such formulas considering a \textit{nonassociative phase space
generalization of the anomalous Master Ward Identity}%
\begin{equation}
\ ^{\shortmid }\mathcal{\aleph }_{0}^{\star }(e_{\ ^{\shortmid }\mathcal{T}%
_{r}}^{i\ :\ ^{\shortmid }V^{\star }:/~\hbar })\equiv \{e_{\ ^{\shortmid }%
\mathcal{T}_{r}}^{i\ :\ ^{\shortmid }V^{\star }:/~\hbar },\ \ ^{\shortmid
}S_{0}^{\star }\}=\frac{i}{\hbar }e_{\ ^{\shortmid }\mathcal{T}_{r}}^{i\ :\
^{\shortmid }V^{\star }:/~\hbar }\cdot _{\mathcal{T}_{r}^{\star }}\left( 
\frac{1}{2}\{\ \ ^{\shortmid }S_{0}^{\star }+\ ^{\shortmid }V^{\star },\ \
^{\shortmid }S_{0}^{\star }+\ ^{\shortmid }V^{\star }\}_{\ ^{\shortmid }%
\mathcal{T}_{r}}-i\hbar \ ^{\shortmid }\widehat{\triangle }_{\star
}^{V}\right) ,  \label{aux03b}
\end{equation}%
where $^{\shortmid }\widehat{\triangle }_{\star }^{V}$ is identified with
the anomaly term. For spacetime bases, similar details are provided in \cite%
{hol08,bd08}), when the necessary formulas and proofs can be extended in
parametric and abstract nonassociative geometric forms. Therefore, \textit{%
the renormalized quantum nonassociative master equation} corresponding to (%
\ref{aux03b}) are%
\begin{equation}
\frac{1}{2}\{\ \ ^{\shortmid }S_{0}^{\star }+\ ^{\shortmid }V^{\star },\ \
^{\shortmid }S_{0}^{\star }+\ ^{\shortmid }V^{\star }\}_{\ ^{\shortmid }%
\mathcal{T}_{r}}-i\hbar \ ^{\shortmid }\widehat{\triangle }_{\star }^{V}=0.
\label{aux04b}
\end{equation}%
Replacing in this formula $^{\shortmid }\widehat{\triangle }_{\star }^{V}(\
^{\shortmid }F^{\star }):=\frac{d}{d\lambda }^{\shortmid }\widehat{\triangle 
}_{\star }^{V+\lambda \ ^{\shortmid }F^{\star }}|_{\lambda =0}$ and
considering that the renormalized nQME hold, we can write such a master
equation using the renormalized nonassociative BV s-operator from (\ref%
{aux03d}), 
\begin{equation*}
\ ^{\shortmid }\widehat{\mathcal{\aleph }}^{\star }\ ^{\shortmid }F^{\star
}=\{\ ^{\shortmid }F^{\star },\ \ ^{\shortmid }S_{0}^{\star }+\ ^{\shortmid
}V^{\star }\}-i\hbar \ ^{\shortmid }\widehat{\triangle }_{\star }^{V}(\
^{\shortmid }F^{\star }).
\end{equation*}%
So, using the renormalized time ordered product $\cdot _{\mathcal{T}%
_{r}^{\star }},$ we obtained an anomaly (which is local on respective h- and
c-components and of order $O(\hbar \ ,\kappa )$) via $\ ^{\shortmid }%
\widehat{\triangle }_{\star }^{V}(\ ^{\shortmid }F^{\star })$ instead of $\
^{\shortmid }\widehat{\triangle }_{\star }(\ ^{\shortmid }F^{\star }).$ In
the renormalized nonassociative gauge gravity theory, $\ ^{\shortmid }%
\widehat{\triangle }_{\star }^{V}$ can be well-defined on local s-vector
fields, in contrast to $\ ^{\shortmid }\widehat{\triangle }_{\star }.$ 

\vskip4pt In section 4 of \cite{rejzner20} and in \cite{bf20} (in the variant with local S-matrices and generating $C^{\ast }$-algebra) possible approaches towards a non-perturbative formulation of the BV-formalism elaborated for the pAQFT  are analyzed. In a general formalism with
nonassociative twisted star products, the formulation of a general variational
formalism is impossible (except effective models with parametric
decompositions). This may consist a program of research and a series of
future works on nonassociative pAQFT using effective S-matrices and
nonassociative Schwinger-Dyson equations. Here we note that using
off-diagonal parametric solutions in nonassociative gauge gravity, with
nonlinear symmetries (\ref{etapolgen}) and gravitational polarizations $\
_{s}^{\shortmid }\eta \ ^{\shortmid }\mathring{g}_{\alpha _{s}}\sim \
^{\shortmid }\zeta _{\alpha _{s}}(1+\kappa \ ^{\shortmid }\chi _{\alpha
_{s}})\ \ ^{\shortmid }\mathring{g}_{\alpha _{s}},$ we can define and
compute $\ ^{\shortmid }\widehat{\triangle }_{\star }^{V}$ as a respective
distortion of $\ ^{\shortmid }\widehat{\triangle }_{\star },$ for
functionals $\ ^{\shortmid }V^{\star }[\ _{s}^{\shortmid }\eta ]\sim \
^{\shortmid }V^{\star }[\ _{s}^{\shortmid }\chi ].$ Such anomaly terms and
their distortions can be computed using nonlinear symmetries (\ref{nonlinsym}%
). So, our approach with nonholonomic deformations and generating classical
and off-diagonal solutions is generic non-perturbative and allow to select
well-defined, for instance, quasi-stationary nonassociative configurations
resulting in renormalized nonassociative BV s-operators. In this work, we do not consider nonassociative and noncommutative extensions of non-perturbative quantum methods with local S-matrices and generating $C^*$-algebras \cite{rejzner20,bf20} because it is not clear how such a formalism can be elaborated in a general form for nonlocal and non-variational theories encoding nonassociativity. For explicit classes of physically important solutions with parametric R-flux nonassociativity existing in the low energy limit of string/ M-theory, an effective variational formulation of corresponding models is possible. This motivates the goals of this work to elaborate on a geometric and quantum BV formalism for some general classes of off-diagonal solutions with parametric nonassociative data.  

\subsection{ BV quantization of nonassociative 8-d modified BH configurations%
}

In this subsection, we consider an example of how the classical and quantum
BV schemes from sections \ref{sec03} and \ref{sec04} can be applied for
quantizing nonassociative BH solutions. The BV formalism is used in explicit
form for quantizing the quasi-stationary off-diagonal R-flux deformations of
regular phase spaces with Dymnikova backgrounds resulting in 8-d s-metrics
of type (\ref{sol4of}).\footnote{%
Similar nonassociative BV constructions can be performed for any
nonassociative BH and WH solutions which may include, or not, singularities,
off-diagonal deformations etc. Such solutions were constructed and studied
in classical form \cite{partner02,partner04,partner05,partner06,partner07}.
It is not possible to elaborate on their BV quantization (on hundred of
pages) in this work.} All formulas are provided in abstract geometric form
which allows an "economic" geometric formulation and straightforward
application of the BV method for nonassociative gauge gravity theories. 

\subsubsection{Generating data for nonassociative star product deformed of
Dymnikova BHs}

We parameterize the quasi-stationary solutions for off-diagonal deformations
of primary metrics (\ref{pm7d8d}) to the target ones (\ref{sol4of}) in the
form 
\begin{equation}
\ _{s}^{\shortmid }\mathbf{\breve{g}}=[\ ^{\shortmid }\breve{g}_{\alpha
_{s}},\ ^{\shortmid }\breve{N}_{i_{s-1}}^{a_{s}}]\rightarrow \ _{\eta
}^{\shortmid }\mathbf{\breve{g}=}\ _{s}^{\shortmid }\eta \lbrack \breve{r}%
_{0},\breve{r}_{g},\ _{s}^{\shortmid }\Lambda ]\ ^{\shortmid }\breve{g}%
_{\alpha _{s}}[\breve{r}_{0},\breve{r}_{g}]\sim \ ^{\shortmid }\zeta
_{\alpha _{s}}[\breve{r}_{0},\breve{r}_{g},\ _{s}^{\shortmid }\Lambda
](1+\kappa \ ^{\shortmid }\chi _{\alpha _{s}}[\breve{r}_{0},\breve{r}_{g},\
_{s}^{\shortmid }\Lambda ])\ ^{\shortmid }\breve{g}_{\alpha _{s}}[\breve{r}%
_{0},\breve{r}_{g}].  \label{auxparam1a}
\end{equation}%
In these formulas, $\breve{r}_{0}$ and $\breve{r}_{g}$ are physical
constants for the prime metric's phase space defined by the generalized
Dymnikova BH solution, see (\ref{sourc2bb}) and (\ref{sourc2}). The
quasi-stationary target metrics $\ _{\eta }^{\shortmid }\breve{g}=\{ _{\eta
}^{\shortmid }\breve{g}_{\alpha _{s}}\}$ (\ref{sol4of}) are of type (\ref%
{offdiagdefr}) involving generating functions (\ref{etapolgen})
parameterized in the form:%
\begin{eqnarray}
\psi &\simeq &\psi (\hbar ,\kappa ,\breve{r}_{0},\breve{r}_{g},\
_{1}^{\shortmid }\Lambda ,x^{k_{1}}),\eta _{4}\ \simeq \eta _{4}(\hbar
,\kappa ,\breve{r}_{0},\breve{r}_{g},\ _{2}^{\shortmid }\Lambda
,x^{k_{1}},y^{3}),  \label{generdatadymn} \\
\ ^{\shortmid }\eta ^{6} &\simeq &\ ^{\shortmid }\eta ^{6}(\hbar ,\kappa ,%
\breve{r}_{0},\breve{r}_{g},\ _{3}^{\shortmid }\Lambda ,x^{i_{2}},p_{5}),\
^{\shortmid }\eta ^{8}\simeq \ ^{\shortmid }\eta ^{8}(\hbar ,\kappa ,\breve{r%
}_{0},\breve{r}_{g},\ _{4}^{\shortmid }\Lambda ,x^{i_{2}},p_{5},p_{7}). 
\notag
\end{eqnarray}%
The effective shell cosmological constants, $\ _{s}^{\shortmid }\Lambda ,$ from (\ref{generdatadymn}) are related to effective matter sources encoding nonassociative data, $\ _{s}^{\shortmid }\mathcal{J}^{\star },$ via nonlinear symmetries of type (\ref{nonlinsym}). For such formulas,  the generating functions, generating source and effective cosmological constants transform respectively as 
\begin{equation}
\ [\eta _{4},\ ^{\shortmid }\eta ^{6},\ ^{\shortmid }\eta ^{8},\
_{s}^{\shortmid }\mathcal{J}^{\star }]\longleftrightarrow \lbrack \ _{s}\Psi
,\ _{s}^{\shortmid }\mathcal{J}^{\star }]\longleftrightarrow \lbrack \
_{s}\Phi ^{\star },\ _{s}^{\shortmid }\Lambda ],  \label{nonlinsymd}
\end{equation}%
with dependencies on respective sets of physical constants $[\hbar ,\kappa ,%
\breve{r}_{0},\breve{r}_{g},\ _{s}^{\shortmid }\Lambda ].$ We can introduce
also other constants and integration functions stating, for instance,
certain ellipsoidal symmetries with an eccentricity $\epsilon ,$ or certain
constants defining toroid/ cylindric off-diagonal deformation etc. We
emphasize that the nonlinear symmetries (\ref{nonlinsym}) and (\ref%
{nonlinsymd}) are not gauge-like symmetries considered for gravitational
gauge theories. They reflect (nonassociative) parametric properties of
certain classes of quasi-stationary solutions of modified Einstein equations
and their lifts or equivalents in total spaces written as YM equations. 

\subsubsection{The BV formalism for nonassociative BH solutions}

Let us explain the main steps for performing BV quantization of $\
_{\eta}^{\shortmid }\mathbf{\breve{g}}$ (\ref{sol4of}) determined by
(nonassociative) data (\ref{generdatadymn}) and (\ref{nonlinsymd}). The
geometric quantization can be formulated in a non-perturbative parametric
form using gravitational $\eta $-polarizations and nonlinear symmetries
transforming generating sources $\ _{s}\mathcal{J}^{\star }$ into effective
cosmological constants $\ _{s}^{\shortmid }\Lambda .$ For certain classes of
well-defined (as effective relativistic theories) off-diagonal
configurations with asymptotic quasi-classical limits (selected by
respective data $[\hbar ,\kappa ,\breve{r}_{0},\breve{r}_{g},\
_{s}^{\shortmid }\Lambda ]$), we can elaborate on perturbative schemes on $%
\hbar $ and linearized on $\kappa $ and possible recurrent higher order
terms. We omit in this section cumbersome perturbative formulas involving $k$%
-linear decompositions of type (\ref{generdatadymn}). 

\vskip4pt For a $\ _{\eta }^{\shortmid }\mathbf{\breve{g}}$ (\ref{sol4of})
we can construct a parametric nonassociative gravitational gauge potential $%
\ _{s}^{\shortmid }\widehat{\mathcal{A}}_{[P]}^{\star }\rightarrow \
_{\eta}^{\shortmid }\widehat{\mathcal{A}}_{[P]}^{\star }$ (\ref{affinpot})
as a R-flux deformation of (\ref{cdSgp}). Such an effective gauge potential
contains all data on quasi-stationary deformations (via $\eta $%
-polarizations) and allows to define an effective gauge gravitational action
(\ref{effectgauge}) when%
\begin{equation*}
\ _{\eta }^{\shortmid }L^{\star }=\ _{gr}^{\shortmid }L^{\star }(\
^{\shortmid }f)[\ \ _{\eta }^{\shortmid }\widehat{\mathcal{A}}^{\star }]=-%
\frac{1}{2}\int_{\ _{\eta }^{\shortmid }\mathcal{U}_{\mathfrak{Cqs}}}\
^{\shortmid }f\quad tr(\ _{\eta }^{\shortmid }\widehat{\mathcal{F}}^{^{\star
}}\wedge (\divideontimes \ _{\eta }^{\shortmid }\widehat{\mathcal{F}}%
^{^{\star }})),
\end{equation*}%
for $\ _{\eta }^{\shortmid }\widehat{\mathcal{F}}^{^{\star }}$ computed as
the nonassociative strength defined by $\ \ _{\eta }^{\shortmid }\widehat{%
\mathcal{A}}^{\star },$ and when the measure $\delta ^{8}\mu $ is defined by
a chosen $\ _{\eta }^{\shortmid }\mathbf{\breve{g}\in }\mathfrak{Cqs}$ for a 
$\eta $-deformed region $\ _{\eta }^{\shortmid }\mathcal{U}_{\mathfrak{Cqs}}.
$ Necessary formulas from subsection \ref{ss32} can be applied in abstract
nonassociative form for geometric objects with $\eta $-labels when $\ _{\eta
}^{\shortmid }L^{\star }\in \ ^{\shortmid }\mathcal{L}^{\star }$ and $\
^{\shortmid }\mathcal{\varphi }^{\star }= \ _{\eta }^{\shortmid }\widehat{%
\mathcal{A}}^{\star }\in \ ^{\shortmid }\mathcal{E}_{\mathfrak{Cqs}}^{\star
}.$ For such quasi-stationary configurations, an effective variational
calculus can be defined in parametric form even though a unique nonassociative
differential and integral calculus can't be defined for a general twist
product. Then, we can introduce in functional effective form the
Euler-Lagrange derivative of a corresponding effective action $\ _{\eta
}^{\shortmid }S^{\star },$ see (\ref{effectelag}) with $\ ^{\shortmid}%
\mathcal{E}_{\mathfrak{\eta }}^{\star }\subset \ ^{\shortmid }\mathcal{E}_{%
\mathfrak{Cqs}}^{\star }$, when 
\begin{eqnarray}
&&\ ^{\shortmid }d\ _{\eta }^{\shortmid }S^{\star }:\ ^{\shortmid }\mathcal{E%
}_{\mathfrak{\eta }}^{\star }\rightarrow \ ^{\shortmid }\mathcal{E}_{%
\mathfrak{\eta }}^{^{\prime }C}\mbox{ defined by }  \label{effectelageta} \\
&&<\ ^{\shortmid }d\ _{\mathfrak{\eta }}^{\shortmid }S^{\star }(\ _{\eta
}^{\shortmid }\widehat{\mathcal{A}}^{\star }),\ \ _{\eta }^{\shortmid }%
\widehat{\mathcal{A}}_{1}^{\star }>:=\lim_{\varsigma \rightarrow 0}\frac{1}{%
\varsigma }\ ^{\shortmid }\delta \ _{\eta }^{\shortmid }L^{\star }(\varsigma
\ _{\eta }^{\shortmid }\widehat{\mathcal{A}}_{1}^{\star })[\ _{\eta
}^{\shortmid }\widehat{\mathcal{A}}^{\star }]=\int \frac{\ ^{\shortmid
}\delta \ _{\eta }^{\shortmid }L^{\star }(\ ^{\shortmid }f)}{\ ^{\shortmid
}\delta \ _{\eta }^{\shortmid }\widehat{\mathcal{A}}^{\star }}\ _{\eta
}^{\shortmid }\widehat{\mathcal{A}}_{1}^{\star }(\ ^{\shortmid }u),\mbox{for
}  \notag \\
&&\ _{\eta }^{\shortmid }\widehat{\mathcal{A}}_{1}^{\star }\in \ ^{\shortmid
}\mathcal{E}_{\eta }^{C},\frac{\ ^{\shortmid }\delta \ _{\eta }^{\shortmid
}L^{\star }(\ ^{\shortmid }f)}{\ ^{\shortmid }\delta \ ^{\shortmid }\mathcal{%
\varphi }^{\star }}\in (\ ^{\shortmid }\mathcal{E}_{\eta }^{\star })\subset
\ ^{\shortmid }\mathcal{E}_{\eta }^{^{\prime }C}; \mbox{when\ } \
^{\shortmid }d\ _{\eta }^{\shortmid }S^{\star }(\ _{\eta }^{\shortmid }%
\widehat{\mathcal{A}})\equiv 0\mbox{ defines the effective field equations };
\notag \\
&&\ _{Sol}^{\shortmid }\mathcal{E}_{\eta }^{\star }%
\mbox{ denotes
the spaces of solutions defined as the zero locus of the 1-form }d\ _{\eta
}^{\shortmid }S^{\star },\ _{Sol}^{\shortmid }\mathcal{E}_{\eta }^{\star
}\subset \ ^{\shortmid }\mathcal{E}_{\eta }^{\star };  \notag \\
&&\ _{Sol}^{\shortmid }\mathcal{F}_{\eta }%
\mbox{ denotes the space
of off-shell functionals in the space of functionals on  }\
_{Sol}^{\shortmid }\mathcal{E}_{\eta }^{\star }.  \notag
\end{eqnarray}

\vskip4pt So, for a more special case of parametric quasi-stationary
configurations (\ref{effectelageta}) with corresponding effective
Lagrangians and actions, the \textit{nonassociative classical BV
differential for off-diagonal phase space Dymnikova BHs is }computed as 
\begin{equation}
\ _{\eta }^{\shortmid }\mathcal{\aleph }^{\star }=\{\bullet ,\ _{\eta
}^{\shortmid }S^{\star }+\ _{\eta }^{\shortmid }\Theta ^{\star
}\}:=\{\bullet ,\ _{\eta }^{\shortmid }S_{ext}^{\star }\},%
\mbox{ with
extended action }\ _{\eta }^{\shortmid }S_{ext}^{\star }.  \label{ncmebh}
\end{equation}%
These functional equations depend on certain subclasses of generating and
integration data (\ref{integrfunctrf}) and (\ref{poas2d}) defining
nonassociative off-diagonal BH solutions. The nilpotent property, $(\
_{\eta}^{\shortmid }\mathcal{\aleph }^{\star })^{2}=0,$ allows to define the 
\textit{nonassociative classical master equation} (nCME) for such BHs, which
modulo terms vanishing in the limit of a constant $\ ^{\shortmid }f$ is
written in the form  $\{\ _{\eta }^{\shortmid }L_{ext}^{\star }(\
^{\shortmid }f),\ _{\eta }^{\shortmid }L_{ext}^{\star }(\ ^{\shortmid }f)\}=0
$. For such nCME, it is used an effective $\ _{\eta }^{\shortmid
}L_{ext}^{\star }$ is used for defining a respective $\ _{\eta
}^{\shortmid}S_{ext}^{\star }.$ 

\vskip4pt The operator $\ _{\eta }^{\shortmid }\mathcal{\aleph }^{\star }$
from (\ref{ncmebh}) also increases the ghost number by one and can be
expressed as a sum, $\ _{\eta }^{\shortmid }\mathcal{\aleph }^{\star }=\
_{\eta }^{\shortmid }\delta ^{\star }+\ _{\eta }^{\shortmid }\gamma ^{\star }
$, for an extension of $\ _{\eta }^{\shortmid }\delta _{S}^{\star }$ denoted 
$\ _{\eta }^{\shortmid }\delta ^{\star }.$ Defining $\ _{\eta
}^{\shortmid}\Theta ^{\star }:=\ _{\eta }^{\shortmid }S_{ext}^{\star } - \
_{\eta}^{\shortmid }S^{\star }$ as in formulas (\ref{nBVdif}), we can
express 
\begin{equation*}
~_{\eta }^{\shortmid }\delta ^{\star }\ =\{\cdot ,\ _{\eta }^{\shortmid
}S^{\star }\}\mbox{
and }\ _{\eta }^{\shortmid }\gamma ^{\star }=\{\cdot ,\ _{\eta }^{\shortmid
}\Theta ^{\star }\}.
\end{equation*}%
In above formulas, the differential $\ _{\eta }^{\shortmid }\delta ^{\star }$
acts trivially both on fields and antifields. Using $\ _{\eta}^{\shortmid
}\delta ^{\star }\ _{\ddag \eta }^{\shortmid }\widehat{\mathcal{A}}^{\star }=%
\frac{^{\shortmid }\delta \ _{\eta }^{\shortmid }S^{\star }}{\delta \
_{\ddag \eta }^{\shortmid }\widehat{\mathcal{A}}^{\star }}$, we derive
nonassociative gauge-fixed equations of motion which are hyperbolic
equations of motion of $\ _{\eta }^{\shortmid }S^{\star }.$ 

\vskip4pt On nonassociative phase space Dymnikova background (\ref{pm7d8d})
and for further parametric decompositions to an off-diagonal solution $\
_{\eta}^{\shortmid }\mathbf{\breve{g}}$ (\ref{sol4of}), and respective $\_{%
\eta}^{\shortmid }\widehat{\mathcal{A}}_{[P]}^{\star }$ (\ref{affinpot}), we
can split and linearize respectively the extended action $\ _{\eta
}^{\shortmid}S_{ext}^{\star }$ in a form similar to  (\ref{interact1}),%
\begin{eqnarray*}
\ _{\eta }^{\shortmid }S_{0}^{\star } &=&\ _{\eta }^{\shortmid }\breve{S}%
_{00}^{\star }+\ _{\eta }^{\shortmid }\Theta _{0}^{\star },\mbox{ the
quadratic term in (anti) fields},\#ta(\ _{\eta }^{\shortmid }S_{00}^{\star
})=0,\#ta(\ _{\eta }^{\shortmid }\Theta _{0}^{\star })=1; \\
\ _{\eta }^{\shortmid }V^{\star } &=&\ ^{\shortmid }\breve{V}_{0}^{\star }+\
_{\eta }^{\shortmid }\Theta ^{\star },\mbox{ the interacton term }; \\
\ ^{\shortmid }S^{\star } &=&\ _{\eta }^{\shortmid }\breve{S}_{00}^{\star
}+\ ^{\shortmid }\breve{V}_{0}^{\star },\mbox{ the total antifield
independent part of the action}.
\end{eqnarray*}%
%

\vskip4pt We can chose such nonholonomic s-adapted distributions, when $\
_{\eta }^{\shortmid }\breve{S}_{00}^{\star }=\ _{\eta }^{\shortmid }\breve{S}%
_{00}$ and $^{\shortmid }\breve{V}_{0}^{\star }=~^{\shortmid }\breve{V}_{0}$
are defined by s-adapted classical decompositions of the primary s-metric (%
\ref{pm7d8d}) and its associative quantum deformations; and when the $\
_{\eta }^{\shortmid}\Theta _{0}^{\star }$ and $\ _{\eta }^{\shortmid }\Theta
^{\star }$ are for quantum nonassociative deformations. For the target
nonassociative BH solutions, the formulas (\ref{aux02c}) transform into
parametric linearized s-operators of BRST and BV type,%
\begin{equation*}
\ _{\eta }^{\shortmid }\gamma ^{\star }\ ^{\shortmid }\mathcal{F}:=\{\
^{\shortmid }\mathcal{F},\ _{\eta }^{\shortmid }\Theta _{0}^{\star }\}%
\mbox{
and }\ _{\eta }^{\shortmid }\mathcal{\aleph }_{0}^{\star }=~_{\eta
}^{\shortmid }\delta _{0}^{\star }+\ _{\eta }^{\shortmid }\gamma _{0}^{\star
},
\end{equation*}%
where $\ _{\eta }^{\shortmid }\delta _{0}^{\star }(~_{\ddag \eta
}^{\shortmid }\widehat{\mathcal{A}}_{[P]}^{\star \check{\alpha}})=-$ $\frac{%
_{\eta }^{\shortmid }\delta \ _{\eta }^{\shortmid }S_{00}^{\star }}{\delta
~_{\eta }^{\shortmid }\widehat{\mathcal{A}}_{\check{\alpha}[P]}^{\star }}$
and respective nonassociative differential operators, $_{\eta }^{\shortmid
}P_{\star }^{\check{\alpha}\breve{\beta}}(\ ^{\shortmid }u)$ and $\ _{\eta
}^{\shortmid }K_{\star \breve{\beta}}^{\check{\alpha}}$ are defined from 
\begin{eqnarray*}
\frac{_{l}^{\shortmid }\delta \ ^{\shortmid }S_{00}^{\star }}{\delta ~_{\eta
}^{\shortmid }\widehat{\mathcal{A}}_{\check{\alpha}[P]}^{\star }}(~_{\eta
}^{\shortmid }\widehat{\mathcal{A}}_{\check{\alpha}[P]}^{\star }) &:=& \
_{\eta }^{\shortmid }P_{\star }^{\check{\alpha}\breve{\beta}}(\ ^{\shortmid
}u)(~_{\eta }^{\shortmid }\widehat{\mathcal{A}}_{\breve{\beta}[P]}^{\star }),%
\mbox{ in brief },=\ _{\eta }^{\shortmid }P_{\star }\ ~_{\eta }^{\shortmid }%
\widehat{\mathcal{A}}_{\breve{\beta}[P]}^{\star };\mbox{ and }\  \\
\frac{_{r}^{\shortmid }\delta _{l}^{\shortmid }\delta \ \ _{\eta
}^{\shortmid }\Theta _{0}^{\star }}{\delta \ _{\eta }^{\shortmid }\widehat{%
\mathcal{A}}_{\check{\alpha}[P]}^{\star }(\ ^{\shortmid }u_{1})\delta \
~_{\ddag \eta }^{\shortmid }\widehat{\mathcal{A}}_{[P]}^{\star \breve{\beta}%
}(\ ^{\shortmid }u_{1})}(~_{\eta }^{\shortmid }\widehat{\mathcal{A}}_{\check{%
\alpha}[P]}^{\star }) &:=& \ _{\eta }^{\shortmid }K_{\star \breve{\beta}}^{%
\check{\alpha}}(\ ^{\shortmid }u)(\ _{\eta }^{\shortmid }\widehat{\mathcal{A}%
}_{\breve{\beta}[P]}^{\star }).
\end{eqnarray*}%
These functional equations depend on respective classes of generating data (%
\ref{integrfunctrf}) and (\ref{poas2d}).

\vskip4pt We chose nonholonomic configurations on nonassociative phase space and assume a gauge fixing in the total nonassociative vector bundle in such a way that $\ _{\eta }^{\shortmid }P_{\star }$ is Green hyperbolic for any $h$- and $c$-component as for (associative and commutative) gauge and gravity
theories. The corresponding double Green function is defined by  the motion operator 
$\ _{\eta }^{\shortmid }P_{\star }$ determined in canonical form (with hat d-operators) by 
$\ _{\eta }^{\shortmid }\mathbf{\breve{g}}$ as a particular case of (\ref{nonassocpj}). This allows us to define the respective nonassociative Pauli-Jordan functionals%
\begin{equation}
\ _{\eta }^{\shortmid }\widehat{\triangle }_{\star }=\ _{\eta }^{\shortmid }%
\widehat{\triangle }_{\star }^{R}- \ _{\eta }^{\shortmid }\widehat{\triangle 
}_{\star }^{A}.  \label{nonassocpjbh}
\end{equation}%
We need additional assumptions to prescribe respective generating data for
the nonassociative BHs to transform such functionals into certain
Pauli-Jordan functions. 

\vskip4pt Using the function (\ref{nonassocpjbh}), we can express in terms of $\eta $-polarization functions the formulas (\ref{aux02a})  and (\ref{aux02b}), for nonassociative off-diagonal BH deformations. Then, we 
define and compute the corresponding classical BV operator (\ref{nBVdif}), 
\begin{eqnarray}
\ _{\eta }^{\shortmid }\mathcal{\aleph }^{\star } &=& _{\eta }^{\shortmid
}r_{\lambda V}^{\star -1}\circ ~_{\eta }^{\shortmid }\mathcal{\aleph }%
_{0}^{\star }~\circ \ ~_{\eta }^{\shortmid }r_{\lambda V}^{\star }, \mbox{
where }  \label{cnassbvsop} \\
&& \ _{\eta }^{\shortmid }r_{\lambda V}^{\star -1}(~_{\eta }^{\shortmid }%
\widehat{\mathcal{A}}_{\check{\alpha}[P]}^{\star }) = ~_{\eta }^{\shortmid }%
\widehat{\mathcal{A}}_{\check{\alpha}[P]}^{\star }+\ ~_{\eta }^{\shortmid }%
\widehat{\triangle }_{\star \check{\alpha}\breve{\beta}}\frac{%
_{l}^{\shortmid }\delta \ \ _{\eta }^{\shortmid }V^{\star }}{\delta \
~_{\eta }^{\shortmid }\widehat{\mathcal{A}}_{\breve{\beta}[P]}^{\star }}(\
~_{\eta }^{\shortmid }\widehat{\mathcal{A}}_{[P]}^{\star })\mbox{ and } 
\notag \\
&& \ ~_{\eta }^{\shortmid }r_{\lambda V}^{\star }(\ ~_{\eta }^{\shortmid }%
\widehat{\mathcal{A}}_{\check{\alpha}[P]}^{\star }) =~_{\eta }^{\shortmid }%
\widehat{\mathcal{A}}_{\check{\alpha}[P]}^{\star }-\ ~_{\eta }^{\shortmid }%
\widehat{\triangle }_{\star \check{\alpha}\breve{\beta}}\frac{%
_{l}^{\shortmid }\delta \ \ _{\eta }^{\shortmid }V^{\star }}{\delta \
~_{\eta }^{\shortmid }\widehat{\mathcal{A}}_{\breve{\beta}[P]}^{\star }}(\
~_{\eta }^{\shortmid }r_{\lambda V}^{\star }(~_{\eta }^{\shortmid }\widehat{%
\mathcal{A}}_{[P]}^{\star })).  \notag
\end{eqnarray}%
Above s-operators allow us to formulate the $nCME(\ _{\eta
}^{\shortmid}S^{\star })$ (i.e. the classical master equations for
nonassociative star product deformation (\ref{starpn}) of generalized
Dymnikova BHS in gauge gravity):%
\begin{eqnarray}
\ ~_{\eta }^{\shortmid }r_{\lambda V}^{\star -1}(\{\ ^{\shortmid }\mathcal{X}%
^{\star },\ ~_{\eta }^{\shortmid }S_{0}^{\star }\}) &=&\{\ ~_{\eta
}^{\shortmid }r_{\lambda V}^{\star -1}(\ ^{\shortmid }\mathcal{X}^{\star
}),\ ~_{\eta }^{\shortmid }S_{0}^{\star }+\ ~_{\eta }^{\shortmid }V^{\star
}\}-  \label{aux04a} \\
&&\int \frac{_{r}^{\shortmid }\delta \ ^{\shortmid }\mathcal{X}^{\star }}{%
\delta \ ~_{\eta }^{\shortmid }\widehat{\mathcal{A}}_{\check{\alpha}%
[P]}^{\star }(\ ^{\shortmid }u^{\prime })}(\ ~_{\eta }^{\shortmid
}r_{\lambda V}^{\star -1}(~_{\eta }^{\shortmid }\widehat{\mathcal{A}}%
_{[P]}^{\star }))\ ~_{\eta }^{\shortmid }\widehat{\triangle }_{\star \check{%
\alpha}\breve{\beta}}(\ ^{\shortmid }u^{\prime },\ ^{\shortmid }u)\frac{%
_{l}^{\shortmid }\delta (nCME(~_{\eta }^{\shortmid }S^{\star }))}{\delta \ \
~_{\eta }^{\shortmid }\widehat{\mathcal{A}}_{\breve{\beta}[P]}^{\star }(\
^{\shortmid }u)}.  \notag
\end{eqnarray}%
%

\vskip4pt Here we emphasize that such functional equations depend on respective classes of generating data (\ref{integrfunctrf}) and (\ref{poas2d}). This can be chosen in certain forms allowing to construct well-defined physical solutions of (\ref{aux04a}) involving respective s-operators (\ref%
{nonassocpjbh}) and (\ref{cnassbvsop}).

\subsubsection{Nonholonomic BV scheme and quantization of nonassociative BH solutions}

The nonassociative BV s-operator for phase space BH R-flux deformations (\ref%
{cnassbvsop}) can be extended in quantum normalized form using the formula (%
\ref{nBVdif1}), 
\begin{equation*}
~_{\eta }^{\shortmid }\mathcal{\aleph }_{0}^{\star }(e_{\ ^{\shortmid }%
\mathcal{T}}^{i\ ^{\shortmid }:\ ~_{\eta }^{\shortmid }V^{\star }:/~\hbar })=%
\frac{i}{\hbar }e_{\ ^{\shortmid }\mathcal{T}}^{i\ ^{\shortmid }:\ ~_{\eta
}^{\shortmid }V^{\star }:/~\hbar }\cdot _{\mathcal{T}^{\star }}\ ^{\shortmid
}\mathcal{T^{\star }}\left( \frac{1}{2}\{\ \ ~_{\eta }^{\shortmid
}S_{0}^{\star }+\ ~_{\eta }^{\shortmid }V^{\star },\ \ ~_{\eta }^{\shortmid
}S_{0}^{\star }+\ ~_{\eta }^{\shortmid }V^{\star }\}-i\hbar \ ~_{\eta
}^{\shortmid }\widehat{\triangle }_{\star }(\ ~_{\eta }^{\shortmid
}S_{0}^{\star }+~_{\eta }^{\shortmid }V^{\star })\right) ,
\end{equation*}%
when $\ _{\eta }^{\shortmid }\triangle _{\star }\ _{\eta
}^{\shortmid}S_{0}^{\star }=0.$ Using the classical master equations nCME (%
\ref{nonassocme}), we obtain for phase space R-flux deformed BHs the
nonassociative quantum master equation, nQME:%
\begin{equation*}
\frac{1}{2}\{\ ~_{\eta }^{\shortmid }S_{0}^{\star }+~_{\eta }^{\shortmid
}V^{\star },\ ~_{\eta }^{\shortmid }S_{0}^{\star }+~_{\eta }^{\shortmid
}V^{\star }\}=i\hbar \ ~_{\eta }^{\shortmid }\widehat{\triangle }_{\star
}(~_{\eta }^{\shortmid }S_{0}^{\star }+~_{\eta }^{\shortmid }V^{\star }).
\end{equation*}%
This equation can be considered as a condition on the R-flux deformed
effective potential $\ _{\eta }^{\shortmid }V^{\star }$ stating the
parametric locality of the corresponding \textit{\ nonassociative quantum BV
s-operator,} 
\begin{equation*}
\ _{\eta }^{\shortmid }\widehat{\mathcal{\aleph }}_{0}^{\star }:=(\
^{\shortmid }\mathcal{T^{\star }})^{-1}\circ \ ~_{\eta }^{\shortmid }%
\mathcal{\aleph }_{0}^{\star }\circ \ ^{\shortmid }\mathcal{T}^{\star }= \
_{\eta }^{\shortmid }\mathcal{\aleph }_{0}^{\star }- i \hbar \ _{\eta
}^{\shortmid }\widehat{\triangle }_{\star }.
\end{equation*}%
Such a s-operator can be generalized on regular functionals using formulas (\ref{aux04a}), 
$ _{\eta }^{\shortmid }\widehat{\mathcal{\aleph }}^{\star }=~_{\eta
}^{\shortmid }r_{\lambda V}^{\star -1}\circ ~_{\eta }^{\shortmid }\widehat{%
\mathcal{\aleph }}_{0}^{\star }~\circ ~_{\eta }^{\shortmid }r_{\lambda
V}^{\star }$.  
This allows to compute in parametric local form (assuming nQME): 
\begin{eqnarray*}
\ ~_{\eta }^{\shortmid }\widehat{\mathcal{\aleph }}^{\star }\ ^{\shortmid
}F^{\star } &=&e_{\ ^{\shortmid }\mathcal{T}}^{i\ :\ ~_{\eta }^{\shortmid
}V^{\star }:/~\hbar }\cdot _{\mathcal{T}^{\star }}\ _{\eta }^{\shortmid }%
\mathcal{\aleph }_{0}^{\star }(e_{\ ^{\shortmid }\mathcal{T}}^{i~:\ ~_{\eta
}^{\shortmid }V^{\star }:/~\hbar }\cdot _{\mathcal{T}^{\star }}:\
^{\shortmid }F^{\star }:) \\
&=&\{\ \ ^{\shortmid }F^{\star },\ ~_{\eta }^{\shortmid }S_{0}^{\star }+\
~_{\eta }^{\shortmid }V^{\star }\}-i\hbar \ ~_{\eta }^{\shortmid }\widehat{%
\triangle }_{\star }(\ \ ^{\shortmid }F^{\star })=~_{\eta }^{\shortmid }%
\mathcal{\aleph }_{0}^{\star }-i\hbar \ ~_{\eta }^{\shortmid }\widehat{%
\triangle }_{\star }(\ \ ^{\shortmid }F^{\star }).
\end{eqnarray*}%
The functional character of such nonassociative quantum BV s-operators and
and nQME allows us to define various types of non-perturbative classical and
quantum deformations determined by respective generating data encoding
nonassociative R-flux deformations. 

\vskip4pt In above quantum formulas, we can chose nonholonomic s-adapted
distributions to separate the terms proportional to $\kappa ^{0}$ and $%
\kappa ^{1},$ when the $\kappa -$linear ones encode nonassociative data in
effective quantum form. So, for phase space R-flux deformed Dymnikova BHs,
one holds such a \textit{nonassociative generalization of the anomalous
Master Ward Identity:}%
\begin{equation}
\ \ ~_{\eta }^{\shortmid }\mathcal{\aleph }_{0}^{\star }(e_{\ ^{\shortmid }%
\mathcal{T}_{r}}^{i\ :\ \ ~_{\eta }^{\shortmid }V^{\star }:/~\hbar })\equiv
\{e_{\ ^{\shortmid }\mathcal{T}_{r}}^{i\ :\ \ ~_{\eta }^{\shortmid }V^{\star
}:/~\hbar },\ \ ~_{\eta }^{\shortmid }S_{0}^{\star }\}=\frac{i}{\hbar }e_{\
^{\shortmid }\mathcal{T}_{r}}^{i\ :\ \ ~_{\eta }^{\shortmid }V^{\star
}:/~\hbar }\cdot _{\mathcal{T}_{r}^{\star }}\left( \frac{1}{2}\{\ ~_{\eta
}^{\shortmid }S_{0}^{\star }+~_{\eta }^{\shortmid }V^{\star },~_{\eta
}^{\shortmid }S_{0}^{\star }+\ ~_{\eta }^{\shortmid }V^{\star }\}_{\
^{\shortmid }\mathcal{T}_{r}}-i\hbar ~_{\eta }^{\shortmid }\widehat{%
\triangle }_{\star }^{V}\right) .  \label{amwibh}
\end{equation}%
The s-operator $~_{\eta }^{\shortmid }\widehat{\triangle }_{\star }^{V}$ is
identified with the anomaly term.

\vskip4pt Then, \textit{the renormalized quantum nonassociative master
equation} corresponding to (\ref{aux03b}), (\ref{aux04b}) and (\ref{amwibh})
are%
\begin{equation*}
\frac{1}{2}\{\ ~_{\eta }^{\shortmid }S_{0}^{\star }+ \ _{\eta }^{\shortmid
}V^{\star },\ _{\eta }^{\shortmid }S_{0}^{\star }+\ _{\eta }^{\shortmid
}V^{\star }\}_{\ ^{\shortmid }\mathcal{T}_{r}}- i\hbar \ _{\eta }^{\shortmid
}\widehat{\triangle }_{\star }^{V}=0.
\end{equation*}%
Assuming the renormalized nQME hold, we can write such a master equation
using the renormalized nonassociative BV s-operator from (\ref{aux03d}), 
\begin{equation*}
\ \ ~_{\eta }^{\shortmid }\widehat{\mathcal{\aleph }}^{\star }\ ^{\shortmid
}F^{\star }=\{\ ^{\shortmid }F^{\star },~_{\eta }^{\shortmid }S_{0}^{\star
}+\ ~_{\eta }^{\shortmid }V^{\star }\}-i\hbar \ \ ~_{\eta }^{\shortmid }%
\widehat{\triangle }_{\star }^{V}(\ ^{\shortmid }F^{\star }).
\end{equation*}%
In this subsection, we use the same renormalized time ordered product $\cdot
_{\mathcal{T}_{r}^{\star }}$ as we computed the anomaly in subsection \ref%
{ss423} but considering $\ ~_{\eta }^{\shortmid }\widehat{\triangle }%
_{\star }^{V}(\ ^{\shortmid }F^{\star })$ instead of $\ _{\eta}^{\shortmid }%
\widehat{\triangle }_{\star }(\ ^{\shortmid }F^{\star }).$

The above formulas allow us to construct quantum versions of nonassociative BHs and analyze possible implications both in perturbative and non-perturbative forms. For instance, we conclude that nonassociative R-flux contributions transform certain classical 4-d BH configurations (for instance, the Dymnikova nonsingular metric) into 8-d phase space quasi-stationary configurations. For small parametric deformations, such extra-dimension quantum BH solutions describe  new QG phenomena defined by the renormalized nonassociative BV s-operator. Such effects may exist also in associative and commutative forms for ellipsoidal phase space configurations but described in a different form by nonholonomic BV operators. So, nonassociativity  "enrich" the landscape of QG and various nonassociative and locally anisotropic effects with quantum and quantum gravitational polarizations can be computed as additional off-diagonal deformations of BH metrics.    

\section{Conclusions and perspectives}

\label{sec05} In this paper, we have focused on conceptual problems and elaborating new geometric methods using the Batalin-Vilkovisky, BV, formalism \cite{brst75,brst76,tyu94,barv18,bv81} for quantizing nonassociative and noncommutative gauge gravity theories \cite{partner07,svnc00,sv00}. Such theories are defined by twisted star products \cite{drinf89,connes97} and R-flux parametric deformations considered in string and nonassociative gravity \cite{blumenhagen16,aschieri17}. Corresponding quantum gravity, QG, models encode  nonassociative and noncommutative data for generic off-diagonal solutions of modified gravitational Yang-Mills, YM, or nonassociative Einstein-Dirac-Maxwell, EDM, equations   \cite{partner02,partner04,partner05,partner06,partner07, partner08}. For projections on phase spaces (modelled as cotangent Lorentz bundles), such solutions describe  nonassociative classical EDM and YM quasi-stationary systems, or locally anisotropic cosmological models. The main goal of this work is to elaborate on quantum models of nonassociative geometric flows, gravity and matter field theories. Quantization of such models is performed using generalized BV schemes and advanced mathematic methods from the algebraic quantum field theory, AQFT, \cite{rejzner20,rejzner16,rejzner14,brunetti22,brunetti09,fr12,fr13,df03,hr20,bf20}. This paper  is the first one in the literature which connects general classes of parametric off-diagonal solutions to QG and BV quantization of nonassociative MGTs. It is a natural development of a recent author’s  work \cite{partner09} on nonassociative QG with Gorrof-Sagnotti terms. In that paper, rigorous mathematical methods of nonassociative BV quantization have  not been considered. 

\vskip4pt Nonassociative star products and R-flux deformations substantially modify the classical general relativity, GR, and  modified gravity theories, MGTs, by   introducing nonlocal configurations on nonassociative phase spaces. New types of star-deformed geometric objects such as symmetric and nonsymmetric metrics, nonlinear and linear connections, defining various types of differential and integral calculi and non-unique variational procedures are also defined. Such geometric constructions become effective local for  parametric decompositions on the Planck and string constants which allows us to extend the principles of locality, deformation and homology on nonassociative phase spaces from \cite{rejzner20,brunetti22}. The first important result of our work consisted in a natural generalization of the classical BV formalism in  the abstract geometric form \cite{misner,partner02,partner07} as an effective gauge gravity theory on phase spaces. For such models, a double  de Sitter structure group is used   in a form encoding consequent nonlinear extensions of the (double) affine structure group  and the Poincar\'{e} group. The first structure group is for the base spacetime Lorentz manifold and the second one is for a typical fiber.

\vskip4pt The second main result of this paper consisted in elaborating on nonholonomic frame geometric  methods for quantizing physically important nonlinear systems of partial differential equations, PDEs, in
GR and MGTs. The constructions are related via corresponding classes of generic off-diagonal solutions to  perturbative and non-perturbative BV schemes with effective actions and Lagrangians encoding nonassociative data from string theory. Such  nonassociative nonlinear systems of PDEs are characterized by certain types of nonlinear symmetries which   are different from the prescribed  gauge type symmetries. The nonlinear symmetries allow us to introduce effective cosmological constants and state certain well-defined physical conditions for applying the BV-scheme, with possible linearizations and  definition of nonassociative classical BV operators and related master equations etc.

\vskip4pt The third main result of this article was in relating the BV-scheme to the quantization of nonassociative gauge de Sitter gravity and analyzing the corresponding renormalization problems. That allowed to  derive nonassociative versions of quantum BV operators and quantum master equations. As an example, we have shown how our generalization of the methods of  BV quantization can be applied for quantizing in a non-perturbative way nonassociative 8-d modified BH solutions on phase spaces. That can be also considered as the fourth important result.  We note that in our approach "non-perturbative" means that the scheme of quantization  works for any type of prime or target (off-diagonal) metrics. It  can be further performed in a perturbative way for respective parametric decompositions on physically important parameters by choosing respective classes of generating functions and generating data, and prescribed effective cosmological constants which result in effective asymptotic renormalizable theories.  

\vskip4pt
Finally, we outline four perspectives (P1-P4) for developing the results of the work:

\begin{itemize}
\item P1:\ To elaborate on perturbation algebraic QFT, pAQFT, methods on phase spaces which will result in asymptotic safe (nonassociative) QG models with star R-flux products, Goroff-Sagnotti counter-terms \cite{partner09}, and renormalization flows.

\item P2:\ To extend the BV quantization and pAQFT methods for quantizing nonassociative Einstein-Dirac-YM-Higgs systems studied in \cite{partner08,partner07}. 

\item P3:\ To generalize and apply the BV and pAQFT methods for quantizing locally anisotropic (nonassociative) cosmological models with nonholonomic variables (\ref{cosmvariables}) and study  quantum effects of off-diagonal solutions in MGTs with nonassociative, nonmetric, generalized Finsler-Lagrange-Hamilton configurations \cite{hlfin,vacaru07,vacaru13} modelled on phase spaces (i.e. on (co) tangent Lorentz bundles and their various possible star product deformations and respective models of deformation quantization).

\item P4:\ To elaborate on  nonassociative geometric and quantum information flow theories and BV quantization of generalized G. Perelman thermodynamic models (extending  the results from \cite{perelman1,partner04,partner05,partner06,partner07,partner08,partner09} for AQFT and MGTs).
\end{itemize}

The author plans to report on progress P1-P4  in his (and co-authors') future works.

\vskip5pt
 {\bf Acknowledgement:}\  The author thanks for kind collaboration  his co-authors (L.   Bubuianu,    J. O. Seti, D.  Singleton, P. Stavrinos and  E.  V. Veliev) of partner works \cite{partner02,partner04,partner05,partner06} on classical nonassociative geometric flows,  gravity and MGTs.   The "nonassociative gravity and strings" research program were supported by  Prof. Douglas Alexander Singleton, as a  host of a Fulbright visit to the USA, and by Prof. Dieter L\"{u}st, as a host of a scientist at risk fellowship at CAS LMU, Munich, Germany.   This work is performed in the framework of  author's volunteer research program in Ukraine; it is  devoted to  generalizations and applications of the BV formalism for nonassociative and noncommutative  QG and classical matter fields, and QFT theories,  elaborated in \cite{svnc00,partner07,partner08,partner09}.  

\appendix

\setcounter{equation}{0} \renewcommand{\theequation}
{A.\arabic{equation}} \setcounter{subsection}{0} 
\renewcommand{\thesubsection}
{A.\arabic{subsection}}

\section{Off-diagonal quasi-stationary solutions on nonassociative phase
spaces}

\label{appendixa} In \cite{partner02,partner04,partner06}, the anholonomic
frame and connection deformation method, AFCDM, was developed for decoupling
and integration in general off-diagonal form various physically important
systems of nonlinear systems of PDE in (nonassociative) geometric flow and
MGTs. There were provided various examples and applications involving
nonassociative black holes, BH, wormholes, WH, and locally anisotropic
solitonic cosmological solutions were considered in \cite{partner07}. Here,
we show how applying the AFCDM we can generate quasi-stationary solutions
for nonassociative YM equations (\ref{nonassocymgreq1}). An example of
regular phase space BH off-diagonal solutions is also provided.

\subsection{Gravitational polarizations and nonlinear symmetries}

The techniques of constructing quasi-stationary solutions for nonassociative
modified Einstein equations (\ref{nonassocaneinst}) is outlined in Appendix
B2, with constructions related to formula (B2), to \cite{partner02}. Here we
provide the formulas for a class of solution defined by nonlinear quadratic
element in terms of $\eta $-polarization functions used in (\ref{ans1qs}): 
\begin{eqnarray}
d\ \ ^{\shortmid }\widehat{s}^{2}(\tau ) &=&\ \ ^{\shortmid }g_{\alpha
_{s}\beta _{s}}(\hbar ,\kappa ,\tau ,x^{k},y^{3},p_{a_{3}},p_{a_{4}};\
^{\shortmid }\mathring{g}_{\alpha _{s}};\eta _{4},\ ^{\shortmid }\eta ^{6},\
^{\shortmid }\eta ^{8};~_{s}^{\shortmid }\mathcal{J}^{\star })d~\
^{\shortmid }u^{\alpha _{s}}d~\ ^{\shortmid }u^{\beta _{s}}
\label{offdiagpolfr} \\
&=&e^{\psi )}[(dx^{1})^{2}+(dx^{2})^{2}]-\frac{[\partial _{3}(\eta _{4}%
\mathring{g}_{4})]^{2}}{|\int dy^{3}~_{2}\mathcal{J}^{\star }\partial
_{3}(\eta _{4}\mathring{g}_{4})|\ (\eta _{4}\mathring{g}_{4})}\{dy^{3}+\frac{%
\partial _{i_{1}}[\int dy^{3}\ _{2}\mathcal{J}^{\star }\partial _{3}(\eta
_{4}\mathring{g}_{4})]}{~_{2}\mathcal{J}^{\star }(\tau )\partial _{3}(\eta
_{4})\mathring{g}_{4}}dx^{i_{1}}\}^{2}  \notag \\
&&+\eta _{4}\mathring{g}_{4})\{dt+[\ _{1}n_{k_{1}}+\ _{2}n_{k_{1}}\int dy^{3}%
\frac{[\partial _{3}(\eta _{4}\mathring{g}_{4})]^{2}}{|\int dy^{3}\ _{2}%
\mathcal{J}^{\star }\partial _{3}(\eta _{4}\mathring{g}_{4})|\ (\eta _{4}%
\mathring{g}_{4})^{5/2}}]dx^{k_{1}}\}  \notag
\end{eqnarray}%
\begin{eqnarray*}
&&-\frac{[\ ^{\shortmid }\partial ^{5}(\ ^{\shortmid }\eta ^{6}\ ^{\shortmid
}\mathring{g}^{6})]^{2}}{|\int dp_{5}~_{3}^{\shortmid }\mathcal{J}^{\star }\
^{\shortmid }\partial ^{5}(\ ^{\shortmid }\eta ^{6}\ \ ^{\shortmid }%
\mathring{g}^{6})\ |\ (\ ^{\shortmid }\eta ^{6}\ ^{\shortmid }\mathring{g}%
^{6})}\{dp_{5}+\frac{\ \ ^{\shortmid }\partial _{i_{2}}[\int dp_{5}\
_{3}^{\shortmid }\mathcal{J}^{\star }\ ^{\shortmid }\partial ^{5}(\
^{\shortmid }\eta ^{6}\ ^{\shortmid }\mathring{g}^{6})]}{~_{3}^{\shortmid }%
\mathcal{J}^{\star }\ \ ^{\shortmid }\partial ^{5}(\ ^{\shortmid }\eta ^{6}\
^{\shortmid }\mathring{g}^{6})}dx^{i_{2}}\}^{2} \\
&&+(\ ^{\shortmid }\eta ^{6}\ ^{\shortmid }\mathring{g}^{6})\{dp_{6}+[\
_{1}^{\shortmid }n_{k_{2}}+\ _{2}^{\shortmid }n_{k_{2}}\int dp_{5}\frac{[\
^{\shortmid }\partial ^{5}(\ ^{\shortmid }\eta ^{6}\ ^{\shortmid }\mathring{g%
}^{6})]^{2}}{|\int dp_{5}~_{3}^{\shortmid }\mathcal{J}^{\star }\ \partial
^{5}(\ ^{\shortmid }\eta ^{6}\ ^{\shortmid }\mathring{g}^{6})|\ (\
^{\shortmid }\eta ^{6}\ ^{\shortmid }\mathring{g}^{6})^{5/2}}]dx^{k_{2}}\}
\end{eqnarray*}%
\begin{eqnarray*}
&&-\frac{[\ ^{\shortmid }\partial ^{7}(\ ^{\shortmid }\eta ^{8}\ ^{\shortmid
}\mathring{g}^{8})]^{2}}{|\int dp_{7}~_{4}^{\shortmid }\mathcal{J}^{\star }\
^{\shortmid }\partial ^{8}(\ ^{\shortmid }\eta ^{7}\ ^{\shortmid }\mathring{g%
}^{7})\ |\ (\ ^{\shortmid }\eta ^{7}\ ^{\shortmid }\mathring{g}^{7})}%
\{dp_{7}+\frac{~\ ^{\shortmid }\partial _{i_{3}}[\int dp_{7}~_{4}^{\shortmid
}\mathcal{J}^{\star }\ ^{\shortmid }\partial ^{7}(\ ^{\shortmid }\eta ^{8}\
^{\shortmid }\mathring{g}^{8})]}{~_{4}^{\shortmid }\mathcal{J}^{\star }\
^{\shortmid }\partial ^{7}(\ ^{\shortmid }\eta ^{8}\ ^{\shortmid }\mathring{g%
}^{8})}d\ ^{\shortmid }x^{i_{3}}\}^{2} \\
&&+(\ ^{\shortmid }\eta ^{8}\ ^{\shortmid }\mathring{g}^{8})\{dE+[\
_{1}n_{k_{3}}+\ _{2}n_{k_{3}}\int dp_{7}\frac{[\ ^{\shortmid }\partial
^{7}(\ ^{\shortmid }\eta ^{8}\ ^{\shortmid }\mathring{g}^{8})]^{2}}{|\int
dp_{7}~_{4}^{\shortmid }\mathcal{J}^{\star }[\ ^{\shortmid }\partial ^{7}(\
^{\shortmid }\eta ^{8}\ ^{\shortmid }\mathring{g}^{8})]|\ [(\ ^{\shortmid
}\eta ^{8}\ ^{\shortmid }\mathring{g}^{8})]^{5/2}}]d\ ^{\shortmid
}x^{k_{3}}\}.
\end{eqnarray*}%
The formulas (\ref{offdiagpolfr}) describe nonholonomic off-diagonal
deformations of a prescribed prime metric into other families of target
ones, $\ _{s}^{\shortmid }\mathbf{\mathring{g}}=[\ ^{\shortmid }\mathring{g}%
_{\alpha _{s}},\ ^{\shortmid }\mathring{N}_{i_{s-1}}^{a_{s}}]\rightarrow \
_{s}^{\shortmid }\mathbf{g}$ (\ref{offdiagdefr}). Certain $\eta $%
-polarizations (involving a $\psi $ as a solution of the Poisson equations,
see bellow formula (\ref{poas2d})), can be used as generating functions%
\begin{equation}
\psi \simeq \psi (\hbar ,\kappa ,x^{k_{1}}),\eta _{4}\ \simeq \eta
_{4}(\hbar ,\kappa ,x^{k_{1}},y^{3}),\ ^{\shortmid }\eta ^{6}\simeq \
^{\shortmid }\eta ^{6}(\hbar ,\kappa ,x^{i_{2}},p_{5}),\ ^{\shortmid }\eta
^{8}\simeq \ ^{\shortmid }\eta ^{8}(\hbar ,\kappa ,x^{i_{2}},p_{5},p_{7}).
\label{etapolgen}
\end{equation}%
Here we note that the generating functions can be prescribed in certain
forms which allow to generate "small" (for instance, on parameter $\kappa )$
off-diagonal deformations of some prime s-metrics into target ones. For such
configurations, we have to choose (\ref{etapolgen}) in such forms when $\
_{s}^{\shortmid }\eta \ ^{\shortmid }\mathring{g}_{\alpha _{s}}\sim \
^{\shortmid }\zeta _{\alpha _{s}}(1+\kappa \ ^{\shortmid }\chi _{\alpha
_{s}})\ \ ^{\shortmid }\mathring{g}_{\alpha _{s}}.$ Such solutions were
studied in \cite{partner02,partner04,partner06}, for instance, with the aim
to construct deformations of BH solutions into nonassociative black
ellipsoid, BE, ones etc.

\vskip4pt Off-diagonal solutions of type (\ref{offdiagpolfr}) \ posses
important nonlinear symmetries which allow to change the generating
functions and generating sources into another types of generating functions
and effective cosmological constants on each shell, $\ _{s}^{\shortmid
}\Lambda ,[\eta _{4},\ ^{\shortmid }\eta ^{6},\ ^{\shortmid }\eta ^{8},\
_{s}^{\shortmid }\mathcal{J}^{\star }]\longleftrightarrow \lbrack \ _{s}\Psi
,\ _{s}^{\shortmid }\mathcal{J}^{\star }]\longleftrightarrow \lbrack \
_{s}\Phi ^{\star },\ _{s}^{\shortmid }\Lambda ].$ We put a star label on $%
\Phi $ to emphasize that such generating functions "absorb" nonlinearly the
nonassociative data encoded in parametric forms in $\ _{s}^{\shortmid }%
\mathcal{J}^{\star }.$ By straightforward computations, we can check that
such nonlinear transforms keep invariant the quasi-stationary configurations
if different types of generating data are related by such differential, or
integral, formulas: 
\begin{eqnarray}
\partial _{3}[(\ _{2}\Psi )^{2}] &=&-\int dy^{3}(~_{2}^{\shortmid }\mathcal{J%
}^{\star })\partial _{3}g_{4}\simeq -\int dy^{3}(~_{2}^{\shortmid }\mathcal{J%
}^{\star })\partial _{3}(\ ^{\shortmid }\eta _{4}\ \mathring{g}_{4})
\label{nonlinsym} \\
&\simeq &-\int dy^{3}(~_{2}^{\shortmid }\mathcal{J}^{\star })\partial _{3}[\
^{\shortmid }\zeta _{4}(1+\kappa \ ^{\shortmid }\chi _{4})\ \mathring{g}%
_{4}],  \notag \\
(\ _{2}\Phi ^{\star })^{2} &=&-4\ _{2}\Lambda g_{4}\simeq -4\ _{2}\Lambda \
^{\shortmid }\eta _{4}\ \mathring{g}_{4}\simeq -4\ _{2}\Lambda \ ^{\shortmid
}\zeta _{4}(1+\kappa \ ^{\shortmid }\chi _{4})\ \mathring{g}_{4};  \notag
\end{eqnarray}%
\begin{eqnarray*}
~\ ^{\shortmid }\partial ^{5}[(\ _{3}^{\shortmid }\Psi )^{2}] &=&-\int
dp_{5}(~_{3}^{\shortmid }\mathcal{J}^{\star })\ ^{\shortmid }\partial ^{5}\
^{\shortmid }g^{6}\simeq -\int dp_{5}(~_{3}^{\shortmid }\mathcal{J}^{\star
})\ ^{\shortmid }\partial ^{5}(\ ^{\shortmid }\eta ^{6}\ ^{\shortmid }%
\mathring{g}^{6}) \\
&\simeq &-\int dp_{5}(~_{3}^{\shortmid }\mathcal{J}^{\star })\ ^{\shortmid
}\partial ^{5}[\ ^{\shortmid }\zeta ^{6}(1+\kappa \ ^{\shortmid }\chi ^{6})\ 
\mathring{g}^{6}], \\
(\ _{3}^{\shortmid }\Phi ^{\star })^{2} &=&-4\ _{3}^{\shortmid }\Lambda \
^{\shortmid }g^{6}\simeq \ -4\ _{3}^{\shortmid }\Lambda \ ^{\shortmid }\eta
^{6}(\tau )\ ^{\shortmid }\mathring{g}^{6}\simeq -4\ _{3}^{\shortmid
}\Lambda \ ^{\shortmid }\zeta ^{6}(1+\kappa \ ^{\shortmid }\chi ^{6})\
^{\shortmid }\mathring{g}^{6};
\end{eqnarray*}%
\begin{eqnarray*}
~\ ^{\shortmid }\partial ^{7}[(\ _{4}^{\shortmid }\Psi )^{2}] &=&-\int
dp_{7}(~_{4}^{\shortmid }\mathcal{J}^{\star })\ ^{\shortmid }\partial ^{7}\
^{\shortmid }g^{8}\simeq -\int dp_{7}(~_{4}^{\shortmid }\mathcal{J}^{\star
})\ ^{\shortmid }\partial ^{7}(\ ^{\shortmid }\eta ^{8}\ ^{\shortmid }%
\mathring{g}^{8}) \\
&\simeq &-\int dp_{7}(~_{4}^{\shortmid }\mathcal{J}^{\star }(\tau ))\
^{\shortmid }\partial ^{7}[\ ^{\shortmid }\zeta ^{8}(1+\kappa \ ^{\shortmid
}\chi ^{8})\ \mathring{g}^{8}], \\
(\ _{4}^{\shortmid }\Phi ^{\star })^{2} &=&-4\ _{4}^{\shortmid }\Lambda \
^{\shortmid }g^{8}\simeq \ -4\ _{4}^{\shortmid }\Lambda \ ^{\shortmid }\eta
^{8}\ ^{\shortmid }\mathring{g}^{8}\simeq -4\ _{4}^{\shortmid }\Lambda \
^{\shortmid }\zeta ^{8}(1+\kappa \ ^{\shortmid }\chi ^{8})\ ^{\shortmid }%
\mathring{g}^{8}.
\end{eqnarray*}%
The nonlinear symmetries (\ref{nonlinsym}) allows us to re-write the
nonassociative equations $\ ^{\shortmid }\widehat{\mathbf{R}}_{\ \ \gamma
_{s}}^{\star \beta _{s}}(\ _{s}\Psi )={\delta }_{\ \ \gamma _{s}}^{\beta
_{s}}\ _{s}^{\shortmid }\mathcal{J}^{\star }\ $ (\ref{nonassocaneinst}) into
a system of nonlinear functional parametric equations with effective
cosmological constants: 
\begin{equation}
\ ^{\shortmid }\widehat{\mathbf{R}}_{\ \ \gamma _{s}}^{\beta _{s}}(\
_{s}\Phi ,\ _{s}^{\shortmid }\mathcal{J}^{\star })={\delta }_{\ \ \gamma
_{s}}^{\beta _{s}}\ _{s}^{\shortmid }\Lambda .  \label{nonassocgeomflefcc}
\end{equation}%
For (\ref{nonassocgeomflefcc}), the solution involve effective cosmological
constants $\ _{s}^{\shortmid }\Lambda $ which are important for computing G.
Perelman variables and elaborating (nonassociative) quantum geometric and
information flow theories \cite{partner04,partner06}. Here we note that it
is not possible completely to transform $\ _{s}^{\shortmid }\mathcal{J}%
^{\star }\rightarrow \ _{s}^{\shortmid }\Lambda $ because $\ _{s}^{\shortmid
}\mathcal{J}^{\star }$ is always present in certain coefficients of the
s-metric and N-connection, but in parametric form it is possible explicitly
to find solutions for models with effective cosmological constants.

Quasi-stationary solutions (\ref{offdiagpolfr}) are characterized by such
classes generating and integration functions related via nonlinear
symmetries to generating sources and effective cosmological constants: 
\begin{eqnarray}
\mbox{generating functions: } &&\psi \simeq \psi (\hbar ,\kappa
,x^{k_{1}});\ _{2}\Psi \simeq \ _{2}\Psi (\hbar ,\kappa ,x^{k_{1}},y^{3});
\label{integrfunctrf} \\
&&\ _{3}^{\shortmid }\Psi \simeq \ _{3}^{\shortmid }\Psi (\hbar ,\kappa
,x^{k_{2}},p_{5});\ _{4}^{\shortmid }\Psi \simeq \ _{4}^{\shortmid }\Psi
(\hbar ,\kappa ,\ ^{\shortmid }x^{k_{3}},p_{7});  \notag \\
\mbox{generating sources: \quad} &&\ _{1}^{\shortmid }\mathcal{J}^{\star
}\simeq \ ~_{1}^{\shortmid }\mathcal{J}^{\star }(\hbar ,\kappa ,x^{k_{1}});\
~_{2}^{\shortmid }\mathcal{J}^{\star }\simeq \ ~_{2}^{\shortmid }\mathcal{J}%
^{\star }(\hbar ,\kappa ,x^{k_{1}},y^{3});  \notag \\
&&\ _{3}^{\shortmid }\mathcal{J}^{\star }\simeq \ ~_{3}^{\shortmid }\mathcal{%
J}^{\star }(\hbar ,\kappa ,x^{k_{2}},p_{5});~_{4}^{\shortmid }\mathcal{J}%
^{\star }\simeq \ ~~_{4}^{\shortmid }\mathcal{J}^{\star }(\hbar ,\kappa ,\
^{\shortmid }x^{k_{3}},p_{7});  \notag \\
\mbox{integrating functions: }\ \ g_{4}^{[0]} &\simeq &g_{4}^{[0]}(\hbar
,\kappa ,x^{k_{1}}),\ _{1}n_{k_{1}}\simeq \ _{1}n_{k_{1}}(\hbar ,\kappa
,x^{j_{1}}),\ _{2}n_{k_{1}}\simeq \ _{2}n_{k_{1}}(\hbar ,\kappa ,x^{j_{1}});
\notag \\
\ ^{\shortmid }g_{[0]}^{6} &\simeq &\ ^{\shortmid }g_{[0]}^{6}(\hbar ,\kappa
,x^{k_{2}}),\ _{1}n_{k_{2}}\simeq \ _{1}n_{k_{2}}(\hbar ,\kappa
,x^{j_{2}}),\ _{2}n_{k_{2}}\simeq \ _{2}n_{k_{2}}(\hbar ,\kappa ,x^{j_{2}});
\notag \\
\ ^{\shortmid }g_{[0]}^{8} &\simeq &\ ^{\shortmid }g_{[0]}^{8}(\hbar ,\kappa
,\ ^{\shortmid }x^{j_{3}}),\ _{1}n_{k_{3}}\simeq \ _{1}^{\shortmid
}n_{k_{3}}(\hbar ,\kappa ,\ ^{\shortmid }x^{j_{3}}),\ _{2}n_{k_{3}}\simeq \
_{2}^{\shortmid }n_{k_{3}}(\hbar ,\kappa ,\ ^{\shortmid }x^{j_{3}}).  \notag
\end{eqnarray}%
The generating functions $\psi (\hbar ,\kappa ,x^{k_{1}})$ are solutions of
the 2-d Poisson equations, 
\begin{equation}
\partial _{11}^{2}\psi +\partial _{22}^{2}\psi =2\ _{1}\mathcal{J}^{\star
}(\hbar ,\kappa ,x^{k_{1}}).  \label{poas2d}
\end{equation}%
For certain subclasses of generating and integration data (\ref%
{integrfunctrf}) and (\ref{poas2d}), we can generate BH or WH solutions with
polarization of physical constants, deformation of horizons (if they exist)
and embedding in nonassociative gravitational vacuum, or subjected to
nonassociative off-diagonal interactions with effective matter fields.

\vskip4pt The quasi-stationary solutions (\ref{offdiagpolfr}) can be
transformed into locally anisotropic cosmological solutions encoding in
off-diagonal form nonassociative data. This is possible if we perform corresponding
re-definitions of nonholonomic variables (and respective coordinate
dependencies) on shells $s=1$ and 2:%
\begin{eqnarray}
\partial _{4} &=&\partial _{t}\rightarrow \partial _{3}=\partial /\partial
y^{3},\mbox{ Killing symmetry };x^{3}=y^{3}\rightarrow x^{4}=y^{4}=t,%
\mbox{
space into time coordinates};  \notag \\
g_{3}(x^{i},y^{3}) &\rightarrow &\underline{g}_{4}(x^{i},t)\mbox{ and }%
g_{4}(x^{i},y^{3})\rightarrow \underline{g}_{3}(x^{i},y);
\label{cosmvariables} \\
N_{k}^{3}(x^{i},y^{3}) &=&w_{k}(x^{i},y^{3})\rightarrow \underline{N}%
_{k}^{4}(x^{i},t)=\underline{n}_{k}(x^{i},t)\mbox{ and }%
N_{k}^{4}(x^{i},y^{3})=n_{k}(x^{i},y^{3})\rightarrow \underline{N}%
_{k}^{3}(x^{i},t)=\underline{w}_{k}(x^{i},t).  \notag
\end{eqnarray}%
In above formulas, we underlined the symbols with explicit dependence on $t$%
-coordinate. Such re-definitions of nonholonomic variables introduce time like dependencies into above formulas (\ref{etapolgen}) - (\ref%
{integrfunctrf}). Such nonholonomic parameterizations in nonassociative
gauge gravity were used for deriving parametric solutions with cosmological
solitonic hierarchies \cite{partner07}. We can consider co-fiber
redefinition of nonholonomic variables like (\ref{cosmvariables}), to
transforms off-diagonal solutions with $p_{8}=E_{0}$ into "rainbow"
s-metrics with variable $E,$ but, for instance, $p_{7}=const.$ In \cite%
{partner02,partner04,partner06}, there are studied examples of phase space
parametric solutions encoding nonassociative data.

\vskip5pt In this work, we restrict our approach only for quasi-stationary
parametric configurations (\ref{offdiagpolfr}). Such results
can be re-defined geometrically in certain dual forms (on time and energy type
coordinates) which positively result in another type of locally anisotropic
cosmological and nonassociative gravitationals models.

\subsection{Nonassociative star product deformations of regular Dymnikova
black holes}

In GR, a very interesting BH regular solution was constructed by I.
Dymnikova \cite{dymn}. For recent higher dimension constructions, we cite 
\cite{macedo} and references therein. In abstract geometric form, those
constructions can be redefined in (nonassociative) nonholonomic forms on 8-d
phase spaces $\ _{s}\mathcal{M}$. Similar details on applications of the
AFCDM we provided in \cite{partner05} for quasi-stationary solutions
describing nonassociative star product deformations of a a $d=5$ dimensional
analog of the Reisner-Nordstr\"{o}m AdS, RN. In this section, we consider a
different type of primary metric using phase space coordinates on a 7-d
phase space with signature $(+++-+++)$ trivially extended to a 8-d one with
a diagonal quadratic element of the prime metric, we considered 
\begin{equation}
d\ \breve{s}_{[7+1]}^{2}=\ ^{\shortmid }\breve{g}_{\alpha _{s}}(\
^{\shortmid }u^{\gamma _{s}})(\mathbf{\breve{e}}^{\alpha _{s}})^{2}=\frac{d%
\breve{r}^{2}}{\breve{f}(\breve{r})}-\breve{f}(\breve{r})dt^{2}+\breve{r}%
^{2}[(d^{2}\hat{x}^{2})^{2}+(d\hat{x}%
^{3})^{2}+(dp_{5})^{5}+(dp_{6})^{2}+(dp_{7})^{2}]-dE^{2}.  \label{pm7d8d}
\end{equation}%
In this formula, the spherical coordinates are for $\hat{x}^{1}=\breve{r}=%
\sqrt{%
(x^{1})^{2}+(x^{2})^{2}+(x^{3})^{2}+(p_{5})^{2}+(p_{6})^{2}+(p_{7})^{2},}$
when $\hat{x}^{2}=\hat{x}^{2}(x^{2},x^{3},p_{5},p_{6},p_{7}),$ $\hat{x}^{3}=%
\hat{x}^{3}(x^{2},x^{3},p_{5},p_{6},p_{7}),...$ $\hat{x}^{7}=\hat{x}%
^{7}(x^{2},x^{3},p_{5},p_{6},p_{7})$ are chosen as coordinates for a
diagonal metric on an effective 7-d Einstein phase space $V_{[7]}.$ For
details and physical motivations, we cite section II of \cite{macedo} (with
that difference that we work with higher dimension coordinates considered as
momentum ones; when the dimension $D=6,\,\ $i.e. $d=6$ following our
conventions). The 7-d phase space generalization of the Dymnikova BH is
given by%
\begin{equation}
\breve{f}(\breve{r})=1-(\frac{\breve{r}_{g}}{\breve{r}})^{4}\{1-\exp [(\frac{%
\breve{r}}{\breve{r}_{\ast }})^{6}]\},\mbox{ for constants }\breve{r}_{\ast
}^{6}=\breve{r}_{0}^{2}\breve{r}_{g}^{4},\mbox{ where }\breve{r}_{0}^{2}=%
\frac{15}{\rho _{0}^{2}};  \label{sourc2bb}
\end{equation}%
and when the nontrivial components of the energy-momentum tensors are 
\begin{equation}
T_{1}^{1}=T_{4}^{4}=-\rho _{0}\exp [-(\breve{r}/\breve{r}_{\ast })^{6}]%
\mbox{  and }T_{2}^{2}=T_{3}^{3}=T_{5}^{5}=T_{6}^{6}=T_{7}^{7}=[\frac{6}{5}(%
\frac{\breve{r}}{\breve{r}_{\ast }})^{6}-1]\exp [-(\breve{r}/\breve{r}_{\ast
})^{6}];  \label{sourc2}
\end{equation}%
being used spherical coordinates of unit 5-d sphere, with constant energy
density $\rho _{0}$ defining the vacuum energy of a phase space.

\vskip4pt The diagonal prime metric (\ref{pm7d8d}) \ define a solution of
the Einstein equations (\ref{nonassocaneinst}) on a commutative and
associative 7-d phase space which yields a de Sitter solution for $\breve{r}%
\ll \breve{r}_{\ast }$ and a higher dimension Schwarzschild solution for $%
\breve{r}\gg \breve{r}_{\ast }.$ Such regular BH solutions can be considered
also for the gauge gravity equations (\ref{ymgrcom}) if the energy-momentum
tensor (\ref{sourc2}) is used for defining the source (\ref{sourcgauge1}).

\vskip4pt The apply the AFCDM and construct nonassociative solutions we
consider certain ( s-adapted coordinate transforms $\ ^{\shortmid }u^{\gamma
_{s}}=\ ^{\shortmid }u^{\gamma _{s}}(\ ^{\shortmid }\hat{u}^{\gamma _{s}})$
of (\ref{pm7d8d}) \ \ into certain data $\ \ _{s}^{\shortmid }\mathbf{\breve{%
g}=\{}^{\shortmid }\breve{g}_{\alpha _{s}}\}=$ $\ _{s}^{\shortmid }\mathbf{%
\mathring{g}}=[\ ^{\shortmid }\mathring{g}_{\alpha _{s}},\ ^{\shortmid }%
\mathring{N}_{i_{s-1}}^{a_{s}}]$ as in (\ref{offdiagpolfr}), when nontrivial
values $\ ^{\shortmid }\mathring{g}_{\alpha _{s}}$ and$\ ^{\shortmid }%
\mathring{N}_{i_{s-1}}^{a_{s}}$ allow to generate nonsingular off-diagonal
solutions. For general star product deformations, it is not clear what
physical interpretation could be provided for such nonassociative
modifications of Dymnikova phase space BH solutions of (\ref{nonassocymgreq1}%
) and (\ref{nonassocaneinst}). In principle, we can assume that certain
stability can be achieved by corresponding nonholonomic constraints on $\eta 
$-polarizations as we considered in section 5.3 of \cite{partner05}. Then,
considering small parametric distortions of type $\ _{s}^{\shortmid }\eta \
^{\shortmid }\mathring{g}_{\alpha _{s}}\sim \ ^{\shortmid }\zeta _{\alpha
_{s}}(1+\kappa \ ^{\shortmid }\chi _{\alpha _{s}})\ ^{\shortmid }\mathring{g}%
_{\alpha _{s}},$ we can model additional locally anisotropic polarization of
the vacuum energy $\rho _{0}$ and respective horizons; and various effective
source parameters encoding nonassociative data. For some nonholonomic
configurations, we can model, for instance, ellipsoidal-type deformations of
horizons and keep a standard interpretation of phase space systems defined
on a regular phase space Dymnikova background, which are $\kappa $-deformed. 
\begin{eqnarray}
d\ _{\shortmid }^{\chi }s_{[7\subset 8d]}^{2} &=&e^{\psi _{0}}(1+\kappa \
^{\psi }\ ^{\shortmid }\chi )[\ \breve{g}_{1}(\breve{r})d\breve{r}^{2}+%
\breve{g}_{2}(\breve{r})(d\hat{x}^{2})]-\{\frac{4[\hat{\partial}_{3}(|\zeta
_{4}\breve{g}_{4}(\breve{r})|^{1/2})]^{2}}{\ \breve{g}_{4}(\breve{r})|\int d%
\hat{x}^{3}\{\ _{2}\mathcal{J}^{\star }\hat{\partial}_{3}(\zeta _{4}\ \breve{%
g}_{4}(\breve{r}))\}|}  \label{sol4of} \\
&&-\kappa \lbrack \frac{\hat{\partial}_{3}(\chi _{4}|\zeta _{4}\ \breve{g}%
_{4}(\breve{r})|^{1/2})}{4\hat{\partial}_{3}(|\zeta _{4}\ \breve{g}_{4}(%
\breve{r})|^{1/2})}-\frac{\int d\hat{x}^{3}\{\ _{2}\mathcal{J}^{\star }\hat{%
\partial}_{3}[(\zeta _{4}\ \breve{g}_{4}(\breve{r}))\chi _{4}]\}}{\int d\hat{%
x}^{3}\{\ _{2}\mathcal{J}^{\star }\hat{\partial}_{3}(\zeta _{4})\ \breve{g}%
_{4}(\breve{r}))\}}]\}\ \breve{g}_{3}(\mathbf{e}^{3})^{2}+\ \zeta
_{4}(1+\kappa \ \chi _{4})\breve{g}_{4}(\breve{r})dt^{2}  \notag
\end{eqnarray}%
\begin{eqnarray*}
&&-\{\frac{4[\hat{\partial}_{5}(|\ ^{\shortmid }\zeta ^{6}\ \breve{g}%
^{6}|^{1/2})]^{2}}{~\breve{g}_{5}(\breve{r})|\int d\hat{x}^{5}\{\
_{3}^{\shortmid }\mathcal{J}^{\star }{}^{\shortmid }\partial ^{7}(\
^{\shortmid }\zeta ^{6}(\tau )~\breve{g}^{6})\}|}-\kappa \lbrack \frac{\ 
\hat{\partial}_{5}(\ ^{\shortmid }\chi ^{6}|\ ^{\shortmid }\zeta ^{6}\ 
\breve{g}^{6}|^{1/2})}{4\hat{\partial}_{5}(|\ ^{\shortmid }\zeta ^{6}\ 
\breve{g}^{6}|^{1/2})}-\frac{\int d\hat{x}^{5}\{\ _{3}^{\shortmid }\mathcal{J%
}^{\star }\ \hat{\partial}_{5}[(\ ^{\shortmid }\zeta ^{6}\breve{g}^{6})\
^{\shortmid }\chi ^{8}]\}}{\int d\hat{x}^{5}\{\ _{3}^{\shortmid }\mathcal{J}%
^{\star }\ \hat{\partial}_{5}[(\ ^{\shortmid }\zeta ^{6}\breve{g}^{6})]\}}%
]\}\ \breve{g}_{5}(\breve{r})(\mathbf{e}^{5})^{2} \\
&&+\ ^{\shortmid }\zeta ^{6}\ (1+\kappa \ ^{\shortmid }\chi
^{6})(dp_{6})^{2}+(dp_{7})^{2}-dE^{2},
\end{eqnarray*}%
where%
\begin{equation*}
\mathbf{e}^{3}=d\hat{x}^{3}+[\frac{\hat{\partial}_{i_{1}}\int d\hat{x}^{3}\
_{2}\mathcal{J}^{\star }\ \hat{\partial}_{3}\zeta _{4}}{\breve{N}%
_{i_{1}}^{3}\ _{2}\mathcal{J}^{\star }\hat{\partial}_{3}\zeta _{4}}+\kappa (%
\frac{\hat{\partial}_{i_{1}}[\int d\hat{x}^{3}\ _{2}\mathcal{J}^{\star }\hat{%
\partial}_{3}(\zeta _{4}\chi _{4})]}{\hat{\partial}_{i_{1}}\ [\int d\hat{x}%
^{3}\ _{2}\mathcal{J}^{\star }\hat{\partial}_{3}\zeta _{4}]}-\frac{\hat{%
\partial}_{3}(\zeta _{4}\chi _{4})}{\hat{\partial}_{3}\zeta _{4}})]\ \breve{N%
}_{i_{1}}^{3}dx^{i_{1}},
\end{equation*}%
\begin{equation*}
\ \ ^{\shortmid }\mathbf{e}^{5}=d\hat{x}^{5}+[\frac{\hat{\partial}_{i_{2}}\
\int d\hat{x}^{5}\ _{3}^{\shortmid }\mathcal{J}^{\star }\ \ \hat{\partial}%
_{5}(\ ^{\shortmid }\zeta ^{6})}{\ ^{\shortmid }\breve{N}_{i_{2}}^{5}\
_{3}^{\shortmid }\mathcal{J}^{\star }\ \ \hat{\partial}_{5}(\ ^{\shortmid
}\zeta ^{6})}+\kappa (\frac{\hat{\partial}_{i_{2}}[\int d\hat{x}^{5}\
_{3}^{\shortmid }\mathcal{J}^{\star }\ \ \hat{\partial}_{5}(\ ^{\shortmid
}\zeta ^{6}\ \ \breve{g}^{6})]}{\hat{\partial}_{i_{2}}\ [\int d\hat{x}^{5}\
_{3}^{\shortmid }\mathcal{J}^{\star }\ \ \hat{\partial}_{5}(\ ^{\shortmid
}\zeta ^{6})]}-\frac{\ \hat{\partial}_{5}(\ ^{\shortmid }\zeta ^{6}\ \breve{g%
}^{6})}{\ \hat{\partial}_{5}(\ ^{\shortmid }\zeta ^{6})})]\ \ ^{\shortmid }%
\breve{N}_{i_{2}}^{5}d\ ^{\shortmid }x^{i_{2}}.
\end{equation*}%
\begin{equation*}
\ \ ^{\shortmid }\mathbf{e}^{7}=d\hat{x}^{7}+[\frac{\hat{\partial}_{i_{3}}\
\int d\hat{x}^{7}\ _{4}^{\shortmid }\mathcal{J}^{\star }\ \ \hat{\partial}%
_{7}(\ ^{\shortmid }\zeta ^{8})}{\ ^{\shortmid }\breve{N}_{i_{3}}^{5}\
_{4}^{\shortmid }\mathcal{J}^{\star }\ \ \hat{\partial}_{7}(\ ^{\shortmid
}\zeta ^{8})}+\kappa (\frac{\hat{\partial}_{i_{3}}[\int d\hat{x}^{7}\
_{4}^{\shortmid }\mathcal{J}^{\star }\ \ \hat{\partial}_{7}(\ ^{\shortmid
}\zeta ^{8}\ \ \breve{g}^{8})]}{\hat{\partial}_{i_{3}}\ [\int d\hat{x}^{7}\
_{4}^{\shortmid }\mathcal{J}^{\star }\ \ \hat{\partial}_{7}(\ ^{\shortmid
}\zeta ^{8})]}-\frac{\ \hat{\partial}_{7}(\ ^{\shortmid }\zeta ^{8}\ \ 
\breve{g}^{8})}{\ \hat{\partial}_{7}(\ ^{\shortmid }\zeta ^{8})})]\ \
^{\shortmid }\breve{N}_{i_{3}}^{7}d\ ^{\shortmid }x^{i_{3}}.
\end{equation*}%
Such solutions are similar to those defined by formulas (95) in \cite%
{partner05}. Nevertheless, in this case they may not involve BH or black ellipsoid 
singularities (because in (\ref{sol4of}) we use regular Dymnikova type
configurations). The nonassociative generating sources $\ _{s}^{\shortmid }%
\mathcal{J}^{\star }$ are different in such cases being defined by star
product deformations of certain primary energy-momentum components (\ref%
{sourc2}).

\end{document}